\documentclass[useAMS,usenatbib, referee]{pse}

\def\bSig\mathbf{\Sigma}

\usepackage{graphicx}
\graphicspath{{./}}
\usepackage{tikz}
\usetikzlibrary{
    decorations.pathmorphing,
    positioning,
    arrows,
    decorations.pathreplacing,
    calc
}
\usepackage{caption}          
\usepackage{subcaption}

\usepackage{hyperref}         
\hypersetup{
    colorlinks=true,          
    linkcolor=blue,           
    citecolor=blue,           
    urlcolor=blue             
}

\usepackage{amsmath, amssymb, amsfonts}
\usepackage{mathtools}
\usepackage{bbm}
\usepackage{commath}

\usepackage{booktabs}
\usepackage{multirow}
\usepackage{multicol}
\usepackage{longtable}
\usepackage{array}
\usepackage{threeparttable}
\usepackage{comment}
\usepackage{xcolor}

\newcommand{\indep}{\perp\!\!\!\perp}

\newcommand{\expit}{\mathrm{expit}}

\newcommand{\logit}{\text{logit}}

\DeclareMathOperator{\pa}{pa}

\newcommand{\hilbert}{\hbox{${\rm I\kern-.2em H}$}}

\newcommand{\openn}{\hbox{${\rm I\kern-.2em N}$}}
\newcommand{\opend}{\hbox{${\rm I\kern-.2em D}$}}

\newcommand{\mt}[1]{\big[\begin{matrix}#1\end{matrix}\big]}
\newcommand\numberthis{\addtocounter{equation}{1}\tag{\theequation}}

\def\P{\mathrm{P}}

\definecolor{babyblue}{rgb}{0.54, 0.81, 0.94}
\newcolumntype{C}[1]{>{ \arraybackslash}p{#1}}

\allowdisplaybreaks
\makeatletter
\newcommand{\pushright}[1]{\ifmeasuring@#1\else\omit\hfill$\displaystyle#1$\fi\ignorespaces}
\newcommand{\pushleft}[1]{\ifmeasuring@#1\else\omit$\displaystyle#1$\hfill\fi\ignorespaces}
\makeatother

\usepackage{titletoc}
\titlecontents{section}
  [0em]{}{\contentslabel{2.3em}}{}
  {\titlerule*[0.5pc]{.}\contentspage}
\titlecontents{subsection}
  [2.3em]{}{\contentslabel{3.2em}}{}
  {\titlerule*[0.5pc]{.}\contentspage}
\hypersetup{hypertexnames=false}
\makeatletter
\let\arxiv@maketitle\maketitle
\let\arxiv@@maketitle\@maketitle
\newcommand{\arxivbeginsupplement}{%
  \clearpage
  \setcounter{page}{1}%
  \@for\arxiv@counter:=section,subsection,subsubsection,paragraph,subparagraph,equation,figure,table,posttable,footnote,example\do{%
    \setcounter{\arxiv@counter}{0}}%
  \let\maketitle\arxiv@maketitle
  \let\@maketitle\arxiv@@maketitle
  \gdef\@thanks{}\gdef\noaffil{yes}%
  \setbox\absbox=\box\voidb@x
  \setbox\keywdbox=\box\voidb@x
}
\makeatother

\title[Inference for Longitudinal Path-Specific Effects]{Inference for Causal Path-Specific Effects in Longitudinal Studies with Repeated Treatments and Multiple Mediators}

\author{
Xiaxian Ou, 
Xinwei He, and 
Razieh Nabi\textsuperscript{\normalfont 1} \\
Department of Biostatistics and Bioinformatics, Emory University, Atlanta, GA, U.S.A.}

\AddToHookNext{env/thebibliography/end}{\label{lastpage}}
\begin{document}
\begingroup

\pagerange{\pageref{firstpage}--\pageref{lastpage}} 
\volume{}
\pubyear{}
\artmonth{}


\doi{}

\label{firstpage}


\begin{abstract}
In longitudinal studies, repeated treatments may be followed by different sequences of mediators and time-varying covariates. We develop semiparametric inference for path-specific effects in this setting, allowing the number, type, and ordering of intermediate variables to vary across treatment occasions. Under consistency, positivity, edge consistency, and the cross-world restrictions of a nonparametric structural equation model with independent errors, path-specific means are identified by the \textit{edge g-formula}. We derive two parameterizations of the efficient influence function of the identifying functional. The first follows the observed-data factorization and uses conditional intermediate-variable densities. The second uses sequential regressions to replace the corresponding integrations and Bayes’ rule to express density ratios through treatment-history regressions, thereby avoiding direct estimation of intermediate-variable densities. Each yields a one-step estimator that permits flexible nuisance estimation. We establish product-rate conditions for asymptotic linearity and semiparametric efficiency. The density-based estimator has \textit{blockwise multiple robustness}, generalizing triply robust estimation, while the regression-based estimator has \textit{sequential multiple robustness}. We further develop a sensitivity analysis for selected violations of cross-world independence assumptions used to identify these effects. Simulations examine finite-sample performance and robustness to model misspecifications. An application to the Framingham Heart Study examines sustained-smoking effects on systolic blood pressure along pathways through repeated body mass index measurements. The proposed methods are implemented in the \textsf{R} package \href{https://github.com/xxou/flexPaths}{\texttt{flexPaths}}. 
\\
\end{abstract}

%

\begin{keywords}
Causal mediation analysis; Edge g-formula; Efficient influence function; Longitudinal data; Multiple robustness; Sensitivity analysis.
\end{keywords}

\footnotetext[1]{Corresponding author: razieh.nabi@emory.edu}


\maketitle


%

\section{Introduction}
\label{sec:intro}

Many scientific questions concern not only whether an exposure affects an outcome, but which causal pathways contribute to that effect. Natural direct and indirect effects and their path-specific generalizations formalize such questions by assigning different treatment values along different causal pathways \citep{robins1992identifiability,pearl2001direct,avin2005identifiability,shpitser2013counterfactual}. In longitudinal studies, these questions may involve repeated treatments and varying sets of discrete, continuous, or multivariate intermediate variables, which may play roles conventionally described as mediators or time-varying confounders. Conventional longitudinal mediation is a structured special case of this more general setting.

Path-specific effects (PSEs) target effects transmitted along specified causal pathways, but identifying them can require stronger assumptions  beyond those needed for total effects \citep{robins2010alternative,shpitser2013counterfactual}. Directed acyclic graphs (DAGs) specify the assumed causal structure and make the pathways of interest explicit. They also help reveal when a PSE is not identified from observed data: the recanting-witness criterion detects an obstruction in fully observed DAGs, while the recanting-district criterion extends this analysis to settings with latent confounding \citep{avin2005identifiability,shpitser2013counterfactual,shpitser2016causal}.

These challenges have prompted several responses. Partial-identification methods characterize what can be learned about the original PSE under weaker restrictions \citep{tchetgen2014bounds,miles2017partial}. Alternative estimands include interventional, randomized, and stochastic effects \citep{vanderweele2014effect,vansteelandt2017interventional,benkeser2021nonparametric,diaz2020causal,miles2023causal} and separable effects \citep{stensrud2021generalized,stensrud2022separable}. These alternatives avoid some cross-world restrictions but answer different questions; for example, a randomized interventional indirect effect can be nonzero even when no individual's outcome is affected through the mediator \citep{miles2023causal}. Another approach retains such cross-world effects while representing them as mixtures of single-world controlled effects under assumptions involving a baseline susceptibility marker \citep{nabi2026formulating}. We focus on PSEs when the scientific question concerns specified causal pathways and the identifying assumptions are defensible, and assess sensitivity to selected violations.

Graphical identification of PSEs is well developed for general causal structures \citep{avin2005identifiability,shpitser2013counterfactual,shpitser2016causal}. For the class considered here, when the identification conditions hold, the PSE is represented by an observed-data functional known as the \textit{edge g-formula}. Semiparametric inference for this functional was first developed in single-treatment settings. \citet{tchetgen2012semiparametric} developed locally efficient, triply robust estimation for a single treatment and mediator, and \citet{miles2017quantifying,miles2020semiparametric} extended this framework to two ordered intermediate variables. \citet{zhou2022semiparametric} subsequently extended it to an arbitrary number of causally ordered, possibly multivariate mediators; related imputation-based weighting methods were developed by \citet{zhou2023tracing}. For a single baseline exposure, \citet{domingo2026path} also formulated PSEs with time-varying mediators and competing risks in a specialized time-to-event setting. Recent proximal approaches identify and estimate PSEs under unmeasured confounding using proxy variables, confounding bridge functions, and completeness conditions \citep{bai2026proximal,wu2026proximal}. These approaches address a complementary identification problem rather than the repeated-treatment structure considered here. 

A separate literature has studied path-specific and related mediation effects with time-varying treatments, mediators, and covariates under more structured longitudinal settings \citep{vanderweele2017mediation,zheng2017longitudinal,lin2017parametric,mittinty2020longitudinal}. \citet{tai2023causal} extended the mediational g-formula to multiple time-varying mediators, while \citet{diaz2023efficient} developed efficient, regression-based inference for longitudinal interventional effects. These longitudinal formulations accommodate repeated treatments, but do not provide general semiparametric inference for edge g-formula PSEs when the number, type, and causal ordering of intermediate variables may vary across treatment occasions.

Our contributions are threefold. \textit{First,} we formulate path-specific effects for longitudinal studies in which repeated treatments may be followed by different sequences of mediators and time-varying covariates. A sequence may contain a single mediator or several ordered intermediate variables, which may be discrete, continuous, or multivariate. We represent each time-indexed treatment and intermediate variable separately and use a node-wise assignment map to specify the treatment history associated with each intermediate variable according to the pathways of interest. Longitudinal mediation with a repeatedly measured mediator is included as a special case. Under the stated identification conditions, the resulting path-specific mean is identified by the edge g-formula.

\textit{Second,} we develop semiparametric estimation and inference for the identification functional. Direct plug-in estimation requires estimating and integrating conditional intermediate-variable laws, which may be difficult when intermediate variables are continuous or multivariate, and may incur first-order bias when the nuisance functions are estimated flexibly. We derive two equivalent representations of the target functional and its efficient influence function. The first follows the natural factorization of the observed-data law and retains the conditional intermediate-variable densities. The second expresses the required integrals through sequential regressions and uses Bayes' rule to replace intermediate-variable density ratios with treatment regressions, thereby avoiding direct density estimation. These representations yield plug-in, weighting, and cross-fitted one-step estimators that permit data-adaptive nuisance estimation. We give product-rate conditions for asymptotic linearity and semiparametric efficiency and establish multiple robustness to nuisance model misspecification, including a longitudinal generalization of the familiar triply robust structure. 

\textit{Third,} we develop a sensitivity analysis for selected violations of the cross-world independence restrictions. We parameterize these violations using functions that describe how the conditional mean of the outcome counterfactual varies across levels of the relevant intermediate-variable counterfactuals. For a specified set of sensitivity functions, we derive an observed-data representation of the PSE and construct an augmented estimator, allowing investigators to examine how estimates vary over scientifically plausible patterns and magnitudes of departure from the identifying assumptions.

We evaluate the proposed methods in simulation studies examining finite-sample performance, efficiency, and robustness to nuisance-model misspecification. An application to the Framingham Heart Study examines sustained-smoking effects on systolic blood pressure along selected pathways through repeated body mass index measurements. An accompanying \textsf{R} package, \href{https://github.com/xxou/flexPaths}{\texttt{flexPaths}}, implements the proposed methods.

The remainder of the paper is organized as follows. Section~\ref{sec:par_id} defines the class of path-specific effects and presents their identifying assumptions and functionals. The estimators and their asymptotic properties are derived in Sections~\ref{sec:est} and~\ref{sec:infer}.  Section~\ref{sec:sensitivity} presents the sensitivity analysis. Sections~\ref{sec:sims} and~\ref{sec:real_data} report the  simulation studies and the empirical application. Section~\ref{sec:disc} concludes. Proofs and additional results appear in the Supplementary Materials. 

\vspace{-0.2cm}
\section{Target parameter and identification}
\label{sec:par_id} 

We consider a longitudinal study with baseline covariates $C$, binary treatments $A_1,\ldots,A_K$, and outcome $Y$. Let $R_k$ denote the number of intermediate variables observed after $A_k$ and before the next treatment, or before $Y$ when $k=K$. For notational simplicity, we assume $R_k > 0$. The observed data are
\begin{align}
O=\{C,A_1,X_{1,1},\ldots,X_{1,R_1},\ldots,
A_K,X_{K,1},\ldots,X_{K,R_K},Y\}. 
\label{eq:data_structure}
\end{align} 

We observe i.i.d. copies $O_1,\ldots,O_n\sim P$, where $P$ belongs to a nonparametric statistical model $\mathcal M$. We represent the causal structure by a DAG $\mathcal G$ respecting the temporal ordering in \eqref{eq:data_structure} and posit a nonparametric structural equation model with independent errors (NPSEM-IE) \citep{pearl2009causality,malinsky2019potential}. Each node $V$ is generated as $V=f_V(\pa_{\mathcal G}(V),\epsilon_V)$, where $f_V$ is unrestricted, $\pa_{\mathcal G} (V)$ denotes its parents in $\mathcal G$, and the exogenous errors are mutually independent. The observed-data law factorizes as $P(O)=\prod_{V\in O}P(V\, | \, \pa_{\mathcal G}(V))$. The NPSEM-IE also encodes the cross-world restrictions used to identify PSEs below.

Let ${\bf A}_k=(A_1,\ldots,A_k)$, and let ${\bf a}_K\in\{0,1\}^K$ denote a treatment-assignment vector, with first $k$ components ${\bf a}_k = (a_1, \ldots, a_k) = {\bf a}_K[1:k]$. Write $\mathbb X=\{X_{k,r}:k=1,\ldots,K,\ r=1,\ldots,R_k\}$. The potential intermediate variable $X_{k,r}({\bf a}_k)$ is defined by setting $A_j=a_j$, $j=1,\ldots,k$, and recursively evaluating the resulting structural equations; the potential outcome $Y({\bf a}_K)$ is defined analogously. 

Let $\rho$ be a collection of \textit{proper} causal pathways of interest, where a path from $A_k$ to $Y$ is proper if it does not pass through another treatment \citep{shpitser2013counterfactual}. Let ${\bf a}_K$ and ${\bf a}'_K$ denote the active and reference treatment-assignment vectors, respectively. We define the nested counterfactual outcome by recursively expanding the structural equations backward from $Y$. Each occurrence of a treatment $A_j$ in this expansion corresponds to a complete proper path from $A_j$ to $Y$ and is assigned $a_j$ if that path belongs to $\rho$, and $a'_j$ otherwise.

For a general path collection, the same intermediate variable may appear multiple times in the nested outcome expression, with different treatment assignments across occurrences. To distinguish them, we index each counterfactual version of a variable by its downstream continuation to $Y$. For example, if both $M\rightarrow Y$ and $M\rightarrow L\rightarrow Y$ are present, then $M^{(Y)}$ denotes the version of $M$ supplied directly to $Y$, whereas $M^{(L,Y)}$ denotes the version supplied to $L$, whose value is then supplied to $Y$. These two versions need not coincide.

Formally, for $Z\in\mathbb X$, a downstream continuation $\xi=(V_1,\ldots,V_m)$ lists the nodes after $Z$ on a directed path $Z\rightarrow V_1\rightarrow\cdots \rightarrow V_m=Y$ containing no treatment nodes. For $Z=Y$, the continuation is the empty sequence $\varnothing$. Write $(Z,\xi)$ for the sequence formed by prepending $Z$ to $\xi$; thus, $(Z,\xi)=(Z,V_1,\ldots,V_m)$, and $(Z,\varnothing)=(Z)$. 

Given $({\bf a}_K,{\bf a}'_K,\rho)$ and a continuation $\xi$, write $Z^\xi=Z({\bf a}_K,{\bf a}'_K;\rho, \xi)$ for the corresponding counterfactual version of $Z$. If an intermediate variable $W$ is a parent of $Z$, the version supplied to this occurrence of $Z$ is $W^{(Z,\xi)} = W({\bf a}_K,{\bf a}'_K;\rho,(Z,\xi))$, whose downstream path is $W\rightarrow Z\rightarrow V_1\rightarrow\cdots\rightarrow Y$. The continuation index therefore records where a variable appears in the nesting; it does not represent an intervention on the downstream variables. 

Define these counterfactual variables recursively by
\begin{align}
Z^\xi
&=
f_Z\Bigl(
C,\,
\{a_j^\rho(Z;\xi):
  A_j\in\pa_{\mathcal G}(Z)\},\,
\{W^{(Z,\xi)}:
  W\in\pa_{\mathcal G}(Z)\cap\mathbb X\},\,
\epsilon_Z
\Bigr),
\label{eq:nested_cf}
\end{align}
where, for a nonempty continuation $\xi=(V_1,\ldots,V_m)$, $a_j^\rho(Z;\xi)=a_j$ if the complete directed path $A_j\rightarrow Z\rightarrow V_1 \rightarrow\cdots\rightarrow V_m=Y$ belongs to $\rho$, and $a_j^\rho(Z;\xi)=a'_j$ otherwise. For $Z=Y$, $\xi=\varnothing$ and $a_j^\rho(Y;\varnothing)=a_j$ if the direct path $A_j\rightarrow Y$ belongs to $\rho$, and $a'_j$ otherwise. 

Starting at $Y$ with the empty continuation, the recursion works backward through intermediate parents, substituting their counterfactual values until only treatment assignments, baseline covariates, and structural errors remain. All versions of a node use the same structural function and the same error $\epsilon_Z$. The resulting counterfactual outcome is 
\begin{align*}
Y^\rho 
= Y({\bf a}_K,{\bf a}'_K;\rho) 
= Y^\varnothing.
\end{align*} 
Appendix~\ref{app:ex_nested_cf} illustrates this recursion.

For fixed active and reference histories $({\bf a}_K,{\bf a}'_K)$ and a proper path set $\rho$, our target is 
\begin{align}
    \psi^\rho=E\left[Y({\bf a}_K,{\bf a}'_K;\rho)\right]. 
    \label{eq:target}
\end{align}

\noindent 
PSEs are contrasts of such means. For example, a contrast relative to the all-zero reference compares $E[Y({\bf 1}_K,{\bf 0}_K;\rho)]$ with $E[Y({\bf 0}_K)]$ \citep{daniel2015causal,steen2017flexible}. The results below apply to each mean separately; additive effect contrasts follow by linearity.

Identification by the \textit{edge g-formula} requires consistency, the NPSEM-IE cross-world restrictions, positivity on the support required by the edge intervention, and \textit{edge consistency} \citep{shpitser2016causal}. These assumptions are stated explicitly in Appendix~\ref{app:proofs_ID}. Edge inconsistency arises when two downstream continuations $\xi$ and $\xi'$ from the same intermediate variable to $Y$ require conflicting values of the same treatment, yielding counterfactual versions $X_{k,r}^{\xi}$ and $X_{k,r}^{\xi'}$ with different treatment inputs. Graphically, this occurs when, for some $A_j$ with $a_j\neq a'_j$, proper paths beginning with $A_j\rightarrow X_{k,r}$ are both included in and excluded from $\rho$. Such a variable is a \textit{recanting witness}, which generally obstructs nonparametric identification of the $\rho$-specific effect without additional restrictions \citep{avin2005identifiability,shpitser2016causal}. See Appendix~\ref{app:ex_nested_cf} for examples.

Let $\mathcal A(Z)$ and $\mathcal X(Z)$ denote the treatment and intermediate histories preceding $Z$, respectively. For a joint data realization, write $\mathcal X(z)$ for the corresponding realized intermediate history preceding $Z$, with dependence on the other components of the realization suppressed. Under edge consistency, the continuation-specific versions of each intermediate variable appearing in the nested outcome coincide. Given $({\bf a}_K,{\bf a}'_K,\rho)$, their common treatment inputs can therefore be encoded by a node-wise assignment map $\tau(Z)={\bf a}_Z$, indexed by the treatments in $\mathcal A(Z)$. Its $j$th component is $a_j$ if $A_j\rightarrow Z$ lies on a path in $\rho$, and $a'_j$ otherwise. Write ${\bf a}_{X_{k,r}}=\tau(X_{k,r})$ and ${\bf a}_Y=\tau(Y)$.

\vspace{-0.25cm}
\begin{lemma}[Identification by the edge g-formula]\label{lem:id}
Suppose consistency, the NPSEM-IE cross-world restrictions, edge consistency, and positivity hold. Then $\psi^\rho$ in \eqref{eq:target} is identified by 
\begin{align}
    \psi(P) 
    &= \int \! y \, dP(y \mid c, \mathcal{A}(Y)={\bf a}_Y, \mathcal{X}(y))  \prod_{k=1}^K \prod_{r=1}^{R_k}  dP(x_{k,r} \mid c, \mathcal{A}(X_{k, r}) = {\bf a}_{X_{k,r}}, \mathcal{X}(x_{k, r}))  \ dP(c).  
    \label{eq:pse_id}
\end{align}
\end{lemma}

\vspace{-0.65cm}
The proof and explicit identification assumptions appear in Appendix~\ref{app:proofs_ID}. 

\section{Statistical representations and estimators}
\label{sec:est} 

Having identified the target $\psi^\rho$ in \eqref{eq:target} as the observed-data functional $\psi(P)$ in \eqref{eq:pse_id}, we now develop statistical representations of this functional that lead to complementary estimators. Throughout, for an integrable function $h$ of $O$, write $Ph=\int h(o)\,dP(o)$ and $P_nh=n^{-1}\sum_{i=1}^n h(O_i)$. For intermediate variables, $X_{j,q}\prec X_{k,r}$ indicates that $X_{j,q}$ precedes $X_{k,r}$ in the ordering in \eqref{eq:data_structure}. We first introduce the nuisance functions used throughout. 

For any treatment history ${\bf b}_k=(b_1,\ldots,b_k)\in\{0,1\}^k$ with $\mathbf b_0=\varnothing$, define 
\begin{align}
\pi_k(b_k\mid{\bf b}_{k-1},\mathcal X(A_k), C)
&=
P(A_k=b_k\mid
\mathcal A(A_k)={\bf b}_{k-1},\mathcal X(A_k), C),
\label{eq:pi_def}
\end{align}
and $\bar\pi_k({\bf b}_k) = \prod_{j=1}^k \pi_j(b_j\mid{\bf b}_{j-1},\mathcal X(A_j), C)$ where dependence of $\bar\pi_k({\bf b}_k)$ on the observed histories is suppressed for notational convenience. We also define treatment-history regressions as the intermediate variables are successively observed. For ${\bf b}_k\in\{0,1\}^k$, let
\begin{align}
g_{k,0}({\bf b}_k\mid \mathcal X(A_k), C)
&=
P({\bf A}_k={\bf b}_k\mid \mathcal X(A_k), C),
\label{eq:g0_def}\\
g_{k,r}({\bf b}_k\mid \mathcal X(X_{k,r}),X_{k,r}, C)
&=
P({\bf A}_k={\bf b}_k\mid \mathcal X(X_{k,r}),X_{k,r}, C),
\qquad r=1,\ldots,R_k.
\label{eq:g_def}
\end{align}
Thus, $g_{k,0}$ conditions on the nontreatment history preceding the intermediate-variable block following $A_k$, while $g_{k,r}$ additionally conditions on intermediate variables through $X_{k,r}$. 

For each intermediate variable $X_{k,r}$, we also let 
\begin{equation}
f_{k,r}(x_{k,r}\mid \mathcal A(X_{k,r})= {\bf b}_k,\mathcal X(X_{k,r}), C)
\label{eq:mediator_density}
\end{equation}
denote the conditional density or probability mass function of $X_{k,r}$ given $\mathcal A(X_{k,r})={\bf b}_k$, $\mathcal X(X_{k,r})$ and $C$, with respect to a dominating measure $\nu_{k,r}$. 

Finally, define the outcome and sequential regressions. The outcome regression is 
\begin{equation}
\mu_{K,R_K}(\mathcal X(Y), C)
=
E[Y\mid\mathcal A(Y)={\bf a}_Y,\mathcal X(Y), C].
\label{eq:outcome_regression}
\end{equation}
For $r=R_k-1,\ldots,1$ define the within-block sequential regressions by
\begin{align}
&\mu_{k,r}(X_{k,r}, \mathcal X(X_{k,r}), C) \notag \\ 
&\hspace{0.5cm}=
E\left[
\mu_{k,r+1}(X_{k,r+1}, \mathcal X(X_{k,r+1}), C)
\mid
\mathcal A(X_{k,r+1})={\bf a}_{X_{k,r+1}}, \mathcal X(X_{k,r+1}), C
\right]
\label{eq:mu_within}
\\ 
&\hspace{0.5cm}=
\int
\mu_{k,r+1}(x_{k,r+1}, \mathcal X(X_{k,r+1}), C) \, 
f_{k,r+1}(x_{k,r+1}\mid{\bf a}_{X_{k,r+1}},\mathcal X(X_{k,r+1}), C)
\,d\nu_{k,r+1}(x_{k,r+1}). \notag 
\end{align}
For $r = R_k$ and $k=K-1,\ldots,1$, define
\begin{equation}\label{eq:mu_between}
\begin{aligned}
&\mu_{k,R_k}(X_{k,R_k}, \mathcal X(X_{k,R_k}), C) \\
&\hspace{1.5cm}=
E\left[
\mu_{k+1,1}(X_{k+1,1}, \mathcal X(X_{k+1,1}), C)
\mid
\mathcal A(X_{k+1,1})={\bf a}_{X_{k+1,1}},\mathcal X(X_{k+1,1}), C
\right].
\end{aligned}
\end{equation}
At block boundaries, we identify $\mu_{k,0}\equiv\mu_{k-1,R_{k-1}},\ k=2,\ldots,K$, so that the recursion proceeds continuously in reverse order from $Y$ through $X_{K,R_K},\ldots,X_{1,1}$. The recursion terminates at $\mu_{1,0}(C) = E\left[ \mu_{1,1}(X_{1,1},C) \mid A_1= {\bf a}_{X_{1,1}},C \right]$, and we obtain $\psi(P)=P\{\mu_{1,0}(C)\}$.  

Let $\pi$, $f$, and $g$ denote the collections of treatment mechanisms, intermediate-variable densities, and treatment-history regressions defined above, respectively, and let $\mu_{\mathrm{seq}}=\{\mu_{1,0}\}\cup\{\mu_{k,r}\}_{k,r}$. We use two nuisance parameterizations throughout. The \textbf{density-based parameterization} $Q_1=(\pi,f,\mu_{K,R_K})$ follows the observed-data factorization, with earlier sequential regressions induced by iterating the terminal outcome regression backward through the fitted intermediate-variable laws. The \textbf{regression-based parameterization} $Q_2=(\pi,g,\mu_{\mathrm{seq}})$ instead estimates the sequential outcome and treatment-history regressions directly. Both parameterizations underlie the plug-in, weighting, and one-step estimators developed below.

\vspace{-0.2cm}
\subsection{Plug-in estimators}

A density-based plug-in estimator of $\psi(P)$ in \eqref{eq:pse_id} replaces the conditional laws by fitted laws:

\vspace{-0.6cm}
\begin{align}
\widehat\psi_{\mathrm{gcomp}}
=
P_n\bigg\{
\int
\widehat\mu_{K,R_K}(\mathcal X(y), C)
\prod_{k=1}^K\prod_{r=1}^{R_k}
\widehat f_{k,r}
(x_{k,r}\mid{\bf a}_{X_{k,r}},\mathcal X(x_{k,r}), C)
\,d\nu(x)
\bigg\},
\label{eq:gcomp_estimator}
\end{align}

\vspace{0.25cm}
\noindent 
where $d\nu(x)=\prod_{k=1}^K\prod_{r=1}^{R_k} d\nu_{k,r}(x_{k,r})$. This estimator is most straightforward for discrete intermediate variables. Continuous or multivariate intermediates require conditional-density estimation and potentially high-dimensional numerical or Monte Carlo integration. Moreover, direct substitution retains first-order sensitivity to nuisance-estimation error and thus does not generally support root-$n$ inference with flexible nuisance estimators.

The representation $\psi(P) = P\{\mu_{1,0}(C)\}$  provides a regression-based plug-in estimator: 
\begin{align}
\widehat\psi_{\mathrm{seq}}
=
P_n\{\widehat\mu_{1,0}(C)\},
\label{eq:sequential_estimator}
\end{align}
where the sequential regressions are fit recursively backward from the outcome regression. This approach avoids estimating intermediate-variable densities directly, but remains a plug-in estimator and thus retains first-order sensitivity to nuisance-estimation error. 

\vspace{-0.2cm}
\subsection{Inverse probability weighted estimators}

The sequential structure of $\psi(P)$ also yields weighting representations indexed by $Z\in\{Y\}\cup\mathbb X$. Each representation weights the observed history through $Z$ and, for intermediate $Z$, uses a sequential regression for the remaining variables. As with the plug-in estimators, the regressions and weights can be estimated under either $Q_1$ or $Q_2$.

To characterize the weights, partition the intermediate variables according to their assigned treatment histories. Let $S$ be the minimum number of distinct full-length treatment assignment vectors other than ${\bf a}_Y$ such that these $S$ vectors, together with ${\bf a}_Y$, cover all node-specific assignments represented in $\mathbb X$. Set ${\bf a}^{(0)}={\bf a}_Y$ and choose ${\bf a}^{(1)},\ldots,{\bf a}^{(S)}$, distinct from ${\bf a}_Y$, achieving this minimum.  For each $X_{k,r}\in\mathbb X$, fix a compatible index $s(X_{k,r})$ satisfying ${\bf a}_{X_{k,r}}={\bf a}^{(s(X_{k,r}))}[1:k]$. Define
\begin{equation}
\mathbb M
=
\{X_{k,r}\in\mathbb X:s(X_{k,r})=0\},
\qquad
\mathbb L=\mathbb X\setminus\mathbb M,
\label{eq:M_partition}
\end{equation}
and for $s \in\{1,\ldots,S\}$
\begin{equation}
\mathbb L_s
=
\{X_{k,r}\in\mathbb X:s(X_{k,r})=s\}, 
\qquad 
\mathbb L
=
\mathbb L_1\mathbin{\dot\cup}\cdots\mathbin{\dot\cup}\mathbb L_S
.
\label{eq:L_partition}
\end{equation}
An example and a backward construction for deciding the minimum $S$ are provided in Appendix~\ref{app:ex_minimal_S}. When $S=0$, $\mathbb L=\varnothing$ and the target reduces to an ordinary treatment-regime mean, yielding the usual total-effect contrast.  When $S > 0$, a node-specific assignment of length $k<K$ may be compatible with more than one full-length vector. Such a node may be assigned to any compatible class.  Any fixed choice among compatible indices gives the same weighting identities, which depend on the node-specific assignment prefixes.

For ${\bf b}_k,{\bf d}_k\in\{0,1\}^k$, define the density-based ratio
\begin{equation}
R_{k,r}^{(1)}( X_{k,r},\mathcal X(X_{k,r}), C ; {\bf b}_k,{\bf d}_k)
=
\frac{
f_{k,r}(X_{k,r}\mid{\bf b}_k,\mathcal X(X_{k,r}), C)
}{
f_{k,r}(X_{k,r}\mid{\bf d}_k,\mathcal X(X_{k,r}), C)
}.
\label{eq:density_ratio}
\end{equation}
Bayes' rule gives the equivalent treatment-regression ratio
\begin{align}
R_{k,r}^{(2)}(X_{k,r},\mathcal X(X_{k,r}), C ; {\bf b}_k,{\bf d}_k)
&=
\frac{
g_{k,r}({\bf b}_k\mid X_{k,r}, \mathcal X(X_{k,r}), C)
}{
g_{k,r-1}({\bf b}_k\mid \mathcal X(X_{k,r}), C)
} \, 
\frac{
g_{k,r-1}({\bf d}_k\mid \mathcal X(X_{k,r}), C)
}{
g_{k,r}({\bf d}_k\mid X_{k,r}, \mathcal X(X_{k,r}), C)
}. 
\label{eq:bayes_ratio}
\end{align}

\vspace{0.2cm} \noindent 
Thus $R_{k,r}^{(1)}=R_{k,r}^{(2)}$ at the population law; $\ell\in\{1,2\}$ indexes the estimation approach. The practical gain is that $R^{(2)}_{k,r}$ can be estimated using regressions for discrete treatment histories even when $X_{k,r}$ is continuous or multivariate, whereas $R^{(1)}_{k,r}$ requires estimating the conditional density of $X_{k,r}$ itself.
 
For either ratio representation, the target has the form
\begin{equation}
\psi(P)=P\{W_Z^{(\ell)}m_Z\}, \qquad Z\in\{Y\}\cup\mathbb X,
\label{eq:weighting_general}
\end{equation} 
where $m_Y=Y$ and $m_{X_{k,r}}=\mu_{k,r}(X_{k,r},\mathcal X(X_{k,r}),C)$. For $Z=Y$, the weight is
\begin{equation}
W_Y^{(\ell)}
=
\frac{
\mathbb I\{\mathcal A(Y)={\bf a}_Y\}
}{
\bar\pi_K({\bf a}_Y)
} \ 
\prod_{X_{j,q}\in\mathbb L}
R_{j,q}^{(\ell)}
\big(X_{j,q},\mathcal X(X_{j,q}), C; {\bf a}_{X_{j,q}},{\bf a}_Y[1:j]\big).
\label{eq:weight_Y}
\end{equation}
Variables in $\mathbb M$ are omitted because their assigned treatment prefixes agree with the corresponding prefixes of ${\bf a}_Y$, so their ratios equal one.

For $Z = X_{k,r} \in \mathbb M$, the weight is
\begin{equation}
W_{k,r}^{(\ell)}
=
\frac{
\mathbb I\{\mathcal A(X_{k,r})={\bf a}_Y[1:k]\}
}{
\bar\pi_k({\bf a}_Y[1:k])
}
\prod_{\substack{
X_{j,q}\prec X_{k,r}\\
X_{j,q}\in\mathbb L
}}
R_{j,q}^{(\ell)}
\big(X_{j,q},\mathcal X(X_{j,q}), C; {\bf a}_{X_{j,q}},{\bf a}_Y[1:j]\big).
\label{eq:weight_M}
\end{equation}
Only preceding variables in $\mathbb L$ may have nontrivial ratios.

For $Z = X_{k,r} \in \mathbb L_s$, the weight is
\begin{align}
W_{k,r}^{(\ell)}
&=
\frac{
\mathbb I\{\mathcal A(X_{k,r})={\bf a}^{(s)}[1:k]\}
}{
\bar\pi_k({\bf a}^{(s)}[1:k])
}
\!\! \prod_{\substack{
X_{j,q}\prec X_{k,r}\\
X_{j,q} \, \in \, \mathbb M \, \cup \, \mathbb L \setminus \mathbb L_s
}}
R_{j,q}^{(\ell)}
\big(X_{j,q},\mathcal X(X_{j,q}), C; {\bf a}_{X_{j,q}}, {\bf a}^{(s)}[1:j]).
\label{eq:weight_L}
\end{align}
Preceding variables in $\mathbb L_s$ are omitted because their ratios equal one. Variables in $\mathbb M$ or another $\mathbb L_h$ remain in the product, although their ratios also equal one whenever their assigned prefixes agree with ${\bf a}^{(s)}$.

To construct estimators, let $\widehat W_Z^{(1)}$ use the fitted treatment mechanisms and density ratios under $Q_1$, and let $\widehat W_Z^{(2)}$ use the fitted treatment mechanisms and treatment-regression ratios under $Q_2$. For an intermediate variable, write $\widehat\mu_{k,r}^{(1)}$ for the regression induced by $\widehat\mu_{K,R_K}$ and the fitted intermediate-variable laws through backward integration. Write $\widehat\mu_{k,r}^{(2)}$ for the directly estimated sequential regression under $Q_2$. The paired estimators are
\begin{align}
\widehat\psi_{\mathrm{IPW},Y}^{(\ell)}
&=
P_n\{\widehat W_Y^{(\ell)}Y\}, \nonumber\\
\widehat\psi_{\mathrm{IPW},X_{k,r}}^{(\ell)}
&=
P_n\!\left\{
\widehat W_{k,r}^{(\ell)}
\widehat\mu_{k,r}^{(\ell)}
(X_{k,r},\mathcal X(X_{k,r}),C)
\right\},
\qquad \ell\in\{1,2\}.
\label{eq:ipw_general}
\end{align}
The population regression $\mu_{k,r}$ is the same in both representations; the superscript distinguishes how its estimate is constructed. These weighting identities do not depend on the choice of a compatible partition, and also enter the one-step estimators developed below. 

\subsection{One-step estimators}
\label{subsec:one_step_estimator} 

The plug-in and weighting estimators can retain first-order nuisance-estimation error. The efficient influence function (EIF) combines the outcome residual and successive sequential-regression residuals, each with its corresponding weight from \eqref{eq:weight_Y}--\eqref{eq:weight_L}.

\begin{theorem}[Efficient influence function]
\label{thm:eif}
Suppose the observed-data law belongs to the nonparametric model, the weights in \eqref{eq:weight_Y}--\eqref{eq:weight_L} are well defined under the stated positivity conditions, and the following expression is square-integrable. For $\ell\in\{1,2\}$, define
\begin{align}
\Phi_\ell(P)(O)
&=
W_Y^{(\ell)}
\left\{
Y-\mu_{K,R_K}(\mathcal X(Y), C)
\right\}
\nonumber\\
&\quad+
\sum_{k=1}^K\sum_{r=1}^{R_k}
W_{k,r}^{(\ell)}
\left\{
\mu_{k,r}(X_{k,r}, \mathcal X(X_{k,r}), C)
-\mu_{k,r-1}(\mathcal X(X_{k,r}), C)
\right\}
\nonumber\\
&\quad+
\mu_{1,0}(C)-\psi(P),
\label{eq:eif_compact}
\end{align}
where $W_Y^{(\ell)}$ is given by \eqref{eq:weight_Y}, and $W_{k,r}^{(\ell)}$ is given by \eqref{eq:weight_M} or \eqref{eq:weight_L} according to the assignment class of $X_{k,r}$. Then $\Phi_1(P)$ is the EIF of $\psi(P)$. By Bayes' rule, $R_{k,r}^{(1)}=R_{k,r}^{(2)}$ on the relevant support, so $W_Z^{(1)}=W_Z^{(2)}$ for every $Z\in\{Y\}\cup\mathbb X$. Consequently, $\Phi_2(P)=\Phi_1(P)$ almost surely and gives an equivalent EIF representation.
\end{theorem}

\vspace{-0.2cm}
See a proof in Appendix~\ref{app:proofs_est}. The two EIF expressions in the above theorem coincide at the population law, but their fitted versions generally differ.

For $\ell\in\{1,2\}$, let $\widehat Q_\ell$ denote the fitted nuisance collection under $Q_\ell$. Write $\Phi_\ell(\widehat Q_\ell)(O)$ for \eqref{eq:eif_compact} with the nuisance functions replaced by their estimates and $\psi(P)$ replaced by the corresponding plug-in estimator $\psi(\widehat Q_\ell)$. For $\ell=1$, the plug-in term is \eqref{eq:gcomp_estimator}; the earlier sequential regressions are induced by the fitted outcome regression and intermediate-variable laws, which also determine the density ratios. For $\ell=2$, the plug-in term is \eqref{eq:sequential_estimator}; the sequential regressions are fitted directly, and the ratios use treatment-history regressions. The one-step estimator is

\vspace{-1.2cm}
\begin{align}
\psi^+(\widehat Q_\ell)
=
\psi(\widehat Q_\ell)
+
P_n\{\Phi_\ell(\widehat Q_\ell)\}.
\label{eq:onestep}
\end{align}

A cross-fitted version of the one-step estimator can be constructed by fitting the nuisance functions outside each observation's validation fold. In the asymptotic linearity argument of Section~\ref{sec:infer}, cross-fitting controls the empirical-process term without Donsker conditions, provided the fitted EIF expression converges in $L^2(P)$. Let $v(i)$ denote the validation fold containing observation $i$, and let $\widehat Q_{\ell,-v(i)}$ denote the nuisance functions fitted without that fold. Define $D_\ell(P)(O)=\Phi_\ell(P)(O)+\psi(P)$, and evaluate this uncentered expression at the fitted nuisances to obtain $\widehat D_{\ell,i} =D_\ell(\widehat Q_{\ell,-v(i)})(O_i)$. The cross-fitted one-step estimator is $\psi^+(\widehat Q_\ell) = \frac{1}{n}\sum_{i=1}^n\widehat D_{\ell,i}$, and its asymptotic linearity follows under the conditions in Section~\ref{sec:infer}.

\vspace{-0.1cm}
\section{Inference and asymptotic properties}
\label{sec:infer}

For $\ell\in\{1,2\}$, the one-step estimator in \eqref{eq:onestep}
admits the von Mises expansion
\begin{align}
\psi^+(\widehat Q_\ell)-\psi(P)
&=
(P_n-P)\Phi_\ell(Q_\ell)
+
(P_n-P)\{\Phi_\ell(\widehat Q_\ell)-\Phi_\ell(Q_\ell)\}
+
R_{2,\ell}(\widehat Q_\ell,Q_\ell),
\label{eq:stochastic_behavior_onestep}
\end{align}
where $R_{2,\ell}(\widehat Q_\ell,Q_\ell) =P\{\Phi_\ell(\widehat Q_\ell) + \psi(\widehat Q_\ell)\}-\psi(P)$. The empirical-process term is $o_p(n^{-1/2})$ under the conditions outlined in Appendix~\ref{app:proofs_inf}. For the cross-fitted version, the corresponding term is negligible when the fitted EIF expression converges in $L^2(P)$, without a Donsker condition. If the remainder is also $o_p(n^{-1/2})$ and $P\{\Phi_\ell(Q_\ell)^2\}<\infty$, the estimator is asymptotically linear and normal with influence function $\Phi_\ell(Q_\ell)$.

We begin with the \textit{\textbf{density-based estimator}} $\psi^+(\widehat Q_1)$. Its fitted sequential regressions are induced by $\widehat\mu_{K,R_K}$ and the fitted conditional intermediate-variable laws. Let $\|h\|_2=(Ph^2)^{1/2}$.

\vspace{-0.25cm}
\begin{theorem}
\label{thm:asymp_lin_1}
Suppose
$\|\widehat\pi_k-\pi_k\|_2=o_p(n^{-1/b_k})$,
$\|\widehat f_{k,r}-f_{k,r}\|_2=o_p(n^{-1/m_{k,r}})$ for any $X_{k,r}\in\mathbb M$,
$\|\widehat f_{j,q}-f_{j,q}\|_2=o_p(n^{-1/l_{j,q}})$ for any $X_{j,q}\in\mathbb L$,
and $\|\widehat\mu_{K,R_K}-\mu_{K,R_K}\|_2=o_p(n^{-1/u})$.
Under the regularity conditions discussed in Appendix~\ref{app:proofs_inf}, if
\begin{enumerate}
\item
$\frac{1}{b_k}+\frac{1}{u}\geq\frac12$ for all $k$, and
$\frac{1}{b_k}+\frac{1}{m_{j,q}}\geq\frac12$ for all $k$ and
$X_{j,q}\in\mathbb M$;
\item
$\frac{1}{b_k}+\frac{1}{l_{j,q}}\geq\frac12$ for all $k$ and
$X_{j,q}\in\mathbb L$;
\item
$\frac{1}{u}+\frac{1}{l_{j,q}}\geq\frac12$ for all
$X_{j,q}\in\mathbb L$, and
$\frac{1}{m_{k,r}}+\frac{1}{l_{j,q}}\geq\frac12$ for all
$X_{k,r}\in\mathbb M$ and $X_{j,q}\in\mathbb L$;
\item
$\frac{1}{l_{j,q}}+\frac{1}{l_{j',q'}}\geq\frac12$ whenever
$X_{j,q}\in\mathbb L_s$ and $X_{j',q'}\in\mathbb L_h$ for all $s, h = 1, \ldots, S$ with $s \ne h$,
\end{enumerate}
then $\sqrt{n}\big\{\psi^+(\widehat Q_1)-\psi(P)\big\} \rightsquigarrow N\!\left(0,P\{\Phi_1(Q_1)^2\}\right)$. 
\end{theorem}

The pairwise rate conditions above reflect products of nuisance-estimation errors in the second-order remainder derived in Appendix~\ref{app:proofs_inf}. Thus, $o_p(n^{-1/4})$ convergence for all nuisances is sufficient, though unequal rates are allowed provided each condition holds.

The remainder calculation in the appendix also yields a \emph{\textbf{blockwise multiple-robustness}} property for the estimator $\psi^+(\widehat Q_1)$. Define the following blocks of nuisance functions: 
\begin{equation}
\begin{aligned}
&\mathcal B_\pi=\{\pi_k:k=1,\ldots,K\}, 
\qquad
\mathcal B_M=\{\mu_{K,R_K},f_{k,r}:X_{k,r}\in\mathbb M\}, 
\\
&\hspace{1.5cm}\mathcal B_{L_s} =\{f_{k,r}:X_{k,r}\in\mathbb L_s\}, \qquad s=1,\ldots,S.
\end{aligned}
\end{equation}

\vspace{0.2cm}
A block is consistently estimated if all of its components are. For the following result, suppose the nuisance estimators converge in $L^2(P)$ to fixed, possibly misspecified limits and that the bounded-weight and moment conditions in Appendix~\ref{app:proofs_inf} hold.

\begin{corollary}
\label{cor:robustness_1}
The density-based one-step estimator $\psi^+(\widehat Q_1)$ is consistent if any $S+1$ of the $S+2$ nuisance blocks $\mathcal B_\pi$, $\mathcal B_M$, $\mathcal B_{L_1},\ldots,\mathcal B_{L_S}$ are consistently estimated.
\end{corollary}

Thus, the density-based one-step estimator $\psi^+(\widehat Q_1)$ remains consistent if any one block is misspecified while all others are consistently estimated. When $S=1$, this is the familiar two-out-of-three, or triply robust, property. These conditions are sufficient but not necessary and may be further relaxed through treatment-mechanism and compatible assignment partitions, as illustrated via examples in Appendix~\ref{app:ex_multiple_robustness}.

We next consider the \textit{\textbf{regression-based estimator}} $\psi^+(\widehat Q_2)$. Note that the functions $\widehat g_{k,0}$ are derived: $\widehat g_{1,0}=\widehat\pi_1$ and, for $k\geq2$, $\widehat g_{k,0}({\bf b}_k\mid \mathcal X(A_k), C) = \widehat g_{k-1,R_{k-1}}({\bf b}_{k-1}\mid \mathcal X(A_k), C) \widehat\pi_k(b_k\mid{\bf b}_{k-1},\mathcal X(A_k), C)$. They therefore require no separate nuisance-rate assumptions.

\begin{theorem}
\label{thm:asymp_lin_2}
Suppose
$\|\widehat\pi_k-\pi_k\|_2=o_p(n^{-1/b_k})$ for $k=1,\ldots,K$,
$\|\widehat g_{k,r}-g_{k,r}\|_2=o_p(n^{-1/v_{k,r}})$ and
$\|\widehat\mu_{k,r}-\mu_{k,r}\|_2=o_p(n^{-1/u_{k,r}})$ for
$k=1,\ldots,K$, $r=1,\ldots,R_k$, and
$\|\widehat\mu_{1,0}-\mu_{1,0}\|_2=o_p(n^{-1/u_{1,0}})$.
Under the regularity conditions discussed in Appendix~\ref{app:proofs_inf}, if
\begin{enumerate}
\item
$\frac{1}{v_{k,r}}+\frac{1}{u_{k,r}}\geq\frac12$ for every
$k=1,\ldots,K$ and $r=1,\ldots,R_k$;
\item
$\frac{1}{b_{k+1}}+\frac{1}{u_{k,R_k}}\geq\frac12$ for every
$k=1,\ldots,K-1$; and
\item
$\frac{1}{b_1}+\frac{1}{u_{1,0}}\geq\frac12$,
\end{enumerate}
then $\sqrt{n}\big\{\psi^+(\widehat Q_2)-\psi(P)\big\} \rightsquigarrow N\!\left(0,P\{\Phi_2(Q_2)^2\}\right)$. 
\end{theorem}

The corresponding second-order remainder also yields a \textit{\textbf{sequential multiple-robustness}} property: consistency of the treatment-history and sequential regressions may vary across the longitudinal ordering. For the following result, suppose the nuisance estimators converge in $L^2(P)$ to fixed, possibly misspecified limits and that the bounded-weight and moment conditions in Appendix~\ref{app:proofs_inf} hold.

\begin{corollary}
\label{cor:robustness_2}
The regression-based one-step estimator $\psi^+(\widehat Q_2)$ is consistent under any one of the following regimes:
\begin{enumerate}
\item
$\widehat\mu_{1,0}$ and all
$\widehat\mu_{k,r}$, $k=1,\ldots,K$, $r=1,\ldots,R_k$, are consistent;
\item
for some $k'\in\{1,\ldots,K\}$ and
$r'\in\{1,\ldots,R_{k'}\}$:
\begin{itemize}
\item
$\widehat\mu_{k,r}$ is consistent for every
$X_{k,r}\succeq X_{k',r'}$;
\item
$\widehat g_{k,r}$ is consistent for every
$X_{k,r}\prec X_{k',r'}$; and
\item
$\widehat\pi_k$ is consistent for every $k\leq k'$;
\end{itemize}
\item
all $\widehat g_{k,r}$, $k=1,\ldots,K$, $r=1,\ldots,R_k$,
and all $\widehat\pi_k$, $k=1,\ldots,K$, are consistent.
\end{enumerate}
\end{corollary}

The regimes in Corollary~\ref{cor:robustness_2} allow at most one switch from treatment-history to sequential regressions. The second-order remainder permits a broader sufficient condition. At each intermediate variable $X_{k,r}$, at least one of $\widehat g_{k,r}$ and $\widehat\mu_{k,r}$ must be consistent. At each transition from block $k$ to block $k+1$, at least one of $\widehat\mu_{k,R_k}$ and $\widehat\pi_{k+1}$ must be consistent; at the initial step, the same holds for $\widehat\mu_{1,0}$ and $\widehat\pi_1$. These conditions allow the correctly estimated regression to alternate across intermediate variables. For example, one may rely on a treatment-history regression at one variable, a sequential regression at the next, and a treatment-history regression again later, provided the boundary and initial conditions also hold.

Appendix~\ref{app:ex_multiple_robustness} illustrates the blockwise and sequential robustness structures via examples. 

\section{Unmeasured confounding and sensitivity analysis}
\label{sec:sensitivity}

We consider sensitivity to violations of the cross-world restrictions, adapting the approach proposed by \citet{tchetgen2012semiparametric} for a single treatment and mediator. For illustration, we focus on restrictions linking intermediate variables to the outcome. To obtain a tractable formulation, we restrict attention to one non-outcome assignment pattern ($S=1$) and discrete nodes in $\mathbb L$, assigning to $\mathbb M$ all mediators whose treatment assignments are compatible with the outcome treatment assignment. Write ${\bf a}^{(1)}$ for the non-outcome assignment and retain the remaining identification conditions from Section~\ref{sec:par_id}. Thus,
\begin{align}
\mathbb L
=
\left\{
X_{k,r}:
{\bf a}_{X_{k,r}}={\bf a}^{(1)}[1:k],\
{\bf a}_{X_{k,r}}\neq{\bf a}_Y[1:k]
\right\},
\qquad
\mathbb M=\mathbb X\setminus\mathbb L.
\label{eq:sens_L_set}
\end{align}

For each $X_{k,r}\in\mathbb L$ and treatment history ${\bf b}_k\in\{0,1\}^k$, the cross-world independence restriction is $Y({\bf a}_Y,\mathcal X(y)) \indep X_{k,r}\{{\bf b}_{k},\mathcal X(x_{k,r})\} \mid C,{\bf A}_k,\mathcal X(X_{k,r})$. We relax these restrictions while maintaining those involving the remaining node pairs. To quantify departures from cross-world independence, define
\begin{align}
&\hspace{-1.5cm} t_{k,r}({\bf b}_{k},\mathcal X(y),c; {\bf a}_Y) \notag  \\
& =
E\!\left[
Y({\bf a}_Y,\mathcal X(y))
\mid
 \mathcal A(X_{k,r})={\bf b}_{k},
\mathcal X(X_{k,r}) = \mathcal X(x_{k,r}),
X_{k,r}=x_{k,r}, c
\right]
\label{sens:ti}\\
& \quad -
E\!\left[
Y({\bf a}_Y,\mathcal X(y))
\mid
 \mathcal A(X_{k,r})={\bf b}_{k},
\mathcal X(X_{k,r}) = \mathcal X(x_{k,r}), 
X_{k,r}\neq x_{k,r}, c
\right].
\notag 
\end{align}
Under cross-world independence, the two conditional means coincide, so that $t_{k,r}=0$.

For fixed known sensitivity functions $t_{k,r}$, define
\begin{align*}
\delta_{t_{k,r}}(\mathcal X(y),c;f)
={}&
t_{k,r}({\bf a}^{(1)}[1:k],\mathcal X(y),c; {\bf a}_Y)\!
\Big[
1-f_{k,r}(x_{k,r}\mid{\bf a}^{(1)}[1:k],\mathcal X(x_{k,r}), c)
\Big]
\\
&-
t_{k,r}({\bf a}_Y[1:k],\mathcal X(y),c; {\bf a}_Y)\!
\Big[
1-f_{k,r}(x_{k,r}\mid{\bf a}_Y[1:k],\mathcal X(x_{k,r}), c)
\Big],
\end{align*}
where $f_{k,r}$ denotes a conditional probability mass function with respect to counting measure, and define the sensitivity-adjusted outcome regression
\[
m_t(\mathcal X(y),c; \mu,f)
=
\mu_{K,R_K}(\mathcal X(y),c)
+
\sum_{X_{k,r}\in\mathbb L}
\delta_{t_{k,r}}(\mathcal X(y),c;f).
\]
Under the retained identification conditions, the sensitivity-indexed target is 
\begin{align}
\psi_t(P)
={}&
\int
m_t(\mathcal X(y),c; \mu,f)
\prod_{k=1}^K\prod_{r=1}^{R_k}
f_{k,r}(x_{k,r}\mid{\bf a}_{X_{k,r}},\mathcal X(x_{k,r}), c)
\,d\nu(\mathcal X(y))\,dP(c),
\label{eq:sens_id}
\end{align}
where $d\nu(\mathcal X(y))=\prod_{k,r}d\nu_{k,r}(x_{k,r})$. Setting every $t$ to zero recovers the identifying functional under the original cross-world restrictions. See the identification proof in Appendix~\ref{app:proofs_sa_ID}. 

To estimate \eqref{eq:sens_id}, let
\begin{align*}
m_t(\mathcal X(y),c; \widehat {\mu},\widehat f)
& =
\widehat\mu_{K,R_K}(\mathcal X(y),c) \\
& \quad +
\sum_{X_{k,r}\in\mathbb L}
\bigg[
t_{k,r}({\bf a}^{(1)}[1:k],\mathcal X(y),c; {\bf a}_Y)\!
\left[
1-\widehat f_{k,r}
(x_{k,r}\mid{\bf a}^{(1)}[1:k],\mathcal X(x_{k,r}), c)
\right]
\\
&\hspace{1.8cm}-
t_{k,r}({\bf a}_Y[1:k],\mathcal X(y),c; {\bf a}_Y)\!
\left[
1-\widehat f_{k,r}
(x_{k,r}\mid{\bf a}_Y[1:k],\mathcal X(x_{k,r}), c)
\right]
\bigg],
\\
h_t(c; \widehat {\mu},\widehat f)
={}&
\int
m_t(\mathcal X(y),c;\widehat\mu,\widehat f)
\prod_{k=1}^K\prod_{r=1}^{R_k}
\widehat f_{k,r}(x_{k,r}\mid{\bf a}_{X_{k,r}},\mathcal X(x_{k,r}), c)
\,d\nu(\mathcal X(y)),
\\
\widehat W_Y
={}&
\frac{
\mathbb I\{\mathcal A(Y)={\bf a}_Y\}
}{
\widehat{\bar\pi}_K({\bf a}_Y)
} \ 
\prod_{X_{k,r}\in\mathbb L}
\widehat R_{k,r}^{(1)}
\big(X_{k,r},\mathcal X(X_{k,r}), C; {\bf a}_{X_{k,r}},{\bf a}_Y[1:k]\big).
\end{align*}

An augmented estimator of the sensitivity functional is
\begin{align}
\widehat\psi_t
=
P_n\left[
\widehat W_Y
\{Y-\widehat\mu_{K,R_K}(\mathcal X(Y), C)\}
+
h_t(C; \widehat {\mu},\widehat f)
\right].
\label{eq:sens_est}
\end{align}

For each fixed collection of sensitivity functions, this estimator is doubly robust with respect to the outcome regression and treatment mechanisms, provided all intermediate-variable laws entering the functional are consistently estimated. Specifically, $\widehat\psi_t$ is consistent if all $\widehat f_{k,r}$ are consistent and either $\widehat\mu_{K,R_K}$ or all $\widehat\pi_k$, $k=1,\ldots,K$, are consistent. A proof is given in Appendix~\ref{app:proofs_sa_DR}. See Appendix~\ref{app:ex_sa} for two worked examples.

\vspace{-0.3cm}
\section{Simulation studies}
\label{sec:sims} 

Code for reproducing all experiments is available on GitHub at \href{https://github.com/xxou/Longitudinal-PSE}{xxou/Longitudinal-PSE}. 

We conducted three simulation experiments to examine the finite-sample behavior of the proposed one-step estimators, the benefit of flexible nuisance estimation, and the robustness guarantees of Corollaries~\ref{cor:robustness_1} and~\ref{cor:robustness_2}. We considered $K=2$ and $R_1=R_2=2$, with temporal ordering $C, A_1, M_1, L_1, A_2, M_2, L_2, Y$, where $A_1$ and $A_2$ are binary treatments, $M_1,L_1,M_2,L_2$ are causally ordered intermediate variables, and $Y$ is continuous. For each simulation, we generated $1000$ Monte Carlo samples at each sample size with $n\in\{500,1000,2000,4000,8000\}$.

The target assignment pattern is ${\bf a}_{X_{1,1}}=0,\, {\bf a}_{X_{1,2}}=1,\, {\bf a}_{X_{2,1}}=(0,0),\, {\bf a}_{X_{2,2}}=(1,1),\, {\bf a}_Y=(0,0)$. Thus, the $M_1, M_2$ variables and the outcome receive the reference treatment assignment, whereas the $L_1, L_2$ variables receive the active treatment assignment. When contrasted with the counterfactual mean under the all-reference treatment regime, the resulting parameter captures treatment effects transmitted through $L_1$ or $L_2$, including paths $A_1\rightarrow L_1\leadsto Y$, $A_1\rightarrow L_2\rightarrow Y$, and $A_2\rightarrow L_2\rightarrow Y$, where $L_1\leadsto Y$ denotes all proper paths from $L_1$ to $Y$.

\vspace{0.25cm}
\noindent \textbf{Simulation~1: Finite-sample behavior under correct specification.} 
The first experiment examined the asymptotic behavior described in Theorems~\ref{thm:asymp_lin_1} and~\ref{thm:asymp_lin_2}. Under correct nuisance-model specification, we expect bias to be $o(n^{-1/2})$ and $n$-scaled variance to approach the efficiency bound $P\{\Phi_1(Q_1)^2\}=P\{\Phi_2(Q_2)^2\}$. The data-generating process (DGP) and the correctly specified nuisance models are given in Appendix~\ref{app:sims}. Nuisance models are consistently estimable via generalized linear models (GLMs). 

As shown in Table~\ref{tab:sim1}, both estimators had negligible bias, decreasing Monte Carlo standard deviations and mean squared errors, and coverage close to the nominal $95\%$ level. The finite-sample performance of the two estimators was nearly indistinguishable. Figure~\ref{fig:theoretical_estimators} further shows that the $\sqrt n$-scaled bias remained close to zero and that the $n$-scaled variance approached the theoretical efficiency bound.  

\begin{table}[!htbp]
\caption{Finite-sample performance under correct nuisance-model specification.}
\label{tab:sim1}
\centering
\setlength{\tabcolsep}{12pt}
\fontsize{11pt}{12pt}\selectfont
\begin{tabular}{lrccccc}
\toprule
& $n$ & Bias & SD & MSE & Coverage & CI width \\ 
\midrule\addlinespace[2.5pt]
\multirow{5}{*}{\rotatebox{90}{$\psi^+(\widehat{Q}_1)$}} 
& 500 & 0.002 & 0.162 & 0.026 & 0.966 & 0.664 \\
& 1000 & -0.002 & 0.120 & 0.014 & 0.952 & 0.466 \\
& 2000 & -0.002 & 0.084 & 0.007 & 0.953 & 0.328 \\
& 4000 & -0.001 & 0.060 & 0.004 & 0.957 & 0.232 \\
& 8000 & 0.000 & 0.041 & 0.002 & 0.958 & 0.164 \\
\cmidrule(lr){1-7}
\addlinespace[2.5pt]
\multirow{5}{*}{\rotatebox{90}{$\psi^+(\widehat{Q}_2)$}} 
& 500 & 0.002 & 0.165 & 0.027 & 0.961 & 0.674 \\
& 1000 & -0.002 & 0.121 & 0.015 & 0.953 & 0.469 \\
& 2000 & -0.002 & 0.084 & 0.007 & 0.949 & 0.329 \\
& 4000 & -0.001 & 0.060 & 0.004 & 0.954 & 0.233 \\
& 8000 & 0.000 & 0.041 & 0.002 & 0.958 & 0.164 \\
\bottomrule
\end{tabular}
\end{table}

\begin{figure}[!htbp]
\centering
\includegraphics[width=0.92\textwidth]
{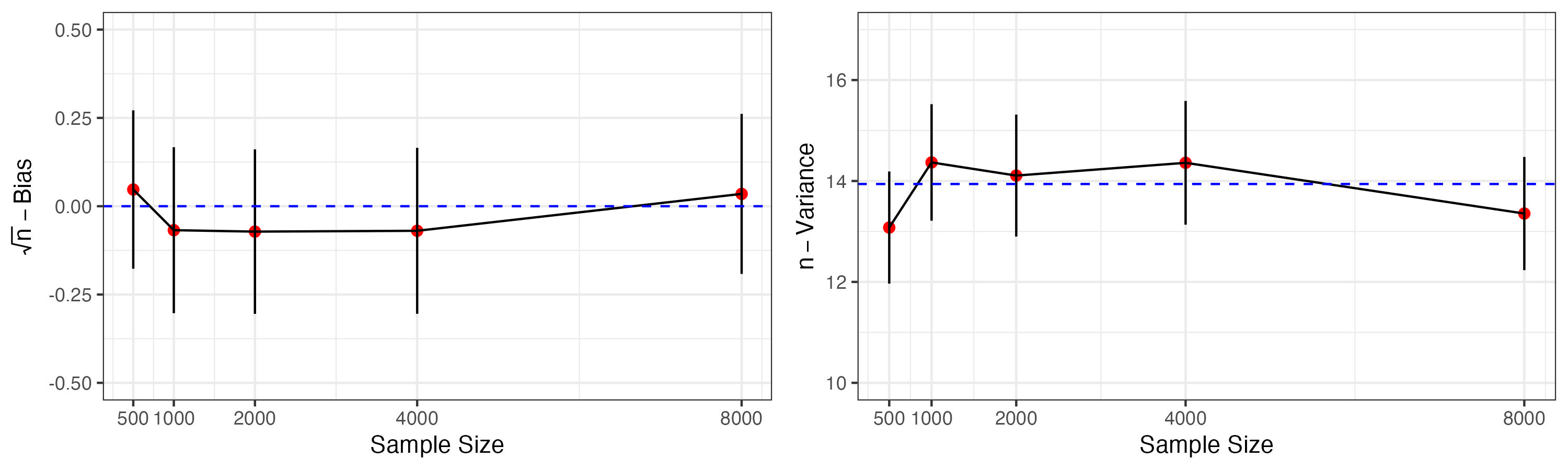}

\vspace{0.5em}

\includegraphics[width=0.92\textwidth]
{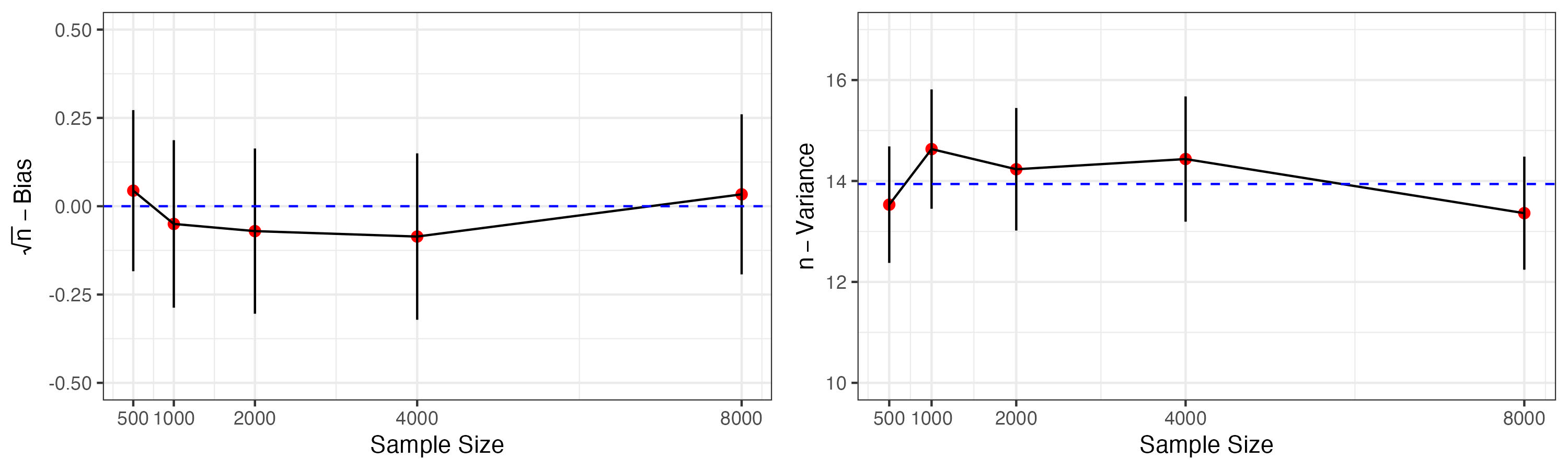}
\caption{Correct-specification simulations for the density-based estimator (upper panel) and regression-based estimator (lower panel). The panels show the $\sqrt n$-scaled bias and $n$-scaled variance, with 95\%confidence interval; horizontal dashed lines indicate zero bias (left) and the theoretical efficiency bound (right).}
\label{fig:theoretical_estimators}
\end{figure}

\noindent \textbf{Simulation~2: Misspecified parametric models vs. flexible estimation.} 
Using the DGP from Simulation~1, we compared two nuisance-estimation strategies. The first used parametric GLMs containing only linear and first-order terms and was deliberately misspecified by omitting the higher-order and interaction terms in the data-generating process. The second used a Super Learner (SL) ensemble containing \texttt{xgboost}, \texttt{ranger}, \texttt{dbarts}, \texttt{glm}, and \texttt{glm.interaction}. Additional implementation details are provided in Appendix~\ref{app:sims}. 

Table~\ref{tab:sim2} shows persistent GLM bias for both estimators. Although their standard deviations decreased with $n$, the bias did not attenuate over the examined sample sizes. In contrast, the SL bias generally decreased toward zero. The two approaches had similar Monte Carlo standard deviations, suggesting that nuisance-model misspecification primarily affected centering in this experiment. 
SL coverage was below nominal at smaller sample sizes (about $92\%$ for $n\le 2000$) and reached $94\%$ at $n=8000$, whereas GLM coverage fell to about $88\%$ as standard errors decreased while bias persisted.

\vspace{0.2cm}

\begin{table}[!htbp]
\caption{Finite-sample performance under misspecified GLMs and Super Learner nuisance estimation.}
\label{tab:sim2}
\centering
\setlength{\tabcolsep}{7pt} 
\fontsize{11pt}{12pt}\selectfont
\begin{tabular}{lrcccccccccc}
\toprule
& & \multicolumn{2}{c}{Bias} & \multicolumn{2}{c}{SD} & \multicolumn{2}{c}{MSE} & \multicolumn{2}{c}{Coverage} & \multicolumn{2}{c}{CI width} \\ 
\cmidrule(lr){3-4} \cmidrule(lr){5-6} \cmidrule(lr){7-8} \cmidrule(lr){9-10} \cmidrule(lr){11-12}
& $n$ & GLM & SL & GLM & SL &  GLM & SL & GLM & SL & GLM & SL \\ 
\midrule\addlinespace[2.5pt]
\multirow{5}{*}{\rotatebox{90}{$\psi^+(\widehat{Q}_1)$}}
& 500 & -0.031 & -0.003 & 0.161 & 0.152 & 0.027 & 0.023 & 0.957 & 0.915 & 0.654 & 0.527 \\ 
& 1000 & -0.033 & -0.006 & 0.118 & 0.114 & 0.015 & 0.013 & 0.950 & 0.922 & 0.460 & 0.391 \\ 
& 2000 & -0.034 & -0.006 & 0.083 & 0.081 & 0.008 & 0.007 & 0.925 & 0.918 & 0.324 & 0.285 \\ 
& 4000 & -0.033 & -0.002 & 0.059 & 0.058 & 0.005 & 0.003 & 0.909 & 0.926 & 0.230 & 0.207 \\ 
& 8000 & -0.031 & 0.001 & 0.040 & 0.040 & 0.003 & 0.002 & 0.883 & 0.940 & 0.162 & 0.149 \\ 
\cmidrule(lr){1-12}
\addlinespace[2.5pt]
\multirow{5}{*}{\rotatebox{90}{$\psi^+(\widehat{Q}_2)$}}
& 500 & -0.030 & -0.006 & 0.161 & 0.155 & 0.027 & 0.024 & 0.957 & 0.923 & 0.656 & 0.545 \\ 
& 1000 & -0.033 & -0.009 & 0.118 & 0.115 & 0.015 & 0.013 & 0.949 & 0.922 & 0.461 & 0.400 \\ 
& 2000 & -0.034 & -0.009 & 0.083 & 0.081 & 0.008 & 0.007 & 0.925 & 0.920 & 0.324 & 0.290 \\ 
& 4000 & -0.033 & -0.005 & 0.059 & 0.058 & 0.005 & 0.003 & 0.909 & 0.931 & 0.230 & 0.210 \\ 
& 8000 & -0.031 & -0.001 & 0.040 & 0.040 & 0.003 & 0.002 & 0.883 & 0.943 & 0.162 & 0.151 \\  
\bottomrule
\end{tabular}
\end{table}

Supplementary Figure~\ref{fig:flex_est} displays the corresponding $\sqrt n$-scaled biases and $n$-scaled variances. The GLM scaled bias moved away from zero as $n$ increased, whereas the SL scaled bias remained substantially closer to zero. 

\vspace{0.25cm}
\noindent \textbf{Simulation~3: Robustness against misspecification.}  
The third experiment examined the consistency regimes in Corollaries~\ref{cor:robustness_1} and~\ref{cor:robustness_2} using a second DGP comprising four independent uniform baseline covariates, two binary treatments, four continuous intermediate variables, and a continuous outcome. Details of the DGP and misspecified nuisance models appear in Appendix~\ref{app:simu_misspecification}.

For the density-based estimator, $\psi^+(\widehat Q_1)$, let
$\mathcal B_\pi=\{\pi_1,\pi_2\}$,
$\mathcal B_M=\{\mu_{2,2},f_{1,1},f_{2,1}\}$, and
$\mathcal B_L=\{f_{1,2},f_{2,2}\}$,
where $f_{1,1}$ and $f_{2,1}$ are the conditional laws of $M_1$ and $M_2$, and $f_{1,2}$ and $f_{2,2}$ those of $L_1$ and $L_2$. Sequential regressions were recursively induced by $\widehat\mu_{2,2}$ and the fitted intermediate-variable laws. We examined the three two-out-of-three regimes, condition 1:
$(\mathcal B_\pi,\mathcal B_L)$,
condition 2: $(\mathcal B_M,\mathcal B_L)$, and
condition 3: $(\mathcal B_\pi,\mathcal B_M)$,
corresponding to misspecification of the $\mathbb M$ block, treatment mechanisms, and $\mathbb L$ block, respectively. 

For the regression-based estimator, $\psi^+(\widehat Q_2)$, the protected regimes consist of the two endpoint cases and the four possible cut points in the ordering $X_{1,1}=M_1\prec X_{1,2}=L_1 \prec X_{2,1}=M_2\prec X_{2,2}=L_2$. Table~\ref{tab:sim_robustness_regimes} gives the nuisance functions that were correctly specified in each regime; all nuisance functions not displayed in a row were misspecified. 


\begin{table}[!htbp]
\caption{Nuisance specifications in the regression-based robustness simulations.}
\label{tab:sim_robustness_regimes}
\centering
\setlength{\tabcolsep}{10pt}
\begin{tabular}{cll}
\toprule
Condition & Regime & Correctly specified nuisance functions \\
\midrule
1 & regression-based endpoint
& $\pi_1,\pi_2,g_{1,1},g_{1,2},g_{2,1},g_{2,2}$ \\[2pt]
2 & Cut at $X_{2,2}$
& $\pi_1,\pi_2,g_{1,1},g_{1,2},g_{2,1},\mu_{2,2}$ \\[2pt]
3 & Cut at $X_{2,1}$
& $\pi_1,\pi_2,g_{1,1},g_{1,2},\mu_{2,1},\mu_{2,2}$ \\[2pt]
4 & Cut at $X_{1,2}$
& $\pi_1,g_{1,1},\mu_{1,2},\mu_{2,1},\mu_{2,2}$ \\[2pt]
5 & Cut at $X_{1,1}$
& $\pi_1,\mu_{1,1},\mu_{1,2},\mu_{2,1},\mu_{2,2}$ \\[2pt]
6 & Sequential-regression endpoint
& $\mu_{1,0},\mu_{1,1},\mu_{1,2},\mu_{2,1},\mu_{2,2}$ \\
\bottomrule
\end{tabular}
\end{table}


Misspecification was induced by replacing the baseline covariates with a set of  ``false covariates'' $C^{\mathrm{false}} = \{C_1^2,\exp(C_2),C_1^{1/2}/C_3,\cos(10C_4)\}$ in the corresponding GLMs. We also included all-correct and all-misspecified benchmarks. 

Figures~\ref{fig:robustness_density} and~\ref{fig:robustness_treg} show that both estimators remained well centered under the nuisance configurations protected by their respective corollaries, whereas appreciable bias remained when all nuisance models were misspecified.  
For the regression-based estimator, this held at both endpoints and at all four cut points, illustrating that the switch from treatment-history to sequential regressions can occur at any intermediate variable.

\begin{figure}[!htbp]
\centering
\includegraphics[width=0.95\textwidth]
{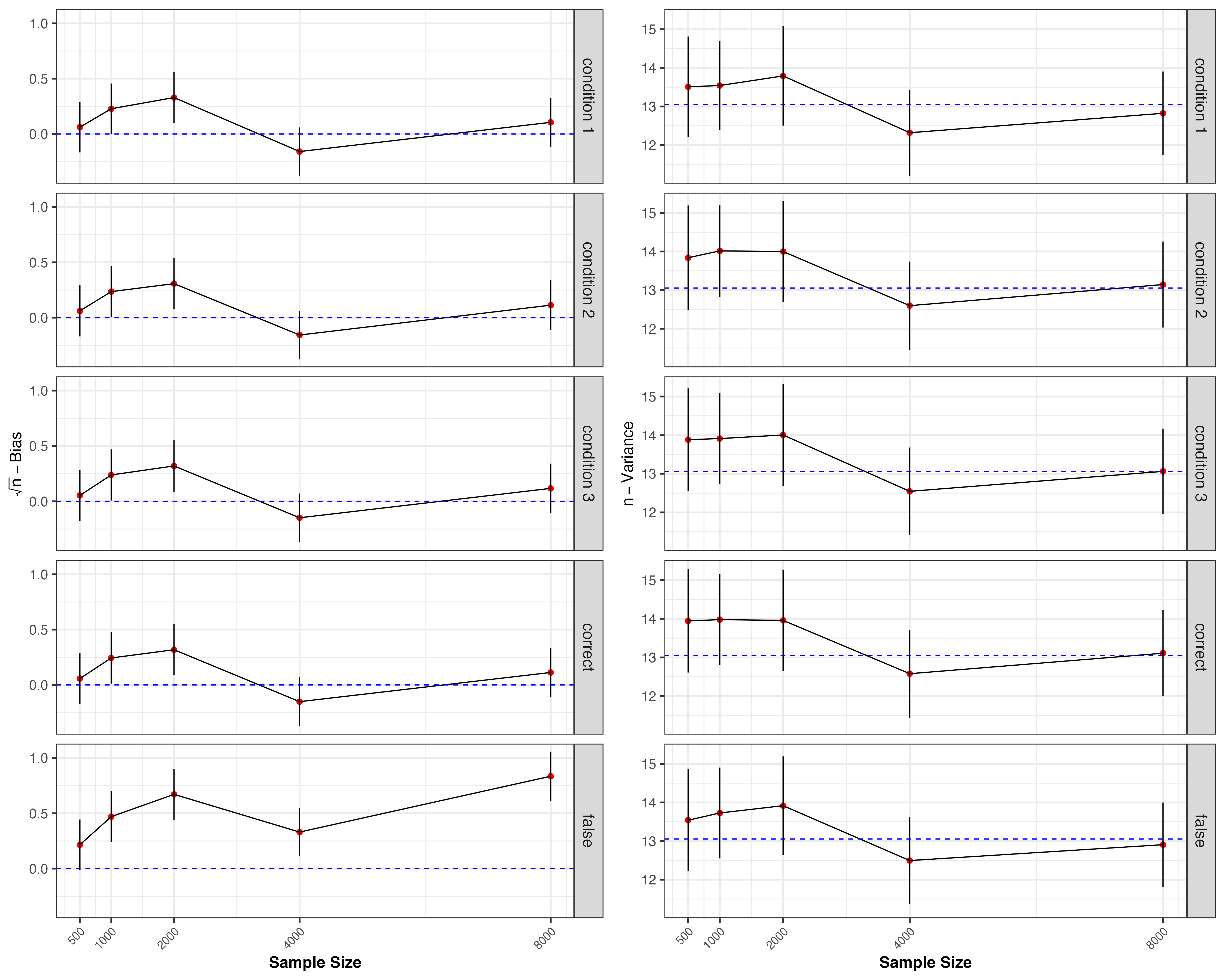}
\caption{Robustness simulations for the density-based estimator. The figure includes the three two-out-of-three regimes, corresponding to correct estimation of $(\mathcal B_\pi,\mathcal B_L)$, $(\mathcal B_M,\mathcal B_L)$, or $(\mathcal B_\pi,\mathcal B_M)$, together with all-correct and all-misspecified benchmarks.}
\label{fig:robustness_density}
\end{figure}

\begin{figure}[!htbp]
\centering
\includegraphics[width=0.95\textwidth]
{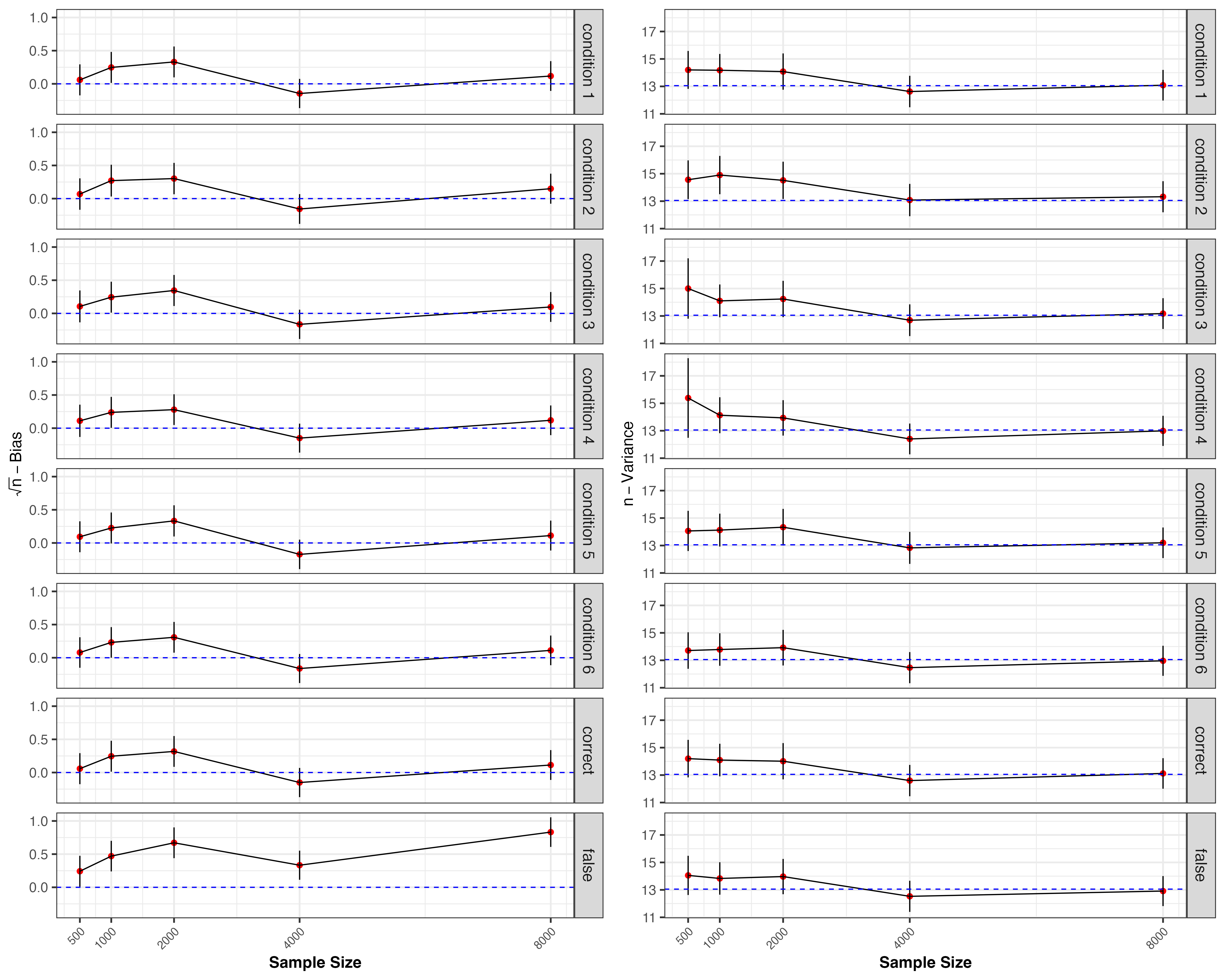}
\caption{Robustness simulations for the regression-based estimator. The figure includes the two endpoint regimes and the four cut-point regimes in Table~\ref{tab:sim_robustness_regimes}, together with all-correct and all-misspecified benchmarks. 
}
\label{fig:robustness_treg}
\end{figure}

\vspace{-1.cm}
\section{Real data application}
\label{sec:real_data} 

We illustrate the regression-based estimator using data from the original cohort of the Framingham Heart Study. The analysis examines how sustained smoking might affect systolic blood pressure (SBP) at Exam~7 along pathways through repeated body-mass-index (BMI) measurements. Baseline covariates came from Exam~1, whereas Exams~2, 4, 5, and 7 supplied repeated smoking indicators, BMI mediators, time-varying clinical covariates, and the outcome. The observed ordering was
$O=\{C_0,M_0,A_1,L_1,M_1,A_2,L_2,M_2,A_3,L_3,M_3,A_4,L_4,Y\}.$
Baseline covariates $C_0$ included sex, age, height, education, occupation, marital status, and baseline tobacco use. For $k \geq 1$, the $M_k$ blocks included earlier SBP, total cholesterol, and antihypertensive treatment whereas $M_0$ included only SBP at Exam 1 and total cholesterol at Exam 2; the full baseline covariates were $C=(C_0,M_0)$; $A_k$ indicated current smoking; $L_k$ was BMI; and $Y$ was Exam~7 SBP. Supplementary Table~\ref{tab:var_structure} gives the measurement schedule. Height was carried forward from Exam~1 to Exam~2 and from Exam~5 to Exam~7; isolated missing BMI or smoking measurements were carried forward by one selected examination. 

The analytic cohort excluded participants who died or were lost before Exam~7, those with diabetes, cancer, or cardiovascular disease at baseline, and those missing analytic variables, leaving 2,850 participants. The reported results therefore pertain to this selected complete-case survivor cohort and should not be interpreted as effects in the baseline cohort without additional assumptions concerning survival and retention. 

Let $\psi_{\mathrm{nonsmoke}}$ and $\psi_{\mathrm{smoke}}$ denote the counterfactual means under sustained nonsmoking and smoking, respectively. To define $\psi_{\mathrm{BMI}}$, the node-specific treatment assignments are ${\bf a}_{L_k}={\bf 1}_k$ for $k=1, \ldots,4$, ${\bf a}_{M_j}={\bf 0}_j$ for $j=1, \ldots,3$, and ${\bf a}_Y={\bf 0}_4$, where ${\bf 1}_k$ and ${\bf 0}_k$  are $k$-vectors of ones and zeros, respectively. For $i \in\{\mathrm{nonsmoke},\mathrm{smoke},\mathrm{BMI}\}$, we estimated $\psi_i$ using the one-step estimator in \eqref{eq:onestep} with the Bayes-rule regression-based representation ($\ell=2$), denoted by $\widehat\psi_i:=\psi_i^+(\widehat Q_{2})$. Then $\psi_{\mathrm{BMI}}-\psi_{\mathrm{nonsmoke}}$ represents the selected pathways through repeated BMI, and $\psi_{\mathrm{smoke}}-\psi_{\mathrm{BMI}}$ the complementary pathways; their sum equals the total-effect contrast $\psi_{\mathrm{smoke}}-\psi_{\mathrm{nonsmoke}}$. 

Table~\ref{tab:FHS_PSE} summarizes the estimated contrasts. Confidence intervals for the total and complementary pathway effects included zero under both nuisance-estimation strategies. In contrast, the estimated BMI-path effect was negative under both GLMs and Super Learner, with reductions of $1.82$ and $0.65$ mm Hg, respectively. The GLM confidence intervals were substantially wider than those obtained with Super Learner.

\begin{table}[!htbp]
\centering
\caption{Estimated effects (95\% CI) on Exam~7 systolic blood pressure (mm Hg) in the selected analytic cohort.}
\label{tab:FHS_PSE}
\setlength{\tabcolsep}{18pt}
\begin{tabular}{lcc}
\toprule
Contrast & GLM & Super Learner \\
\midrule
$\psi_{\mathrm{smoke}}-\psi_{\mathrm{nonsmoke}}$ & $-4.654$ ($-16.532$, $7.224$) & $0.126$ ($-0.910$, $1.163$) \\
$\psi_{\mathrm{BMI}}-\psi_{\mathrm{nonsmoke}}$ & $-1.824$ ($-3.178$, $-0.471$) & $-0.653$ ($-1.188$, $-0.118$) \\
$\psi_{\mathrm{smoke}}-\psi_{\mathrm{BMI}}$ & $-2.830$ ($-13.959$, $8.299$) & $0.780$ ($-0.142$, $1.702$) \\
\bottomrule
\end{tabular}
\begin{tablenotes}
\footnotesize
\item \textit{Note:} The Super Learner library contains \texttt{earth}, \texttt{xgboost}, \texttt{dbarts}, \texttt{ranger}, \texttt{ksvm} and \texttt{glm}, where we used $1000$ trees for \texttt{ranger} and the default settings for the remaining candidate learners.The parametric GLMs contain only linear and first-order terms.
\end{tablenotes}
\end{table}

\vspace{-1.2cm}
\section{Discussion}
\label{sec:disc}

This work extends semiparametric inference for path-specific effects beyond settings with a single treatment or fixed longitudinal mediation structure. The node-wise formulation indexes the statistical problem by treatment assignments to individual intermediate variables, rather than their labels as mediators or time-varying confounders, accommodating changing numbers, types, and orderings of intermediate variables across treatment occasions.

The two EIF representations lead to different nuisance-estimation requirements and robustness properties. The density-based estimator has blockwise multiple robustness, with blocks defined by treatment-assignment patterns. For the regression-based estimator, consistency can instead depend on different combinations of treatment-history and sequential outcome regressions across the longitudinal ordering. The regression-based approach avoids conditional-density estimation, which is particularly useful for continuous or multivariate intermediate variables. The density-based approach remains useful when the conditional laws are readily estimable or its blockwise robustness conditions are easier to satisfy.

Two directions for future work are broader sensitivity analyses and more stable estimation in longer longitudinal studies. The sensitivity analysis is restricted to a single non-outcome assignment pattern and discrete intermediate variables. Extending it to multiple assignment patterns and continuous or multivariate variables would broaden its applicability. Longer longitudinal sequences also require estimating treatment-history regressions for many assignment patterns, some of which may be sparsely represented in the data. Methods that pool information across
related regressions could improve stability, but would need to preserve the robustness properties established here.



\newpage

\begingroup
\renewcommand{\baselinestretch}{1.66}
\selectfont  
\setlength{\bibsep}{12pt}    

\endgroup



\endgroup
\arxivbeginsupplement
\begingroup
\title[Supplementary Materials]{Supplementary Materials for \\ \textit{Inference for Causal Path-Specific Effects in Longitudinal Studies with Repeated Treatments and Multiple Mediators}}

\author{Xiaxian Ou, Xinwei He, and Razieh Nabi$^*$ \\
Department of Biostatistics and Bioinformatics, Emory University, Atlanta, GA, U.S.A.\\ 
$^*$Corresponding author: razieh.nabi@emory.edu}

\renewcommand{\thesection}{S\arabic{section}}
\renewcommand{\theequation}{S\arabic{equation}}
\renewcommand{\thetable}{S\arabic{table}}
\renewcommand{\theposttable}{S\arabic{posttable}}
\setcounter{section}{0}
\setcounter{equation}{0}

\begingroup
\let\newpage\relax
\let\clearpage\relax
\maketitle
\endgroup

\label{supp:firstpage}

\vspace{-1.5cm}

The supplementary materials are structured as follows. Appendix~\ref{app:notation} provides a summary of the notation used throughout the manuscript. Appendix~\ref{app:proofs} contains all proofs. Appendix~\ref{app:examples} provides additional illustrative examples of the robustness conditions and sensitivity analysis. Appendix~\ref{app:sims} presents additional details and results for the simulation studies. Appendix~\ref{app:real_data} provides additional details for the real-data application.

\vspace{1.5cm}
\startcontents[supp]
{ 
\renewcommand{\baselinestretch}{0.8}\selectfont
\printcontents[supp]{}{1}{}
}

\newpage
\section{Glossary of terms and notations} 
\label{app:notation} 

To ease navigation of the notations, we provide a comprehensive list in Table~\ref{tab:notations}. Additional notations introduced in the supplementary materials are provided in Table~\ref{tab:supp_notations}. 

\begingroup
\renewcommand{\baselinestretch}{1}\selectfont
\fontsize{9}{10.5}\selectfont
\setlength{\tabcolsep}{2pt}
\renewcommand{\arraystretch}{1.1}
\setlength{\parindent}{0pt}
\setlength{\parskip}{0pt}
\setlength{\columnsep}{14pt}
\setlength{\columnseprule}{0.4pt}
\setlength{\multicolsep}{4pt}
\setlength{\emergencystretch}{1em}
\raggedcolumns
\newcommand{\glossaryentry}[2]{%
  \noindent
  \begin{tabular}[t]{@{}>{\raggedright\arraybackslash}p{.30\linewidth}>{\raggedright\arraybackslash}p{\dimexpr.70\linewidth-2\tabcolsep\relax}@{}}
    $#1$ & #2\\
  \end{tabular}\par}
\newcommand{\glossaryheading}[1]{%
  \par\addvspace{4pt}\noindent\textbf{\textit{#1}}\par\nopagebreak[4]}
\newcommand{\glossarycolumnheader}{%
  \begin{tabular}{@{}>{\raggedright\arraybackslash}p{.30\linewidth}>{\raggedright\arraybackslash}p{\dimexpr.70\linewidth-2\tabcolsep\relax}@{}}
    \textbf{Symbol} & \textbf{Definition}\\
  \end{tabular}}
\newcommand{\glossaryheaders}{%
  \par\hrule\vspace{3pt}\noindent
  \begin{minipage}[t]{\dimexpr.5\linewidth-.5\columnsep\relax}
    \glossarycolumnheader
  \end{minipage}\hfill
  \begin{minipage}[t]{\dimexpr.5\linewidth-.5\columnsep\relax}
    \glossarycolumnheader
  \end{minipage}\par\vspace{3pt}\hrule}
\begin{multicols}{2}[{%
  \captionof{table}{Glossary of terms and notations}
  \label{tab:notations}
  \glossaryheaders
}][100pt]
\glossaryheading{Observed data and causal notation}
\glossaryentry{O,\allowbreak O_i,\allowbreak n}{Observed data $O$; data for subject $i$ and sample size $n$.}
\glossaryentry{P,\allowbreak\mathcal M}{Observed-data law and nonparametric model; $P\in\mathcal M$.}
\glossaryentry{C,\allowbreak Y}{Baseline covariates and final outcome; $c$ is a realization of $C$.}
\glossaryentry{k,\allowbreak r,\allowbreak R_k}{Treatment occasion $k=1,\ldots,K$; intermediate-node index $r=1,\ldots,R_k$; $R_k$ is the number of intermediate nodes in block $k$.}
\glossaryentry{X_{k,r},\allowbreak\mathbb X}{Intermediate node $r$ in block $k$; the set of all intermediate nodes.}
\glossaryentry{A_k,\allowbreak{\bf A}_k}{Binary treatment at occasion $k$, and observed treatment history ${\bf A}_k=(A_1,\ldots,A_k)$.}
\glossaryentry{{\bf a}_K,\allowbreak{\bf a}'_K}{Active and reference treatment-assignment vectors.}
\glossaryentry{{\bf a}_k,\allowbreak{\bf a}_K[1:k]}{First $k$ components of ${\bf a}_K$.}
\glossaryentry{{\bf b}_k,\allowbreak{\bf d}_k}{Generic binary treatment histories of length $k$.}
\glossaryentry{{\bf 0}_k,\allowbreak{\bf 1}_k}{Length-$k$ vectors of zeros and ones.}
\glossaryentry{\mathcal G,\allowbreak\pa_{\mathcal G}(Z)}{Causal DAG and parents of $Z$.}
\glossaryentry{f_Z,\allowbreak\epsilon_Z}{Structural function and exogenous error for node $Z$.}
\glossaryentry{\rho}{Selected proper pathways.}
\glossaryentry{\xi}{Treatment-free downstream continuation from a node to $Y$; empty for $Y$, $Y^\xi = Y^{\varnothing}$.}
\glossaryentry{(Z,\xi)}{Continuation formed by prepending $Z$ to $\xi$.}
\glossaryentry{Z^\xi}{Continuation-specific counterfactual $Z({\bf a}_K,{\bf a}'_K;\rho,\xi)$.}
\glossaryentry{W^{(Z,\xi)}}{Counterfactual of intermediate parent $W$ supplied to $Z^\xi$.}
\glossaryentry{a_j^\rho(Z;\xi)}{Input $a_j$ if the complete proper path through $(Z,\xi)$ is in $\rho$; $a'_j$ otherwise.}
\glossaryentry{X_{k,r}({\bf a}_k),\allowbreak Y({\bf a}_K)}{Potential intermediate variable under treatment ${\bf a}_k$ and potential outcome under ${\bf a}_K$.}
\glossaryentry{Y^\rho}{Nested outcome $Y({\bf a}_K,{\bf a}'_K;\rho)=Y^\varnothing$.}
\glossaryentry{\psi^\rho}{Counterfactual mean $E(Y^\rho)$.}
\glossaryentry{\mathcal A(Z)}{Treatment history preceding node $Z$.}
\glossaryentry{\mathcal X(Z),\allowbreak\mathcal X(z)}{Intermediate history before $Z$ and its realized value.}
\glossaryentry{\tau(Z),\allowbreak{\bf a}_Z}{Node-wise assignment map and its value, $\tau(Z)={\bf a}_Z$.}
\glossaryentry{{\bf a}_{X_{k,r}},\allowbreak{\bf a}_Y}{Node-specific treatment assignments of lengths $k$ and $K$.}
\glossaryentry{{\bf a}^{(s)},\allowbreak S}{Full-length non-outcome assignment patterns and their number.}
\glossaryentry{\mathbb M}{Nodes with ${\bf a}_{X_{k,r}}={\bf a}_Y[1:k]$.}
\glossaryentry{\mathbb L}{Collection of remaining nodes: $\mathbb X\setminus\mathbb M$.}
\glossaryentry{\mathbb L_s,\allowbreak s(Z)}{Disjoint assignment class and its fixed compatible class index.}
\glossaryentry{\prec,\allowbreak\succ,\allowbreak\succeq}{Before, after, or at/after in the intermediate-node ordering.}
\glossaryheading{Nuisance functions and weighting}
\glossaryentry{\pi_k}{$P(A_k=b_k\mid\mathcal A(A_k)={\bf b}_{k-1},\mathcal X(A_k),C)$.}
\glossaryentry{\bar\pi_k({\bf b}_k)}{$\prod_{j=1}^k\pi_j(b_j\mid{\bf b}_{j-1},\mathcal X(A_j),C)$.}
\glossaryentry{f_{k,r}}{$f_{k,r}(x_{k,r}\mid\mathcal A(X_{k,r})={\bf b}_k,\mathcal X(X_{k,r}),C)$.}
\glossaryentry{\nu_{k,r},\allowbreak\nu}{Node-specific dominating measure and product measure; $d\nu(x)=\prod_{k,r}d\nu_{k,r}(x_{k,r})$.}
\glossaryentry{g_{k,0}}{$P({\bf A}_k={\bf b}_k\mid\mathcal X(A_k),C)$.}
\glossaryentry{g_{k,r}}{$P({\bf A}_k={\bf b}_k\mid\mathcal X(X_{k,r}),X_{k,r},C)$.}
\glossaryentry{\mu_{K,R_K}}{$E\{Y\mid\mathcal A(Y)={\bf a}_Y,\mathcal X(Y),C\}$.}
\glossaryentry{\mu_{k,r}}{$E[\mu_{k,r+1}\mid{\bf a}_{X_{k,r+1}},\mathcal X(X_{k,r+1}),C]$}
\glossaryentry{\mu_{k,R_k}}{$E[\mu_{k+1,1}\mid{\bf a}_{X_{k+1,1}},\mathcal X(X_{k+1,1}),C]$}
\glossaryentry{\mu_{k,0}}{Boundary alias $\mu_{k-1,R_{k-1}}$ for $k\geq2$.}
\glossaryentry{\mu_{1,0}(C)}{$E[\mu_{1,1}(X_{1,1},C)\mid A_1={\bf a}_{X_{1,1}},C]$.}
\glossaryentry{\pi,\allowbreak f,\allowbreak g}{Collections of treatment mechanisms, node densities, and treatment-history regressions.}
\glossaryentry{\mu_{\mathrm{seq}}}{Sequential-regression collection $\{\mu_{1,0}\}\cup\{\mu_{k,r}\}_{k,r}$.}
\glossaryentry{Q_\ell}{Nuisance parameterization:\newline $Q_1=(\pi,f,\mu_{K,R_K})$; $Q_2=(\pi,g,\mu_{\mathrm{seq}})$.}
\glossaryentry{\ell\in\{1,\allowbreak2\}}{Density-based or regression-based parameterization.}
\glossaryentry{R_{k,r}^{(1)}}{Conditional density ratio under ${\bf b}_k$ versus ${\bf d}_k$, at $(X_{k,r},\mathcal X(X_{k,r}),C)$.}
\glossaryentry{R_{k,r}^{(2)}}{The probability ratio expressed using $g_{k,r}$ and $g_{k,r-1}$ by Bayes' rule.}
\glossaryentry{W_Y^{(\ell)}}{Outcome weight combining inverse treatment probabilities and ratios $R_{j,q}^{(\ell)}$ for $X_{j,q}\in\mathbb L$.}
\glossaryentry{W_{k,r}^{(\ell)}}{Weight for $X_{k,r}$ combining inverse treatment probabilities under ${\bf a}_{X_{k,r}}$ and ratios $R_{j,q}^{(\ell)}$ for $X_{j,q}\prec X_{k,r}$.}
\glossaryentry{m_Z}{$m_Y=Y$ and $m_{X_{k,r}}=\mu_{k,r}$.}
\glossaryentry{\psi(P)}{Identified edge g-formula}
\glossaryentry{I\{\cdot\}}{Indicator of the event in braces.}
\glossaryheading{Estimation, cross-fitting, and inference}
\glossaryentry{Ph}{Population average $\int h(o)\,dP(o)$.}
\glossaryentry{P_nh}{Empirical average $n^{-1}\sum_{i=1}^n h(O_i)$.}
\glossaryentry{\widehat\mu_{k,r}^{(1)}}{Regression induced by $\widehat\mu_{K,R_K}$ and fitted intermediate laws through backward integration.}
\glossaryentry{\widehat\mu_{k,r}^{(2)}}{Directly fitted sequential regression under $Q_2$.}
\glossaryentry{\widehat\psi_{\mathrm{gcomp}}}{Plug-in estimator integrating fitted intermediate laws and outcome regression.}
\glossaryentry{\widehat\psi_{\mathrm{seq}}}{Sequential-regression plug-in estimator $P_n\widehat\mu_{1,0}$.}
\glossaryentry{\psi(\widehat Q_\ell)}{The plug-in estimator for $\ell$}
\glossaryentry{\widehat\psi_{\mathrm{IPW},Z}^{(\ell)}}{$P_n\{\widehat W_Z^{(\ell)}\widehat m_Z\}$, with $\widehat m_Y=Y$ and $\widehat m_{X_{k,r}}=\widehat\mu_{k,r}^{(\ell)}$.}
\glossaryentry{\psi^+(\widehat Q_\ell)}{One-step estimator $\psi(\widehat Q_\ell)+P_n\Phi_\ell(\widehat Q_\ell)$.}
\glossaryentry{\Phi_\ell(P),\allowbreak\Phi_\ell(Q_\ell)}{Efficient influence function of $\psi(P)$.}
\glossaryentry{D_\ell(P)}{Uncentered representation $\Phi_\ell(P)(O)+\psi(P)$.}
\glossaryentry{\widehat Q_\ell}{Estimated nuisance collection.}
\glossaryentry{v(i)}{Validation fold with observation $i$.}
\glossaryentry{\widehat Q_{\ell,-v(i)}}{Nuisance functions trained outside observation $i$'s validation fold.}
\glossaryentry{\widehat D_{\ell,i}}{$D_\ell(\widehat Q_{\ell,-v(i)})(O_i)$; its average is the cross-fitted estimate.}
\glossaryentry{R_{2,\ell}}{Second-order remainder.}
\glossaryentry{\|h\|_2}{$L^2(P)$ norm $(Ph^2)^{1/2}$.}
\glossaryentry{b_k,\allowbreak u}{Rate indices for $\widehat\pi_k$ and $\widehat\mu_{K,R_K}$.}
\glossaryentry{m_{k,r},\allowbreak l_{k,r}}{Rate indices for $\widehat f_{k,r}$ at nodes in $\mathbb M$ and $\mathbb L$.}
\glossaryentry{v_{k,r},\allowbreak u_{k,r}}{Rate indices for $\widehat g_{k,r}$ and $\widehat\mu_{k,r}$.}
\glossaryentry{u_{1,0}}{Rate index for $\widehat\mu_{1,0}$.}
\glossaryentry{\mathcal B_\pi}{Treatment-mechanism block $\{\pi_k:k=1,\ldots,K\}$.}
\glossaryentry{\mathcal B_M}{Terminal regression $\mu_{K,R_K}$ and densities of nodes in $\mathbb M$.}
\glossaryentry{\mathcal B_{L_s},\allowbreak\mathcal B_L}{Density block for $\mathbb L_s$; written $\mathcal B_L$ when $S=1$.}
\glossaryheading{Sensitivity analysis}
\glossaryentry{t_{k,r}}{Node-specific sensitivity function.}
\glossaryentry{\delta_{t_{k,r}}}{Node-specific correction $\delta_{t_{k,r}}(\mathcal X(y),c;f)$, formed from the difference of the two assigned $t_{k,r}(1-f_{k,r})$ terms.}
\glossaryentry{m_t(\mathcal X(y),c;\mu, f)}{The sensitivity-adjusted outcome regression}
\glossaryentry{h_t(c;\mu, f)}{Integral of $m_t(\mathcal X(y),c;\mu,f)$ under the assigned intermediate laws.}
\glossaryentry{\psi_t(P)}{The sensitivity target $P\{h_t(C;\mu,f)\}$.}
\glossaryentry{\widehat\psi_t}{An augmented estimator of the sensitivity functional $P_n[\widehat W_Y\{Y-\widehat\mu_{K,R_K}\}+h_t(C;\widehat\mu,\widehat f)]$.}
\glossaryentry{\widehat W_Y}{Fitted density-based outcome weight $\widehat W_Y^{(1)}$ used in sensitivity analysis.}
\glossaryheading{Application}
\glossaryentry{M_j,\allowbreak L_j}{Names of intermediate variables in examples and applications.}
\glossaryentry{C_0,\allowbreak M_0}{Baseline demographic covariates and baseline clinical measurements in FHS; $C=(C_0,M_0)$.}
\glossaryentry{A_k}{In FHS, current smoking status at Exams~2, 4, 5, and 7, for $k=1,\ldots,4$.}
\glossaryentry{L_k}{In FHS, BMI at the four selected examinations, $k=1,\ldots,4$.}
\glossaryentry{M_k}{In FHS, clinical-covariate blocks between treatments, $k=1,\ldots,3$.}
\glossaryentry{Y}{In FHS, Exam~7 SBP.}
\glossaryentry{\psi_{\mathrm{nonsmoke}}}{Counterfactual mean under sustained nonsmoking.}
\glossaryentry{\psi_{\mathrm{smoke}}}{Counterfactual mean under sustained smoking.}
\glossaryentry{\psi_{\mathrm{BMI}}}{Counterfactual mean with only active treatment assignments to BMI nodes}
\glossaryentry{\widehat\psi_i,\allowbreak\psi_i^+(\widehat Q_2)}{Equivalent application estimators for $i\in\{\mathrm{nonsmoke},\mathrm{smoke},\mathrm{BMI}\}$, using $\ell=2$.}
\end{multicols}
\par\nobreak
\vspace{\dimexpr4pt-\multicolsep\relax}
\hrule height 0.4pt width \linewidth
\par\clearpage
\setlength{\postmulticols}{5pt}
\begin{multicols}{2}[{%
  \captionof{table}{Glossary of notation appearing only in the Supplementary Materials}
  \label{tab:supp_notations}
  \glossaryheaders
}][100pt]
\glossaryheading{Identification and estimation proofs}
\glossaryentry{Z^\rho}{Common continuation-specific counterfactual under edge consistency.}
\glossaryentry{\mathcal X(Z^\rho)}{Counterfactual intermediate history preceding $Z$.}
\glossaryentry{{\bf b}_{X_{k,r}}}{Generic treatment history for node $X_{k,r}$ in the identification assumptions.}
\glossaryentry{\mathcal H(Z)}{Observed history $\{C,\mathcal A(Z),\mathcal X(Z)\}$.}
\glossaryentry{P_\epsilon,\allowbreak S(O)}{Regular parametric submodel through $P$ and its score at $\epsilon=0$.}
\glossaryentry{S_{(\cdot)}}{Specific score components.}
\glossaryentry{\Phi_Y,\allowbreak\Phi_{k,r},\allowbreak\Phi_C}{Outcome, intermediate, and baseline contributions to the pathwise derivative.}
\glossaryentry{\nu_{\preceq(k,r)}}{Product measure over $X_{k,r}$ and preceding intermediate nodes.}
\glossaryentry{\widetilde W_Y^{(1)},\allowbreak\widetilde W_{k,r}^{(1)}}{Generic proof weights.}
\glossaryheading{Inference proofs}
\glossaryentry{f_{k,r}^{\bf b},\allowbreak\widehat f_{k,r}^{\bf b}}{$f_{k,r}\{X_{k,r}\mid{\bf b}[1:k],\mathcal X(X_{k,r}),C\}$ and its fitted counterpart.}
\glossaryentry{f_{k,r}^{\star},\allowbreak\widehat f_{k,r}^{\star}}{$f_{k,r}\{X_{k,r}\mid{\bf a}_{X_{k,r}},\mathcal X(X_{k,r}),C\}$ and its fitted counterpart.}
\glossaryentry{F_{\mathcal V}^{\bf b},\allowbreak F_{\mathcal V}^{\star}}{$F_{\mathcal V}^{\bf b}{=}\prod_{X_{k,r}\in\mathcal V}f_{k,r}^{\bf b}, F_{\mathcal V}^{\star}{=}\prod_{X_{k,r}\in\mathcal V}f_{k,r}^{\star}$, \newline Hats denote fitted products.}
\glossaryentry{\Lambda_{r:s}({\bf b})}{$\prod_{j=r}^s\frac{\pi_j\{b_j\mid{\bf b}[1:j-1],\mathcal X(A_j),C\}}{\widehat\pi_j\{b_j\mid{\bf b}[1:j-1],\mathcal X(A_j),C\}}$; equals one if $r>s$.}
\glossaryentry{m_{\mathbb Y},m_{\mathbb L},\allowbreak m_{\mathbb{YL}}}{Integrands in the density-based remainder decomposition.}
\glossaryentry{\mathbb L_{<s},\mathbb L_{>s}}{$\mathbb L_{<s}=\bigcup_{h<s}\mathbb L_h; \mathbb L_{>s}=\bigcup_{h>s}\mathbb L_h$.}
\glossaryentry{e_{\pi,k}}{$\widehat\pi_k-\pi_k$.}
\glossaryentry{e_\mu}{$\widehat\mu_{K,R_K}-\mu_{K,R_K}$.}
\glossaryentry{e_{f,k,r}}{$\widehat f_{k,r}-f_{k,r}$.}
\glossaryentry{\kappa}{Multiplier in a remainder term $P(\kappa e_1e_2)$.}
\glossaryentry{\delta}{Positive lower bound for relevant true and fitted treatment probabilities.}
\glossaryentry{G_{k,r},\widehat G_{k,r}}{$G_{k,r}({\bf b}_k){=}g_{k,r}({\bf b}_k{\mid} X_{k,r},\mathcal X(X_{k,r}),C)$; $\widehat G_{k,r}$ is its fitted counterpart.}
\glossaryentry{{\bf a}_{X_{k,r}}^+}{equals $\mathbf a_{X_{k,r+1}}$ within a block; $\mathbf a_{X_{k+1,1}}[1:k]$ at a block boundary; $\mathbf a_Y$ for the final transition.}
\glossaryentry{p_{k,r}^+,\widehat p_{k,r}^+}{Next-treatment conditional probability at a block boundary and its fitted counterpart; both equal one otherwise.}
\glossaryentry{B_{k,r}}{Product of fitted inverse treatment probabilities and preceding $\widehat G_{j,q}({\bf a}_{X_{j,q}})/\widehat G_{j,q}({\bf a}_{X_{j,q}}^+)$ ratios.}
\glossaryentry{h(Q_2)}{$D_2(Q_2)$.}
\glossaryentry{\widetilde Q_2}{$(\widehat\pi,\widehat g,\mu_{\mathrm{seq}})$: fitted $\pi,g$ with true sequential regressions.}
\glossaryentry{\Gamma_{k,r}}{$G_{k,r}({\bf a}_{X_{k,r}})\widehat G_{k,r}({\bf a}_{X_{k,r}}^+)\widehat p_{k,r}^+$\newline $-\widehat G_{k,r}({\bf a}_{X_{k,r}})G_{k,r}({\bf a}_{X_{k,r}}^+)p_{k,r}^+$.}
\glossaryheading{Sensitivity-analysis proofs and examples}
\glossaryentry{J,L_j}{$J$ is the number of nodes within $\mathbb L$; $L_j=X_{k_j,r_j}$ is the $j$th node in temporal order.}
\glossaryentry{(k_j,r_j),l_j}{Node indices and realization $l_j=x_{k_j,r_j}$.}
\glossaryentry{{\bf b}^{(1)}_j,{\bf b}^{y}_j}{${\bf b}^{(1)}_j={\bf a}^{(1)}[1:k_j]; {\bf b}^{y}_j={\bf a}_Y[1:k_j]$.}
\glossaryentry{t_j,\delta_{t_j}}{$t_j=t_{k_j,r_j}; \delta_{t_j}=\delta_{t_{k_j,r_j}}$.}
\glossaryentry{m_j,m_j^{\neq}}{Conditional means of $Y({\bf a}_Y,\mathcal X(y))$ under the same treatment and preceding history, additionally conditioning on $L_j=l_j$ or $L_j\neq l_j$.}
\glossaryentry{\overline m_j}{Conditional mean of $Y({\bf a}_Y,\mathcal X(y))$ without conditioning on $L_j$.}
\glossaryentry{\Delta_t}{$\Delta_t(\mathcal X(y),c;f){=}\sum_{X_{k,r}\in\mathbb L}\delta_{t_{k,r}}(\mathcal X(y),c;f)$; Replacing $f$ by $\widehat f$ gives the fitted one.}
\glossaryentry{P_f^\rho,P_{\widehat f}^\rho}{Conditional counterfactual intermediate-history law and its fitted counterpart.}
\glossaryentry{\pi^*,\mu_{K,R_K}^*}{Possibly misspecified limits of fitted treatment mechanisms and the terminal outcome regression.}
\glossaryentry{{\bf x},Y_{\bf x}}{In Case~2, ${\bf x}=(m_1,l_1,m_2,l_2)$ and $Y_{\bf x}=Y(a'_1,m_1,l_1,a'_2,m_2,l_2)$.}
\glossaryheading{Simulation specifications}
\glossaryentry{\beta_V, \alpha,\beta,\theta}{Coefficients.}
\glossaryentry{\eta(C)}{$\beta_{A_1}C_{A_1}^{\top}$.}
\glossaryentry{\nu(C)}{$\beta_{M_1C}C_{M_1}^{\top}$.}
\glossaryentry{\gamma(C)}{$\beta_{M_1A_1}[1,C_3]^{\top}$.}
\glossaryentry{Z(C)}{Normalizing constant in the odds-ratio representation.}
\glossaryentry{\zeta_{A_2}}{$-\log\!\left\{\frac{P(A_2\mid A_1=0,L_1,M_1,C)}{P(A_2\mid A_1=1,L_1,M_1,C)}\right\}$.}
\glossaryentry{g_{A_j,k,r}}{Component regression for $A_j$ used to construct $g_{k,r}$.}
\glossaryentry{\sigma_{k,r}}{Conditional standard deviation in a normal node-density model.}
\glossaryentry{C^{\mathrm{false}}}{$\{C_1^2,\exp(C_2),C_1^{1/2}/C_3,\cos(10C_4)\}$.}
\glossaryentry{Q_2^{\mathrm{false}}}{Nuisance collection obtained using $C^{\mathrm{false}}$ in the specified GLMs.}
\glossaryentry{\pi_j^{\mathrm{false}}\!,\! f_{k,r}^{\mathrm{false}}\!,\! \mu_{2,2}^{\mathrm{false}}}{Misspecified treatment probabilities, intermediate-variable densities and terminal outcome regression.}
\glossaryheading{Application table}
\glossaryentry{\overline M}{$(M_0,M_1,M_2,M_3)$.}
\glossaryentry{{\bf A}}{$(A_1,A_2,A_3,A_4)$.}
\end{multicols}
\par\nobreak
\vspace{\dimexpr4pt-\multicolsep\relax}
\hrule height 0.4pt width \linewidth
\par
\endgroup
\setcounter{posttable}{\value{table}}
\clearpage

\section{Proofs}
\label{app:proofs}

\subsection{Identification results}
\label{app:proofs_ID}

Fix active and reference treatment histories $({\bf a}_K,{\bf a}'_K)$ and a collection $\rho$ of proper causal pathways. Recall that $\tau(Z)$ denotes the node-wise treatment assignment induced by $\rho$, with ${\bf a}_{X_{k,r}}=\tau(X_{k,r})$ and ${\bf a}_Y=\tau(Y)$. For any node $Z$, $\mathcal A(Z)$ and $\mathcal X(Z)$ denote, respectively, the treatment and intermediate-variable history preceding $Z$. The identification result relies on the following assumptions.

\renewcommand{\labelenumi}{A\arabic{enumi}} 
\renewcommand{\labelenumii}{\alph{enumii})} 

\begin{enumerate}

\item \textit{Consistency.}
For any non-treatment node $Z\in \mathbb X \cup\{Y\}$, if the observed treatment and non-treatment histories preceding $Z$ equal those specified by a given counterfactual intervention, then the corresponding counterfactual value equals the observed value. In particular, $X_{k,r}({\bf b}_k, \mathcal{X}(x_{k,r}))=X_{k,r}$ whenever $\mathcal A(X_{k,r})={\bf b}_k$ and $\mathcal{X}(X_{k,r}) = \mathcal{X}(x_{k,r})$; and $Y({\bf b}_K, \mathcal{X}(y))=Y$ whenever $\mathcal A(Y)={\bf b}_K$ and $\mathcal{X}(Y) = \mathcal{X}(y)$, with the corresponding preceding non-treatment variables taking their observed values.

\item \textit{Positivity.}
For every $Z\in \mathbb X\cup\{Y\}$ and node-specific assignment ${\bf a}_Z$ appearing under the edge intervention, $P\{\mathcal A(Z)={\bf a}_Z\mid C,\mathcal X(Z)\}>0$ almost surely on the relevant support. For the regression-based parameterization, we additionally require $g_{k,r}({\bf b}_k\mid \mathcal X(X_{k,r}), C,X_{k,r})>0$ and $g_{k,r-1}({\bf b}_k\mid \mathcal X(X_{k,r}), C)>0$ almost surely for each treatment history ${\bf b}_k$ appearing in the Bayes-ratio representation. The latter conditions are required only for the regression-based reparameterization, not for identification by the edge $g$-formula.

\item \textit{NPSEM-IE.}
Each observed node satisfies $Z=f_Z\{\pa_{\mathcal G}(Z),\epsilon_Z\}$, with unrestricted structural functions and mutually independent  exogenous errors. 

\item \textit{Edge consistency.}
The intervention induced by $({\bf a}_K,{\bf a}'_K;\rho)$ is edge consistent if no intermediate variable takes incompatible counterfactual values along included and excluded proper pathways. Equivalently, the DAG contains no recanting witness for $\rho$.

\end{enumerate}

Under the NPSEM-IE, the mutually independent exogenous errors imply the conditional independence relations needed to identify the nested counterfactuals. In the present longitudinal ordering, these include the following.

For any generic treatment history ${\bf b}_k \in \{0,1 \}^k$, any admissible treatment history ${\bf b}_Z$ for the intermediate variable $Z$, and any realization values $\mathcal X(z)$ with $Z \in \mathbb X \cup \{Y\}$, given $A_k$ and any admissible collection of downstream counterfactual assignments,
\begin{align}
A_k
\indep
\Big\{
X_{k,1}({\bf b}_{X_{k,1}}, \mathcal X(x_{k,1})),\ldots,
X_{K,R_K}({\bf b}_{X_{K,R_K}}, \mathcal{X}(x_{K,R_K})),
Y({\bf b}_Y, \mathcal{X}(y))
\Big\}
\;\Big|\;
C,{\bf A}_{k-1},\mathcal X(A_k),
\label{eq:supp_treat_ignorability}
\end{align}
where ${\bf A}_{0} = \varnothing$. Likewise, for each intermediate variable $X_{k,r}$,
\begin{align}
X_{k,r}({\bf b}_{k}, \mathcal{X}(x_{k,r}))
\indep
\Big\{
X_{j,q}({\bf b}_{X_{j,q}}, \mathcal{X}(x_{j,q})):
X_{k,r}\prec X_{j,q}\Big\} 
\cup \Big\{Y({\bf b}_Y, \mathcal{X}(y)) \Big\}
\;\Big|\;
C,{\bf A}_k,\mathcal X(X_{k,r}),
\label{eq:supp_intermediate_ignorability}
\end{align}
with the downstream counterfactuals evaluated at fixed values of their preceding intermediate-variable arguments as required by the recursive counterfactual construction. These conditional independence relations are not additional assumptions beyond A3; they are consequences of the NPSEM-IE and are stated only to make the identification argument explicit. 

Under edge consistency, two downstream continuations, $\xi$ and $\xi'$, from the same intermediate variable to $Y$ must involve the same treatment values. Thus, the counterfactuals $X_{k,r}^{\xi}$ and $X_{k,r}^{\xi'}$ receive the unique treatment inputs ${\bf a}_Z=\tau(Z)$ and therefore coincide. For simplicity, we use $X_{k,r}^{\rho}$ to denote their common value, and then
$$
X_{k,r}^\rho =
X_{k,r}\{ {\bf a}_{X_{k,r}},\mathcal X(X_{k,r}^\rho) \},
\qquad
Y^\rho = Y\{{\bf a}_Y,\mathcal X(Y^\rho)\},
$$
where the counterfactual history  $ \mathcal X(Z^\rho) = \{X_{j,q}^\rho:X_{j,q}\prec Z\}$ for $Z \in \mathbb X \cup \{Y\}$.


Thus, conditional on $C=c$ and on the values of the counterfactual non-treatment history preceding a node, the structural equation for that node is obtained from the observed structural equation by replacing its treatment inputs with ${\bf a}_Z$.

We first establish the conditional distribution of each counterfactual intermediate variable. Consider $X_{k,r}$ and suppose the counterfactual non-treatment history preceding it takes value $\mathcal X(x_{k,r})$. By the recursive definition of the edge intervention,
\[
X_{k,r}^{\rho}
=
f_{X_{k,r}}
\left\{
c,
{\bf a}_{X_{k,r}},
\mathcal X(x_{k,r}),
\epsilon_{X_{k,r}}
\right\},
\]
where the notation suppresses the precise ordering of the arguments of the structural equation. Because $\epsilon_{X_{k,r}}$ is independent of the exogenous errors generating the preceding variables under the NPSEM-IE,
\begin{align}
&dP\left\{
X_{k,r}^{\rho}=x_{k,r}
\mid
C=c,\,
\mathcal X(X_{k,r}^{\rho})=\mathcal X(x_{k,r})
\right\}
=
dP\left[
f_{X_{k,r}}
\left\{
c,
{\bf a}_{X_{k,r}},
\mathcal X(x_{k,r}),
\epsilon_{X_{k,r}}
\right\}
=x_{k,r}
\right].
\label{eq:supp_cf_kernel}
\end{align}
By consistency and modularity of the structural equations, the right-hand side is the same conditional law that is observed among units with the corresponding treatment and non-treatment histories. Hence, under positivity,
\begin{align}
&dP\left\{
X_{k,r}^{\rho}=x_{k,r}
\mid
C=c,\,
\mathcal X(X_{k,r}^{\rho})=\mathcal X(x_{k,r})
\right\}
\nonumber\\
&\qquad=
dP\left\{
x_{k,r}
\mid
c,\,
\mathcal A(X_{k,r})={\bf a}_{X_{k,r}},\,
\mathcal X(X_{k,r})=\mathcal X(x_{k,r})
\right\}.
\label{eq:supp_X_kernel}
\end{align}

The same argument applies to the outcome. Conditional on $C=c$ and on the complete counterfactual non-treatment history taking value $\mathcal X(y)$,
\[
Y^\rho
=
f_Y
\left\{
c,
{\bf a}_Y,
\mathcal X(y),
\epsilon_Y
\right\}.
\]
Since $\epsilon_Y$ is independent of the exogenous errors generating the preceding variables,
\begin{align}
&E\left\{
Y^\rho
\mid
C=c,\,
\mathcal X(Y^\rho)=\mathcal X(y)
\right\}
=
E\left[
f_Y
\left\{
c,
{\bf a}_Y,
\mathcal X(y),
\epsilon_Y
\right\}
\right].
\label{eq:supp_Y_structural}
\end{align}
Consistency and positivity therefore give
\begin{align}
&E\left\{
Y^\rho
\mid
C=c,\,
\mathcal X(Y^\rho)=\mathcal X(y)
\right\}
\nonumber\\
&\qquad=
E\left\{
Y
\mid
C=c,\,
\mathcal A(Y)={\bf a}_Y,\,
\mathcal X(Y)=\mathcal X(y)
\right\}.
\label{eq:supp_Y_kernel}
\end{align}

We now obtain the joint distribution of the counterfactual intermediate variables recursively. Order the intermediate variables according to \eqref{eq:data_structure}. By the chain rule,
\begin{align}
&dP\big(
X_{1,1}^{\rho} {=} x_{1,1},\ldots,X_{K,R_K}^{\rho} {=} x_{K,R_K}
\mid c
\big)
{=}
\prod_{k=1}^K
\prod_{r=1}^{R_k}
dP\left\{
X_{k,r}^{\rho}{=}x_{k,r}
\mid
c,\,
\mathcal X(X_{k,r}^{\rho})=\mathcal X(x_{k,r})
\right\}.
\label{eq:supp_cf_chain}
\end{align}
Applying \eqref{eq:supp_X_kernel} to each factor gives
\begin{align}
&dP\big(
X_{1,1}^{\rho} {=} x_{1,1},\ldots,X_{K,R_K}^{\rho} {=} x_{K,R_K}
\mid c
\big)
{=}
\prod_{k=1}^K
\prod_{r=1}^{R_k}
dP\left\{
x_{k,r}
\mid
c,\,
\mathcal A(X_{k,r}){=}{\bf a}_{X_{k,r}},\,
\mathcal X(X_{k,r}){=}\mathcal X(x_{k,r})
\right\}.
\label{eq:supp_cf_joint}
\end{align}

Combining Equations~\eqref{eq:supp_Y_kernel} and \eqref{eq:supp_cf_joint}, the joint edge-intervention distribution of the intermediate variables and outcome satisfies
\begin{align}
& dP\big(
Y^\rho = y,
X_{1,1}^{\rho} = x_{1,1},\ldots,X_{K,R_K}^{\rho} =x_{K,R_K}
\mid c
\big) \nonumber \\
& \quad =
dP\left\{
y
\mid
c,\,
\mathcal A(Y)={\bf a}_Y,\,
\mathcal X(Y)=\mathcal X(y)
\right\}
\nonumber\\
&\qquad\quad\times
\prod_{k=1}^K
\prod_{r=1}^{R_k}
dP\left\{
x_{k,r}
\mid
c,\,
\mathcal A(X_{k,r})={\bf a}_{X_{k,r}},\,
\mathcal X(X_{k,r})=\mathcal X(x_{k,r})
\right\}.
\label{eq:supp_edge_factorization}
\end{align}
This is the edge $g$-formula for the intervention defined by $({\bf a}_K,{\bf a}'_K;\rho)$. Therefore,
\begin{align}
\psi^\rho
&=
E\left\{
Y({\bf a}_K,{\bf a}'_K;\rho)
\right\}
\nonumber\\
&=
\int
y\,
dP\left\{
y
\mid
c,\,
\mathcal A(Y)={\bf a}_Y,\,
\mathcal X(Y)=\mathcal X(y)
\right\}
\nonumber\\
&\qquad\times
\prod_{k=1}^K
\prod_{r=1}^{R_k}
dP\left\{
x_{k,r}
\mid
c,\,
\mathcal A(X_{k,r})={\bf a}_{X_{k,r}},\,
\mathcal X(X_{k,r})=\mathcal X(x_{k,r})
\right\}
\,dP(c)
\nonumber\\
&=
\psi(P),
\label{eq:supp_edge_gformula}
\end{align}
which is precisely the observed-data functional in \eqref{eq:pse_id}. This proves Lemma~\ref{lem:id}.

\subsection{Estimation results}
\label{app:proofs_est}

We derive the efficient influence function under the density-based parameterization $Q_1=(\pi,f,\mu_{K,R_K})$ and then show that the regression-based representation $Q_2=(\pi,g,\mu_{\text{seq}})$ is algebraically equivalent. 

Let $\{P_\epsilon:\epsilon\in(-\delta,\delta)\}$ be an arbitrary regular parametric submodel through $P=P_0$, with score
\[
S(O)
=
\left.
\frac{\partial}{\partial\epsilon}
\log p_\epsilon(O)
\right|_{\epsilon=0}.
\]
Under the observed-data factorization, the score can be decomposed as
\begin{equation}
S(O)
=
S_C(C)
+
\sum_{k=1}^K S_{A_k}(A_k\mid \mathcal H(A_k))
+
\sum_{k=1}^K\sum_{r=1}^{R_k}
S_{X_{k,r}}(X_{k,r}\mid \mathcal H(X_{k,r}))
+
S_Y(Y\mid \mathcal H(Y)),
\label{eq:score_decomp}
\end{equation}
where $\mathcal H(Z) = \{C,\mathcal A(Z),\mathcal X(Z)\}$ denotes the observed history preceding node $Z$. Each conditional score has conditional mean zero given its corresponding history.

The identifying functional in \eqref{eq:pse_id} can be written as
\begin{align}
\psi(P)
&=
\int
\mu_{K,R_K}(\mathcal X(y), c)
\prod_{k=1}^K\prod_{r=1}^{R_k}
f_{k,r}(x_{k,r}\mid {\bf a}_{X_{k,r}}, \mathcal X(x_{k,r}), c)
\,d\nu(x)\,dP(c),
\label{eq:psi_factorized_proof}
\end{align}
where the histories appearing in the integrand are evaluated at the corresponding integration variables. Notice that the treatment mechanism does not enter the identifying functional directly. Consequently, its pathwise derivative has contributions from the conditional law of $Y$, the conditional laws of the intermediate variables, and the marginal law of $C$, but not from perturbations of the treatment mechanism.

Differentiating \eqref{eq:psi_factorized_proof} along $P_\epsilon$ gives
\begin{equation}
\left.
\frac{\partial}{\partial\epsilon}
\psi(P_\epsilon)
\right|_{\epsilon=0}
=
\Phi_Y
+
\sum_{k=1}^K\sum_{r=1}^{R_k}\Phi_{k,r}
+
\Phi_C,
\label{eq:pathwise_decomp}
\end{equation}
where $\Phi_Y$, $\Phi_{k,r}$, and $\Phi_C$ denote the contributions from perturbing the outcome law, the conditional law of $X_{k,r}$, and the baseline law, respectively. We derive each term in turn.

\paragraph{Contribution from the outcome law}

Perturbing the conditional law of $Y$ while holding the remaining factors fixed yields
\begin{align}
\Phi_Y
&=
\int
y\,
S_Y\{y\mid {\bf a}_Y,\mathcal X(y), c\}
\,dP(y\mid {\bf a}_Y,\mathcal X(y), c)
\prod_{k=1}^K\prod_{r=1}^{R_k}
f_{k,r}(x_{k,r}\mid{\bf a}_{X_{k,r}}, \mathcal X(x_{k,r}), c)
\,d\nu(x)\,dP(c).
\label{eq:TY_initial}
\end{align}
To express this integral under the observed-data law, first consider the generic density-ratio form of the outcome weight,
\begin{align}
\widetilde W_Y^{(1)}
&=
\frac{
I\{\mathcal A(Y)={\bf a}_Y\}
}{
\bar\pi_K({\bf a}_Y)
}
\prod_{k=1}^K\prod_{r=1}^{R_k}
R_{k,r}^{(1)}
\{X_{k,r},\mathcal X(X_{k,r}),C; {\bf a}_{X_{k,r}},{\bf a}_Y[1:k]\}.
\label{eq:WY_generic_proof}
\end{align}
If $X_{k,r}\in\mathbb M$, then ${\bf a}_{X_{k,r}}={\bf a}_Y[1:k]$, so the corresponding ratio equals one. Hence $\widetilde W_Y^{(1)}=W_Y^{(1)}$ as defined in \eqref{eq:weight_Y}. Therefore, $\Phi_Y = E\left[ W_Y^{(1)} \, Y \, S_Y\{Y\mid {\bf a}_Y,\mathcal X(Y), C\} \right]$.  We subtract the outcome regression $\mu_{K,R_K}(\mathcal X(Y), C)$ without changing the expectation, as $E\left[ S_Y(Y{\mid}\mathcal H(Y)) {\mid}\mathcal H(Y) \right]{=}0$:
\begin{align}
\Phi_Y
&=
E\left[
W_Y^{(1)}
\{Y-\mu_{K,R_K}(\mathcal X(Y), C)\}
S_Y(Y\mid\mathcal H(Y))
\right].
\label{eq:TY_centered}
\end{align}
Moreover, $W_Y^{(1)} \{Y-\mu_{K,R_K}(\mathcal X(Y), C)\}$ has conditional mean zero given $\mathcal H(Y)$. It is therefore orthogonal to all score components corresponding to variables preceding $Y$. Hence
\begin{align}
\Phi_Y
&=
E\left[
W_Y^{(1)}
\{Y-\mu_{K,R_K}(\mathcal X(Y), C)\}
S(O)
\right].
\label{eq:TY_fullscore}
\end{align}

\paragraph{Contribution from an intermediate-variable law}

Fix an intermediate variable $X_{k,r}$. Perturbing its conditional law gives
\begin{align}
\Phi_{k,r}
&=
\int
\mu_{k,r}(x_{k,r}, \mathcal X(x_{k,r}), c)
S_{X_{k,r}}
\{x_{k,r}\mid{\bf a}_{X_{k,r}}, \mathcal X(x_{k,r}), c\}
\nonumber\\
&\qquad\qquad\times
f_{k,r}(x_{k,r}\mid{\bf a}_{X_{k,r}}, \mathcal X(x_{k,r}), c)
\prod_{X_{j,q}\prec X_{k,r}}
f_{j,q}(x_{j,q}\mid{\bf a}_{X_{j,q}}, \mathcal X(x_{j,q}), c)
\,d\nu_{\preceq(k,r)}\,dP(c),
\label{eq:Tkr_initial}
\end{align}
where all variables subsequent to $X_{k,r}$ have already been integrated out into $\mu_{k,r}(x_{k,r}, \mathcal X(x_{k,r}), c)$, and $d\nu_{\preceq(k,r)}$ denotes integration over $X_{k,r}$ and all intermediate variables preceding it.

Define the generic weight associated with $X_{k,r}$ by
\begin{align}
\widetilde W_{k,r}^{(1)}
&=
\frac{
I\{\mathcal A(X_{k,r})={\bf a}_{X_{k,r}}\}
}{
\bar\pi_k({\bf a}_{X_{k,r}})
}
\prod_{X_{j,q}\prec X_{k,r}}
R_{j,q}^{(1)}
\{X_{j,q},\mathcal X(X_{j,q}),C; {\bf a}_{X_{j,q}},{\bf a}_{X_{k,r}}[1:j]\}.
\label{eq:Wkr_generic_proof}
\end{align}
Changing measure from the node-specific intervention law in \eqref{eq:Tkr_initial} to the observed-data law gives
\begin{align}
\Phi_{k,r}
&=
E\left[
\widetilde W_{k,r}^{(1)}
\mu_{k,r}(X_{k,r}, \mathcal X(X_{k,r}), C)
S_{X_{k,r}}
\{X_{k,r}\mid\mathcal H(X_{k,r})\}
\right].
\label{eq:Tkr_weighted}
\end{align}

By the sequential-regression definition,
\begin{align}
\mu_{k,r-1}( \mathcal X(X_{k,r}), C)
&=
E\left[
\mu_{k,r}(X_{k,r},  \mathcal X(X_{k,r}), C)
\mid
\mathcal A(X_{k,r})={\bf a}_{X_{k,r}},
 \mathcal X(X_{k,r}), C
\right],
\label{eq:mu_conditional_proof}
\end{align}
with the block-boundary convention introduced in Section~\ref{sec:est}. Since the conditional score for $X_{k,r}$ has conditional mean zero, we may center the continuation regression:
\begin{align}
\Phi_{k,r}
&=
E\left[
\widetilde W_{k,r}^{(1)}
\left\{
\mu_{k,r}(X_{k,r},  \mathcal X(X_{k,r}), C)
-
\mu_{k,r-1}(\mathcal X(X_{k,r}), C)
\right\}
S_{X_{k,r}}
\{X_{k,r}\mid\mathcal H(X_{k,r})\}
\right].
\label{eq:Tkr_centered}
\end{align}
The weighted residual in braces has conditional mean zero given the history preceding $X_{k,r}$ under the observed law after reweighting. Consequently, it is orthogonal to score components for variables preceding $X_{k,r}$, while dependence only on variables through $X_{k,r}$ makes it orthogonal to scores for subsequent variables. Therefore,
\begin{align}
\Phi_{k,r}
&=
E\left[
\widetilde W_{k,r}^{(1)}
\left\{
\mu_{k,r}(X_{k,r},  \mathcal X(X_{k,r}), C)
-
\mu_{k,r-1}(\mathcal X(X_{k,r}), C)
\right\}
S(O)
\right].
\label{eq:Tkr_fullscore}
\end{align}

It remains to relate the generic weight $\widetilde W_{k,r}^{(1)}$ to the piecewise weights in Equations~\eqref{eq:weight_M} and~\eqref{eq:weight_L}.

If $X_{k,r}\in\mathbb M$, then ${\bf a}_{X_{k,r}}={\bf a}_Y[1:k]$. For every preceding $X_{j,q}\in\mathbb M$, ${\bf a}_{X_{j,q}} = {\bf a}_Y[1:j] = {\bf a}_{X_{k,r}}[1:j]$, so $R_{j,q}^{(1)} \{X_{j,q},\mathcal X(X_{j,q}),C; {\bf a}_{X_{j,q}},{\bf a}_{X_{k,r}}[1:j]\}=1$. Thus only preceding variables in $\mathbb L$ can contribute nontrivial ratios, and \eqref{eq:Wkr_generic_proof} reduces exactly to \eqref{eq:weight_M}.

If instead $X_{k,r}\in\mathbb L_s$, then ${\bf a}_{X_{k,r}} = {\bf a}^{(s)}[1:k]$. For a preceding variable $X_{j,q}\in\mathbb L_s$, its node-specific assignment satisfies ${\bf a}_{X_{j,q}} = {\bf a}^{(s)}[1:j] = {\bf a}_{X_{k,r}}[1:j]$, and hence the corresponding ratio is one. The remaining potentially nontrivial ratios arise from preceding variables in $\mathbb M$ and from variables assigned to other classes $\mathbb L_h$, $h\neq s$. Consequently, \eqref{eq:Wkr_generic_proof} reduces to \eqref{eq:weight_L}.

If the fixed full-length completion used to define the classes $\mathbb L_s$ is nonunique for a shorter node-specific history, this conclusion is unchanged. Indeed, any two compatible full-length vectors agree on the assignment prefix through that node; therefore the corresponding density ratio equals one whenever the relevant prefixes coincide. Hence the weight does not depend on the arbitrary fixed completion used to define the partition.

We may therefore replace $\widetilde W_{k,r}^{(1)}$ by $W_{k,r}^{(1)}$ in \eqref{eq:Tkr_fullscore} and obtain
\begin{align}
\Phi_{k,r}
&=
E\left[
W_{k,r}^{(1)}
\left\{
\mu_{k,r}(X_{k,r},  \mathcal X(X_{k,r}), C)
-
\mu_{k,r-1}( \mathcal X(X_{k,r}), C)
\right\}
S(O)
\right].
\label{eq:Tkr_final}
\end{align}

\paragraph{Contribution from the baseline law}

Finally, perturbing the marginal distribution of $C$ gives $\Phi_C = \int \mu_{1,0}(c) S_C(c) \,dP(c)$. Since $E\{S_C(C)\}=0$ and $\psi(P)=E\{\mu_{1,0}(C)\}$, $\Phi_C = E\left[ \{\mu_{1,0}(C)-\psi(P)\} S_C(C) \right]$.  Because $\mu_{1,0}(C)-\psi(P)$ is a function of $C$ alone and all conditional score components have mean zero given their past,
\begin{align}
\Phi_C
&=
E\left[
\{\mu_{1,0}(C)-\psi(P)\}
S(O)
\right].
\label{eq:TC_fullscore}
\end{align}

\paragraph{Combining the pathwise derivative terms}

Substituting Equations~\eqref{eq:TY_fullscore}, \eqref{eq:Tkr_final}, and~\eqref{eq:TC_fullscore} into \eqref{eq:pathwise_decomp} yields
\begin{align}
\left.
\frac{\partial}{\partial\epsilon}
\psi(P_\epsilon)
\right|_{\epsilon=0}
&=
E\Bigg[
\Bigg\{
W_Y^{(1)}
\left[
Y-\mu_{K,R_K}(\mathcal X(Y), C)
\right]
\nonumber\\
&\qquad+
\sum_{k=1}^K\sum_{r=1}^{R_k}
W_{k,r}^{(1)}
\left[
\mu_{k,r}(X_{k,r}, \mathcal X(X_{k,r}), C)
-
\mu_{k,r-1}(\mathcal X(X_{k,r}), C)
\right]
\nonumber\\
&\qquad+
\mu_{1,0}(C)-\psi(P)
\Bigg\}
S(O)
\Bigg].
\label{eq:pathwise_phi1}
\end{align}
The term in braces is precisely $\Phi_1(P)(O)$.

\paragraph{Equivalent regression-based representation}

It remains to establish the second representation. For any ${\bf b}_k,{\bf d}_k\in\{0,1\}^k$, Bayes' rule gives
\begin{align}
f_{k,r}(X_{k,r}\mid{\bf b}_k,\mathcal X(X_{k,r}), C)
&=
\frac{
P({\bf A}_k={\bf b}_k\mid X_{k,r}, \mathcal X(X_{k,r}), C)
}{
P({\bf A}_k={\bf b}_k\mid \mathcal X(X_{k,r}), C)
}
f_{k,r}(X_{k,r}\mid \mathcal X(X_{k,r}), C),
\label{eq:bayes_f_b}
\end{align}
and analogously
\begin{align}
f_{k,r}(X_{k,r}\mid{\bf d}_k,\mathcal X(X_{k,r}), C)
&=
\frac{
P({\bf A}_k={\bf d}_k\mid X_{k,r}, \mathcal X(X_{k,r}), C)
}{
P({\bf A}_k={\bf d}_k\mid \mathcal X(X_{k,r}), C)
}
f_{k,r}(X_{k,r}\mid \mathcal X(X_{k,r}), C).
\label{eq:bayes_f_c}
\end{align}
Taking the ratio and canceling the common marginal conditional density gives
\begin{align}
R_{k,r}^{(1)}( X_{k,r},\mathcal X(X_{k,r}), C ; {\bf b}_k,{\bf d}_k)
&=
\frac{
P({\bf A}_k={\bf b}_k\mid X_{k,r}, \mathcal X(X_{k,r}), C)
}{
P({\bf A}_k={\bf b}_k\mid \mathcal X(X_{k,r}), C)
}
\frac{
P({\bf A}_k={\bf d}_k\mid \mathcal X(X_{k,r}), C)
}{
P({\bf A}_k={\bf d}_k\mid X_{k,r}, \mathcal X(X_{k,r}), C)
}
\nonumber\\
&=
\frac{
g_{k,r}({\bf b}_k\mid X_{k,r}, \mathcal X(X_{k,r}), C)
}{
g_{k,r-1}({\bf b}_k\mid \mathcal X(X_{k,r}), C)
}
\frac{
g_{k,r-1}({\bf d}_k\mid \mathcal X(X_{k,r}), C)
}{
g_{k,r}({\bf d}_k\mid X_{k,r}, \mathcal X(X_{k,r}), C)
}
\nonumber\\
&=
R_{k,r}^{(2)}( X_{k,r},\mathcal X(X_{k,r}), C ; {\bf b}_k,{\bf d}_k),
\label{eq:bayes_ratio_proof}
\end{align}
whenever the denominators are positive. For $r=1$, the pre-$X_{k,1}$ regression in the denominator is understood as $g_{k,0}(\cdot\mid C,\mathcal X(A_k))$, according to the convention in Section~\ref{sec:est}.

Equation~\eqref{eq:bayes_ratio_proof} shows that $R_{k,r}^{(1)} = R_{k,r}^{(2)}$ almost surely for every $(k,r)$ and every pair of treatment histories for which the ratios are defined. Replacing each density ratio in $W_Y^{(1)}$ and $W_{k,r}^{(1)}$ by its regression-based representation therefore gives $W_Y^{(1)}=W_Y^{(2)}$ and $W_{k,r}^{(1)}=W_{k,r}^{(2)}$. Consequently, $\Phi_2(P)(O)=\Phi_1(P)(O)$ almost surely. 

\subsection{Inference results}
\label{app:proofs_inf}

We give conditions under which the empirical-process and second-order remainder terms in \eqref{eq:stochastic_behavior_onestep} are negligible. All arguments below may be applied conditionally on the training sample in each validation fold. Because the number of folds is fixed, the same conclusions then hold after averaging across folds.

For $\ell\in\{1,2\}$, sufficient conditions for the empirical-process term to be $o_p(n^{-1/2})$ without sample splitting are
\begin{enumerate}
\item $\Phi_\ell(\widehat Q_\ell)-\Phi_\ell(Q_\ell)$ belongs to a $P$-Donsker class with probability tending to one; and
\item $\|\Phi_\ell(\widehat Q_\ell)-\Phi_\ell(Q_\ell)\|_2=o_p(1)$. 
\end{enumerate}
Under cross-fitting, the nuisance functions used in each validation fold are estimated from the remaining observations and are therefore independent of that fold. Conditional on the corresponding training sample,
\[
(P_n-P)\{\Phi_\ell(\widehat Q_\ell)-\Phi_\ell(Q_\ell)\}
=
O_p\!\left[
n^{-1/2}
\|\Phi_\ell(\widehat Q_\ell)-\Phi_\ell(Q_\ell)\|_2
\right],
\]
where fold indices are suppressed. Consequently, $\|\Phi_\ell(\widehat Q_\ell)-\Phi_\ell(Q_\ell)\|_2=o_p(1)$ makes this term $o_p(n^{-1/2})$ without requiring a Donsker condition.

We further impose positivity and moment conditions sufficient to control the weights and remainder terms. There exists $\delta>0$ such that the true treatment probabilities appearing in the denominators of the EIF are at least $\delta$ almost surely, and their fitted counterparts are at least $\delta$ with probability tending to one. For the regression-based parameterization, the same condition applies to the relevant true and fitted treatment-history probabilities. For the density-based parameterization, we do not require continuous conditional densities to be globally bounded away from zero. Instead, we assume that the likelihood ratios appearing in the EIF are well defined and that, for every remainder term of the form $P(\kappa e_1e_2)$, its multiplier satisfies $|P(\kappa e_1e_2)| \lesssim_p \|e_1\|_2\|e_2\|_2$. A simple sufficient condition is that the multipliers are uniformly bounded in absolute value with probability tending to one. We also assume the square-integrability conditions required for the EIF and its estimated counterpart. All $L^2(P)$ errors below are understood uniformly over the finite collection of treatment histories at which each nuisance function is evaluated.

We repeatedly use the finite-product identity
\begin{equation}
\prod_{t=1}^T \widehat \lambda_t-\prod_{t=1}^T \lambda_t
=
\sum_{t=1}^T(\widehat \lambda_t-\lambda_t)
\left(\prod_{s<t}\widehat \lambda_s\right)
\left(\prod_{s>t}\lambda_s\right).
\label{eq:product_difference}
\end{equation}
Thus, a difference between two finite products can be written as a sum containing one componentwise estimation error per term.

\paragraph{\textcolor{blue}{Density-based estimator}}
For this estimator, all sequential regressions in the EIF are recursively induced by $\widehat\mu_{K,R_K}$ and the fitted conditional intermediate-variable laws. In particular, within each block,
\[
\widehat\mu_{k,r-1}(\mathcal X(X_{k,r}), C)
=
\int \widehat\mu_{k,r}(x_{k,r}, \mathcal X(X_{k,r}), C)
\widehat f_{k,r}(x_{k,r}\mid {\bf a}_{X_{k,r}},\mathcal X(X_{k,r}), C)
\,d\nu_{k,r}(x_{k,r}),
\]
with the analogous recursion at block boundaries. Any numerical integration error is assumed negligible at the rate under consideration. The remainder calculation below does not apply to sequential regressions fitted independently of $\widehat\mu_{K,R_K}$ and the $\widehat f_{k,r}$'s.

We first give the explicit remainder and then isolate its product-error structure. For a full treatment history ${\bf b}$, define
\[
f_{k,r}^{\bf b}
=f_{k,r}\{X_{k,r}\mid {\bf b}[1:k],\mathcal X(X_{k,r}), C\},
\qquad
\widehat f_{k,r}^{\bf b}
=\widehat f_{k,r}\{X_{k,r}\mid {\bf b}[1:k],\mathcal X(X_{k,r}), C\},
\]
and let
\[
f_{k,r}^{\star}
=f_{k,r}(X_{k,r}\mid {\bf a}_{X_{k,r}},\mathcal X(X_{k,r}), C),
\qquad
\widehat f_{k,r}^{\star}
=\widehat f_{k,r}(X_{k,r}\mid {\bf a}_{X_{k,r}},\mathcal X(X_{k,r}), C).
\]
For a set of intermediate variables $\mathcal V\subseteq\mathbb X$, write
\[
F_{\mathcal V}^{\bf b}=\prod_{X_{k,r}\in\mathcal V}f_{k,r}^{\bf b},
\qquad
F_{\mathcal V}^{\star}=\prod_{X_{k,r}\in\mathcal V}f_{k,r}^{\star},
\]
with analogous notation for fitted products. Subscripts such as $\mathcal V \, {\prec}X_{k,r}$ and $\mathcal V \, {\succ}X_{k,r}$ restrict the product to variables in $\mathcal V$ preceding or following $X_{k,r}$. Empty products equal one. Finally, define
\[
\Lambda_{r:s}({\bf b})
=
\prod_{j=r}^s
\frac{\pi_j\{b_j\mid {\bf b}[1:j-1], \mathcal X(A_j), C \}}
     {\widehat\pi_j\{b_j\mid {\bf b}[1:j-1],\mathcal X(A_j), C\}},
\]
where $\Lambda_{r:s}=1$ when $r>s$, and abbreviate $\mu=\mu_{K,R_K}$ and $\widehat\mu=\widehat\mu_{K,R_K}$.

Starting from $R_{2,1}(\widehat Q_1,Q_1)=P\{D_1(\widehat Q_1)\}-\psi(P)$ and using the recursive definitions of the sequential regressions, direct calculation gives
\begin{equation}
R_{2,1}(\widehat Q_1,Q_1)
=
\iint\{m_{\mathbb Y}(c,\mathcal X(y))+m_{\mathbb L}(c,\mathcal X(y))
             +m_{\mathbb{YL}}(c,\mathcal X(y))\}\,d\nu(\mathcal X(y))\,dP(c),
\label{eq:remainder_density_explicit}
\end{equation}
where
{\small
\begin{align}
m_{\mathbb Y}
={}&
\Lambda_{1:K}({\bf a}_Y)\widehat F_{\mathbb L}^{\star}
(\mu-\widehat\mu)F_{\mathbb M}^{{\bf a}_Y}
\left(\frac{F_{\mathbb L}^{{\bf a}_Y}}
           {\widehat F_{\mathbb L}^{{\bf a}_Y}}-1\right)
\nonumber\\
&+
\sum_{X_{k,r}\in\mathbb M}
\Lambda_{1:k}({\bf a}_Y)\widehat F_{\mathbb L}^{\star}\widehat\mu
\widehat F_{\mathbb M,{\succ}X_{k,r}}^{{\bf a}_Y}
(f_{k,r}^{{\bf a}_Y}-\widehat f_{k,r}^{{\bf a}_Y})
F_{\mathbb M,{\prec}X_{k,r}}^{{\bf a}_Y}
\left(
\frac{F_{\mathbb L,{\prec}X_{k,r}}^{{\bf a}_Y}}
     {\widehat F_{\mathbb L,{\prec}X_{k,r}}^{{\bf a}_Y}}-1
\right)
\nonumber\\
&+
\sum_{X_{k,r}\in\mathbb M}
\Lambda_{1:k}({\bf a}_Y)
\{1-\Lambda_{k+1:K}({\bf a}_Y)\}
\widehat F_{\mathbb L}^{\star}\widehat\mu
\widehat F_{\mathbb M,{\succ}X_{k,r}}^{{\bf a}_Y}
(f_{k,r}^{{\bf a}_Y}-\widehat f_{k,r}^{{\bf a}_Y})
F_{\mathbb M,{\prec}X_{k,r}}^{{\bf a}_Y}
\nonumber\\
&+
\{\Lambda_{1:K}({\bf a}_Y)-1\}\widehat F_{\mathbb L}^{\star}
\{\mu F_{\mathbb M}^{{\bf a}_Y}
  -\widehat\mu\widehat F_{\mathbb M}^{{\bf a}_Y}\},
\label{eq:mY_density}
\end{align}
\begin{align}
m_{\mathbb L}
={}&
\sum_{s=1}^S\sum_{X_{k,r}\in\mathbb L_s}
\Lambda_{1:k}({\bf a}^{(s)})
\widehat F_{\mathbb L\setminus\mathbb L_s}^{\star}
\widehat F_{\mathbb M}^{\star}\widehat\mu
\widehat F_{\mathbb L_s,{\succ}X_{k,r}}^{\star}
(f_{k,r}^{\star}-\widehat f_{k,r}^{\star})
F_{\mathbb L_s,{\prec}X_{k,r}}^{\star}
\left\{
\frac{F_{\mathbb M,{\prec}X_{k,r}}^{{\bf a}^{(s)}}}
     {\widehat F_{\mathbb M,{\prec}X_{k,r}}^{{\bf a}^{(s)}}}
\prod_{h\ne s}
\frac{F_{\mathbb L_h,{\prec}X_{k,r}}^{{\bf a}^{(s)}}}
     {\widehat F_{\mathbb L_h,{\prec}X_{k,r}}^{{\bf a}^{(s)}}}
-1
\right\}
\nonumber\\
&+
\sum_{s=1}^S\sum_{X_{k,r}\in\mathbb L_s}
\Lambda_{1:k}({\bf a}^{(s)})
\{1-\Lambda_{k+1:K}({\bf a}^{(s)})\}
\widehat F_{\mathbb L\setminus\mathbb L_s}^{\star}
\widehat F_{\mathbb M}^{\star}\widehat\mu
\widehat F_{\mathbb L_s,{\succ}X_{k,r}}^{\star}
(f_{k,r}^{\star}-\widehat f_{k,r}^{\star})
F_{\mathbb L_s,{\prec}X_{k,r}}^{\star}
\nonumber\\
&+
\sum_{s=1}^S
\{\Lambda_{1:K}({\bf a}^{(s)})-1\}
\widehat F_{\mathbb L\setminus\mathbb L_s}^{\star}
\widehat F_{\mathbb M}^{\star}\widehat\mu
\{F_{\mathbb L_s}^{\star}-\widehat F_{\mathbb L_s}^{\star}\},
\label{eq:mL_density}
\end{align}
and, writing
$\mathbb L_{<s}=\bigcup_{h<s}\mathbb L_h$ and
$\mathbb L_{>s}=\bigcup_{h>s}\mathbb L_h$,
\begin{align}
m_{\mathbb{YL}}
={}&
(\widehat F_{\mathbb L}^{\star}-F_{\mathbb L}^{\star})
\{\mu F_{\mathbb M}^{\star}
  -\widehat\mu\widehat F_{\mathbb M}^{\star}\}
\nonumber\\
&+
\widehat\mu\widehat F_{\mathbb M}^{\star}
\sum_{s=1}^{S-1}
\widehat F_{\mathbb L_{<s}}^{\star}
(\widehat F_{\mathbb L_s}^{\star}-F_{\mathbb L_s}^{\star})
(F_{\mathbb L_{>s}}^{\star}
 -\widehat F_{\mathbb L_{>s}}^{\star}).
\label{eq:mYL_density}
\end{align}
}
The use of the full extension ${\bf a}^{(s)}$ in \eqref{eq:mL_density} is only notational: for $X_{k,r}\in\mathbb L_s$, ${\bf a}^{(s)}[1:k]={\bf a}_{X_{k,r}}$ by construction.

To display the resulting pairwise error structure, write
\[
e_{\pi,k}=\widehat\pi_k-\pi_k,
\qquad
e_\mu=\widehat\mu_{K,R_K}-\mu_{K,R_K},
\qquad
e_{f,k,r}=\widehat f_{k,r}-f_{k,r},
\]
with arguments and treatment histories suppressed. Applying \eqref{eq:product_difference} to \eqref{eq:mY_density}--\eqref{eq:mYL_density} gives
\begin{align}
R_{2,1}(\widehat Q_1,Q_1)
={}&
\sum_{k=1}^K
P\{\kappa_{\pi\mu,k}e_{\pi,k}e_\mu\}
\nonumber\\
&+
\sum_{k=1}^K\sum_{X_{j,q}\in\mathbb M}
P\{\kappa_{\pi M,k,j,q}e_{\pi,k}e_{f,j,q}\}
\nonumber\\
&+
\sum_{k=1}^K\sum_{X_{j,q}\in\mathbb L}
P\{\kappa_{\pi L,k,j,q}e_{\pi,k}e_{f,j,q}\}
\nonumber\\
&+
\sum_{X_{j,q}\in\mathbb L}
P\{\kappa_{\mu L,j,q}e_\mu e_{f,j,q}\}
\nonumber\\
&+
\sum_{X_{k,r}\in\mathbb M}
\sum_{X_{j,q}\in\mathbb L}
P\{\kappa_{ML,k,r,j,q}e_{f,k,r}e_{f,j,q}\}
\nonumber\\
&+
\sum_{1\leq s<h\leq S}
\sum_{X_{j,q}\in\mathbb L_s}
\sum_{X_{j',q'}\in\mathbb L_h}
P\{\kappa_{LL,j,q,j',q'}e_{f,j,q}e_{f,j',q'}\}.
\label{eq:remainder_density_factorization}
\end{align}
Here each $\kappa$ is an explicit finite product of true and fitted nuisance functions, inverse treatment probabilities, and likelihood ratios produced by \eqref{eq:product_difference}; its particular value may differ across treatment histories. Importantly, every summand contains errors from two distinct nuisance blocks. There are no products involving two errors from within the same $\mathbb M$ block or within the same $\mathbb L_s$ block.

Under the multiplier conditions above,
\[
|P(\kappa e_1e_2)|
\leq \|\kappa\|_\infty\|e_1\|_2\|e_2\|_2.
\]
Consequently, \eqref{eq:remainder_density_factorization} is $o_p(n^{-1/2})$ under the product-rate conditions in Theorem~\ref{thm:asymp_lin_1}. More generally, essential boundedness of $\kappa$ may be replaced by weighted $L^2(P)$ rates or by H\"older-compatible higher-moment conditions.

The same factorization establishes Corollary~\ref{cor:robustness_1}. Indeed, every term in \eqref{eq:remainder_density_factorization} vanishes in the limit if all but at most one of
\[
\{\pi_k:k=1,\ldots,K\},\qquad
\{\mu_{K,R_K},f_{k,r}:X_{k,r}\in\mathbb M\},\qquad
\{f_{j,q}:X_{j,q}\in\mathbb L_s\},\quad s=1,\ldots,S,
\]
are consistently estimated. This gives the claimed $(S+1)$-out-of-$(S+2)$ robustness.

\paragraph{\textcolor{blue}{regression-based estimator}}
We now derive the remainder after the sequential regressions and Bayes-rule reparameterization. Define
\[
G_{k,r}({\bf b}_k)
=g_{k,r}({\bf b}_k\mid X_{k,r}, \mathcal X(X_{k,r}), C),
\qquad
\widehat G_{k,r}({\bf b}_k)
=\widehat g_{k,r}({\bf b}_k\mid X_{k,r}, \mathcal X(X_{k,r}), C).
\]
For every $(k,r)$, let ${\bf a}_{X_{k,r}}^{+}$ denote the treatment history assigned to the next non-treatment variable in the ordering:
\[
{\bf a}_{X_{k,r}}^{+}
=
\begin{cases}
{\bf a}_{X_{k,r+1}}, & r<R_k,\\
{\bf a}_{X_{k+1,1}}[1:k], & r=R_k,\ k<K,\\
{\bf a}_Y, & r=R_K,\ k=K.
\end{cases}
\]
Also define
\[
p_{k,r}^{+}
=
\begin{cases}
\pi_{k+1}\{{\bf a}_{X_{k+1,1}}[k+1]
\mid {\bf a}_{X_{k+1,1}}[1:k], \mathcal X(A_{k+1}), C\},
& r=R_k,\ k<K,\\
1, & \text{otherwise},
\end{cases}
\]
and let $\widehat p_{k,r}^{+}$ be its fitted counterpart. Thus, $p_{k,r}^{+}$ accounts for the additional treatment mechanism only when the successor lies in the next intermediate-variable block.

For notational convenience, set
\begin{align*}
B_{k,r}
={}&
\left[
\prod_{j=1}^k
\widehat\pi_j\{{\bf a}_{X_{j,1}}[j]
\mid {\bf a}_{X_{j,1}}[1:j-1], \mathcal X(A_{j}), C\}
\right]^{-1}
\prod_{X_{j,q}\prec X_{k,r}}
\frac{\widehat G_{j,q}({\bf a}_{X_{j,q}})}
{\widehat G_{j,q}({\bf a}_{X_{j,q}}^{+})}.
\end{align*}
The convention for $\widehat g_{k,0}$ in the main text ensures that no additional nuisance function is introduced at a block boundary.

Let $h(Q_2)=D_2(Q_2)$. If $\widetilde Q_2$ replaces all estimated sequential regressions in $\widehat Q_2$ by their true values while retaining $\widehat\pi$ and $\widehat g$, the sequential conditional-expectation identities give
\[
P\{h(\widetilde Q_2)\}=\psi(P).
\]
It follows that
\[
R_{2,2}(\widehat Q_2,Q_2)
=P\{h(\widehat Q_2)-h(\widetilde Q_2)\}.
\]
Collecting the coefficients of each sequential-regression error gives the following explicit decomposition into within-block, between-block, final, and initial terms:
{\small
\begin{align}
&\hspace{-1cm}R_{2,2}(\widehat Q_2,Q_2) \nonumber\\
&=
\sum_{k=1}^K\sum_{r=1}^{R_k-1}
P\Bigg[
\frac{B_{k,r}}
{\widehat G_{k,r}({\bf a}_{X_{k,r+1}})}
\Big\{
G_{k,r}({\bf a}_{X_{k,r}})
\widehat G_{k,r}({\bf a}_{X_{k,r+1}})
-
\widehat G_{k,r}({\bf a}_{X_{k,r}})
G_{k,r}({\bf a}_{X_{k,r+1}})
\Big\}
\{\widehat\mu_{k,r}-\mu_{k,r}\}
\Bigg]
\nonumber\\
&+
\sum_{k=1}^{K-1}
P\Bigg[
\frac{B_{k,R_k}}
{\widehat G_{k,R_k}({\bf a}_{X_{k+1,1}}[1:k])
 \widehat p_{k,R_k}^{+}}
\Big\{
G_{k,R_k}({\bf a}_{X_{k,R_k}})
\widehat G_{k,R_k}({\bf a}_{X_{k+1,1}}[1:k])
\widehat p_{k,R_k}^{+}
\nonumber\\
&\hspace{15em}-
\widehat G_{k,R_k}({\bf a}_{X_{k,R_k}})
G_{k,R_k}({\bf a}_{X_{k+1,1}}[1:k])
p_{k,R_k}^{+}
\Big\}
\{\widehat\mu_{k,R_k}-\mu_{k,R_k}\}
\Bigg]
\nonumber\\
&+
P\Bigg[
\frac{B_{K,R_K}}
{\widehat G_{K,R_K}({\bf a}_Y)}
\Big\{
G_{K,R_K}({\bf a}_{X_{K,R_K}})
\widehat G_{K,R_K}({\bf a}_Y)
-
\widehat G_{K,R_K}({\bf a}_{X_{K,R_K}})
G_{K,R_K}({\bf a}_Y)
\Big\}
\{\widehat\mu_{K,R_K}-\mu_{K,R_K}\}
\Bigg]
\nonumber\\
&+
P\left[
\frac{\widehat\pi_1(a_{X_{1,1}}\mid C)-\pi_1(a_{X_{1,1}}\mid C)}
{\widehat\pi_1(a_{X_{1,1}}\mid C)}
\{\widehat\mu_{1,0}(C)-\mu_{1,0}(C)\}
\right].
\label{eq:remainder_treg_factorization}
\end{align}
}
An empty within-block sum, arising when $R_k=1$, is understood to be zero. The second line involving $k=1,\ldots,K-1$ is the contribution at block boundaries; the third contribution is the final transition to the outcome, so no $(K+1)$st treatment mechanism is invoked. The arguments of each sequential-regression error in \eqref{eq:remainder_treg_factorization} are those of the corresponding $\mu_{k,r}$ defined in the main text.

For the rate calculation, the bracketed numerator in each of the first three contributions can be written as
\begin{align}
\Gamma_{k,r}
={}&
G_{k,r}({\bf a}_{X_{k,r}})
\widehat G_{k,r}({\bf a}_{X_{k,r}}^{+})
\widehat p_{k,r}^{+}
-
\widehat G_{k,r}({\bf a}_{X_{k,r}})
G_{k,r}({\bf a}_{X_{k,r}}^{+})
p_{k,r}^{+},
\label{eq:gamma_treg}
\end{align}
with $p_{K,R_K}^{+}=1$ and ${\bf a}_{X_{K,R_K}}^{+}={\bf a}_Y$ for the final transition.

For a within-block transition and the final transition to the outcome, $p_{k,r}^{+}=\widehat p_{k,r}^{+}=1$, so
\begin{align*}
\Gamma_{k,r}
={}&
G_{k,r}({\bf a}_{X_{k,r}})
\{\widehat G_{k,r}({\bf a}_{X_{k,r}}^{+})
  -G_{k,r}({\bf a}_{X_{k,r}}^{+})\}
  -
G_{k,r}({\bf a}_{X_{k,r}}^{+})
\{\widehat G_{k,r}({\bf a}_{X_{k,r}})
  -G_{k,r}({\bf a}_{X_{k,r}})\}.
\end{align*}
At a block boundary $r=R_k$, $k<K$,
\begin{align*}
\Gamma_{k,R_k}
={}&
G_{k,R_k}({\bf a}_{X_{k,R_k}})\widehat p_{k,R_k}^{+}
\{\widehat G_{k,R_k}({\bf a}_{X_{k,R_k}}^{+})
  -G_{k,R_k}({\bf a}_{X_{k,R_k}}^{+})\}\\
&-
G_{k,R_k}({\bf a}_{X_{k,R_k}}^{+})p_{k,R_k}^{+}
\{\widehat G_{k,R_k}({\bf a}_{X_{k,R_k}})
  -G_{k,R_k}({\bf a}_{X_{k,R_k}})\}\\
&+
G_{k,R_k}({\bf a}_{X_{k,R_k}})
G_{k,R_k}({\bf a}_{X_{k,R_k}}^{+})
\{\widehat p_{k,R_k}^{+}-p_{k,R_k}^{+}\}.
\end{align*}
Therefore, under the multiplier conditions above,
\begin{align*}
|R_{2,2}(\widehat Q_2,Q_2)|
\lesssim_p{}&
\sum_{k=1}^K\sum_{r=1}^{R_k}
\|\widehat g_{k,r}-g_{k,r}\|_2
\|\widehat\mu_{k,r}-\mu_{k,r}\|_2\\
&+
\sum_{k=1}^{K-1}
\|\widehat\pi_{k+1}-\pi_{k+1}\|_2
\|\widehat\mu_{k,R_k}-\mu_{k,R_k}\|_2\\
&+
\|\widehat\pi_1-\pi_1\|_2
\|\widehat\mu_{1,0}-\mu_{1,0}\|_2.
\end{align*}
The three lines correspond respectively to within-block (including final), between-block, and initial remainder terms. The product-rate conditions in Theorem~\ref{thm:asymp_lin_2} consequently imply $R_{2,2}(\widehat Q_2,Q_2)=o_p(n^{-1/2})$.

The same display gives the more general sequential robustness conditions:
\begin{enumerate}
\item for every $(k,r)$, at least one of $\widehat g_{k,r}$ and
      $\widehat\mu_{k,r}$ is consistent;
\item for every $k=1,\ldots,K-1$, at least one of
      $\widehat\pi_{k+1}$ and $\widehat\mu_{k,R_k}$ is consistent; and
\item at least one of $\widehat\pi_1$ and $\widehat\mu_{1,0}$ is
      consistent.
\end{enumerate}
These conditions jointly imply the regimes in Corollary~\ref{cor:robustness_2}. For example, if $\widehat\mu_{k,R_k}$ is inconsistent, the first two conditions require both $\widehat g_{k,R_k}$ and $\widehat\pi_{k+1}$ to be consistent. The cut-point regime follows by taking all sequential regressions at or after the cut to be consistent, all treatment-history regressions before the cut to be consistent, and all treatment mechanisms through the occasion containing the cut to be consistent.

Finally, under either theorem, the corresponding remainder is $o_p(n^{-1/2})$ and the cross-fitted empirical-process term is $o_p(n^{-1/2})$. Hence
\[
\psi^+(\widehat Q_\ell)- \psi(P)
=(P_n-P)\Phi_\ell(Q_\ell)+o_p(n^{-1/2}),
\]
and the asserted asymptotic-normality and efficiency conclusions follow from the central limit theorem and the fact that $\Phi_\ell(Q_\ell)$ is the EIF.

\textbf{Further robustness considerations}. The robustness regimes in Corollaries~\ref{cor:robustness_1} and~\ref{cor:robustness_2} provide sufficient but not exhaustive conditions for consistency. For the density-based estimator, additional relaxations may be obtained by exploiting the temporal ordering of misspecified nuisance components to relax treatment-mechanism requirements and the flexibility of compatible assignment partitions. In particular, if the last node associated with a misspecified block $\mathcal B_M$ or $\mathcal B_{L_s}$ precedes $A_k$, the treatment mechanisms $\pi_k,\ldots,\pi_K$ may also be misspecified, provided all remaining nuisance functions are consistently estimated. Here, the terminal regression $\mu_{K,R_K}$ is associated with the outcome node $Y$. For the regression-based estimator, further relaxations may arise from algebraic cancellations in the EIF. These possibilities are illustrated via examples in Appendix~\ref{app:ex_multiple_robustness}.

\subsection{Sensitivity analysis: Identification proof}
\label{app:proofs_sa_ID}

Let $\mathbb L=\{L_1,\ldots,L_J\}$, where the nodes are indexed according to their temporal ordering. Write
$L_j=X_{k_j,r_j},\ l_j=x_{k_j,r_j}$. Let
\[
{\bf b}^{(1)}_j={\bf a}^{(1)}[1:k_j],
\qquad
{\bf b}^{y}_j={\bf a}_Y[1:k_j],
\]
denote the two treatment histories at which the sensitivity function for $L_j$ is evaluated. 

For ${\bf b}_{k}\in\{0,1\}^k$ and $X_{k,r}\in\mathbb L$, define
\begin{align}
& t_{k,r}({\bf b}_{k},\mathcal X(y),c; {\bf a}_Y) \notag\\
& =
E\!\left[
Y({\bf a}_Y,\mathcal X(y))
\mid
\mathcal{A}(X_{k,r})={\bf b}_{k},
X_{k,r}=x_{k,r}, \mathcal{X}(X_{k,r}) = \mathcal{X}(x_{k,r})  , C =c
\right]
\notag\\
&\qquad -
E\!\left[
Y({\bf a}_Y,\mathcal X(y))
\mid
\mathcal{A}(X_{k,r})={\bf b}_{k},
X_{k,r}\neq x_{k,r}
, \mathcal{X}(X_{k,r}) = \mathcal{X}(x_{k,r})  , C =c
\right].
\label{supp:sens:ti}
\end{align}
Under consistency and the retained treatment-exchangeability conditions, $t_{k,r}({\bf b}_{k},\mathcal X(y),c; {\bf a}_Y)\neq0$ for some $({\bf b}_{k},\mathcal X(y),c)$ implies
\begin{align*}
Y({\bf a}_Y,\mathcal X(y))
\not\!\!\indep
X_{k,r}\{{\bf b}_{k},
          \mathcal X(x_{k,r})\}
\ \big|\ 
C,  \mathcal{X}(X_{k,r}), 
\mathcal{A}(X_{k,r})={\bf b}_{k}.
\end{align*}
Thus, a nonzero sensitivity function represents residual cross-world dependence between the counterfactual outcome and $X_{k,r}$ under ${\bf b}_{k}$, conditional on the preceding history. Conversely, the corresponding cross-world independence implies $t_{k,r}=0$.

A nonzero $t_{k,r}$ represents a departure from the cross-world conditional independence between $Y({\bf a}_Y,\mathcal X(y))$ and $X_{k,r}({\bf b}_{k}, \mathcal{X}(x_{k,r}))$ given the preceding history. The functions $t_{k,r}$ are fixed sensitivity parameters rather than identified features of the observed-data law.

For any treatment history ${\bf b}_{k_j}\in\{0,1\}^{k_j}$, define
\begin{align*}
m_j({\bf b}_{k_j};\mathcal X(y),c)
&=
E\!\left[
Y({\bf a}_Y,\mathcal X(y))
\mid
\mathcal{A}(L_j)={\bf b}_{k_j},L_j=l_j, \mathcal{X}(L_j) = \mathcal{X}(l_j), C = c
\right],\\
m_j^{\neq}({\bf b}_{k_j};\mathcal X(y),c)
&=
E\!\left[
Y({\bf a}_Y,\mathcal X(y))
\mid
\mathcal{A}(L_j)={\bf b}_{k_j},L_j\neq l_j, \mathcal{X}(L_j) = \mathcal{X}(l_j), C = c
\right],\\
\overline m_j({\bf b}_{k_j};\mathcal X(y),c)
&=
E\!\left[
Y({\bf a}_Y,\mathcal X(y))
\mid
\mathcal{A}(L_j)={\bf b}_{k_j},\mathcal{X}(L_j) = \mathcal{X}(l_j), C = c
\right].
\end{align*}
By iterated expectation,
\begin{align}
& \overline m_j({\bf b}_{k_j};\mathcal X(y),c)
 \notag \\
& ={} m_j({\bf b}_{k_j};\mathcal X(y),c)
f_{k_j,r_j}(l_j\mid{\bf b}_{k_j}, \mathcal{X}(l_j), c)
+
m_j^{\neq}({\bf b}_{k_j};\mathcal X(y),c)
\{1-f_{k_j,r_j}(l_j\mid{\bf b}_{k_j}, \mathcal{X}(l_j), c)\}
\notag\\
& ={}
m_j({\bf b}_{k_j};\mathcal X(y),c)
-
t_{k_j,r_j}({\bf b}_{k_j},\mathcal X(y),c; {\bf a}_Y)
\{1-f_{k_j,r_j}(l_j\mid{\bf b}_{k_j}, \mathcal{X}(l_j), c)\}.
\label{app:eq:binary_t_identity}
\end{align}
Equivalently,
\begin{align}
m_j({\bf b}^{(1)}_j;\mathcal X(y),c)
-
m_j({\bf b}^{y}_j;\mathcal X(y),c)
 =
\overline m_j({\bf b}^{(1)}_j;\mathcal X(y),c)
-
\overline m_j({\bf b}^{y}_j;\mathcal X(y),c)
+
\delta_{t_j}(\mathcal X(y),c;f),
\label{app:eq:diff_E_t}
\end{align}
where
\begin{align}
\delta_{t_j}(\mathcal X(y),c;f)
=& 
t_{k_j,r_j}({\bf b}^{(1)}_j,\mathcal X(y),c;{\bf a}_Y)
\left[
1-f_{k_j,r_j}
\{l_j\mid{\bf b}^{(1)}_j, \mathcal{X}(l_j), c\}
\right]  \notag \\
& \quad -
t_{k_j,r_j}({\bf b}^{y}_j,\mathcal X(y),c; {\bf a}_Y)
\left[
1-f_{k_j,r_j}
\{l_j\mid{\bf b}^{y}_j, \mathcal{X}(l_j), c\}
\right].
\label{app:eq:sens_delta}
\end{align}

We now apply \eqref{app:eq:diff_E_t} successively over the ordered nodes $L_1,\ldots,L_J$. For $j\geq2$, $L_{j-1}$ is the immediately preceding node in $\mathbb L$. The variables occurring strictly between $L_{j-1}$ and $L_j$ belong to $\mathbb M$ or are treatment nodes. Because the sensitivity model relaxes only the cross-world restrictions between $\mathbb L$ and $Y$, the retained cross-world and treatment-ignorability conditions allow these intervening variables to be removed from the conditional contrast. Consequently,
\begin{align}
&\overline m_j({\bf b}^{(1)}_j;\mathcal X(y),c)
-
\overline m_j({\bf b}^{y}_j;\mathcal X(y),c)
 =
m_{j-1}({\bf b}^{(1)}_{j-1};\mathcal X(y),c)
-
m_{j-1}({\bf b}^{y}_{j-1};\mathcal X(y),c).
\label{app:eq:accumulate_E}
\end{align}
This is the conditional-independence step underlying the recursive argument. It is not an additional sensitivity restriction: it invokes precisely the cross-world relations retained under the proposed sensitivity model.

For the earliest node $L_1$, all preceding intermediate variables belong to $\mathbb M$. The retained cross-world restrictions and treatment ignorability therefore give the base case
\begin{align}
\overline m_1({\bf b}^{(1)}_1;\mathcal X(y),c)
-
\overline m_1({\bf b}^{y}_1;\mathcal X(y),c)
=0.
\label{app:eq:sens_base}
\end{align}
Combining \eqref{app:eq:diff_E_t}--\eqref{app:eq:sens_base} yields
recursively
\begin{align*}
m_1({\bf b}^{(1)}_1;\mathcal X(y),c)
-
m_1({\bf b}^{y}_1;\mathcal X(y),c)
&=\delta_{t_1}(\mathcal X(y),c;f),\\
m_2({\bf b}^{(1)}_2;\mathcal X(y),c)
-
m_2({\bf b}^{y}_2;\mathcal X(y),c)
&=\delta_{t_1}(\mathcal X(y),c;f)+\delta_{t_2}(\mathcal X(y),c;f),
\end{align*}
and, more generally,
\begin{align}
m_j({\bf b}^{(1)}_j;\mathcal X(y),c)
-
m_j({\bf b}^{y}_j;\mathcal X(y),c)
=
\sum_{q=1}^j\delta_{t_q}(\mathcal X(y),c;f).
\label{app:eq:sens_induction}
\end{align}
Applying the retained restrictions once more to the variables occurring after the final node $L_J$, together with treatment-ignorability conditions and consistency for the outcome, gives
\begin{align}
& m_{J}({\bf b}^{(1)}_{J};\mathcal X(y),c)
-
m_{J}({\bf b}^{y}_{J};\mathcal X(y),c) \label{app:eq:outcome_mJ_diff} \\
& \quad =  
E\!\Big[Y({\bf a}_Y,\mathcal X(y)) \mid
{\bf a}^{(1)},\mathcal X(Y)=\mathcal X(y),C=c \Big] -
E\!\Big[ Y \mid {\bf a}_Y,\mathcal X(Y)=\mathcal X(y),C=c \Big], \notag
\end{align}
and 
\begin{align}
E\!\left[
Y({\bf a}_Y,\mathcal X(y))
\mid
\mathcal X(Y^\rho)=\mathcal X(y),C=c
\right]
 = E\!\Big[Y({\bf a}_Y,\mathcal X(y)) \mid
{\bf a}^{(1)},\mathcal X(Y)=\mathcal X(y),C=c \Big].
\label{app:eq:outcome_a1_set}
\end{align}
Therefore, we have
\begin{align}
&E\!\left[
Y({\bf a}_Y,\mathcal X(y))
\mid
\mathcal X(Y^\rho)=\mathcal X(y),C=c
\right]
\notag\\
&\quad =
E\!\left[
Y
\mid
\mathcal{A}(Y)={\bf a}_Y,\mathcal X(Y)=\mathcal X(y),C=c
\right]
+
\sum_{j=1}^J\delta_{t_j}(\mathcal X(y),c;f)
\notag\\
&\quad =
\mu_{K,R_K}(\mathcal X(y),c)
+
\sum_{X_{k,r}\in\mathbb L}
\Bigg[
t_{k,r}({\bf a}^{(1)}[1:k],\mathcal X(y),c; {\bf a}_Y)
\Big[
1-f_{k,r}
(x_{k,r}\mid{\bf a}^{(1)}[1:k],\mathcal X(x_{k,r}), c)
\Big]
\notag\\
&\hspace{4.7cm}-
t_{k,r}({\bf a}_Y[1:k],\mathcal X(y),c; {\bf a}_Y)
\Big[
1-f_{k,r}
(x_{k,r}\mid{\bf a}_Y[1:k],\mathcal X(x_{k,r}), c)
\Big]
\Bigg].
\label{app:eq:sens_adjusted_outcome}
\end{align}

Finally, under the retained identifying conditions and by equation~\ref{eq:supp_cf_joint}, the conditional law of the counterfactual intermediate history, given $C=c$, is
\[
dP_f^\rho(\mathcal X(y)\mid c)
=
\prod_{k=1}^K\prod_{r=1}^{R_k}
f_{k,r}(x_{k,r}\mid{\bf a}_{X_{k,r}}, \mathcal X(x_{k,r}), c)
\,d\nu(\mathcal X(y)),
\]
where we write $dP\big(
X_{1,1}^{\rho} {=} x_{1,1},\ldots,X_{K,R_K}^{\rho} {=} x_{K,R_K} \mid c
\big)$ as $dP_f^\rho(\mathcal X(y)\mid c)$ for simplicity, and
$
d\nu(\mathcal X(y))
=
\prod_{k=1}^K\prod_{r=1}^{R_k}
d\nu_{k,r}(x_{k,r}).
$
Integrating \eqref{app:eq:sens_adjusted_outcome} with respect to the counterfactual intermediate-history law and $P(c)$ gives
\begin{align}
\psi_t(P)
=
\int
\Big[
\mu_{K,R_K}(\mathcal X(y),c)
+
\sum_{X_{k,r}\in\mathbb L}
\delta_{t_{k,r}}(\mathcal X(y),c;f)
\Big]
\,dP_f^\rho(\mathcal X(y)\mid c)\,dP(c),
\end{align}
which is the identifying functional in \eqref{eq:sens_id}. When all $t_{k,r}$ are zero, the adjustment vanishes and the original identifying functional is recovered. 

\subsection{Sensitivity analysis: Doubly robust estimator}
\label{app:proofs_sa_DR}

We establish the robustness property of the estimator in \eqref{eq:sens_est} for a fixed collection of sensitivity functions $t=\{t_{k,r}:X_{k,r}\in\mathbb L\}$. These functions are specified by the analyst and are not estimated from the observed data.

For notational convenience, define
\begin{align*}
\Delta_t(\mathcal X(y),c;f)
={}&
\sum_{X_{k,r}\in\mathbb L}
\Bigg[
t_{k,r}({\bf a}^{(1)}[1:k],\mathcal X(y),c; {\bf a}_Y)
\left[
1-f_{k,r}
(x_{k,r}\mid{\bf a}^{(1)}[1:k],\mathcal X(x_{k,r}), c)
\right]
\\
&\hspace{2.5cm}-
t_{k,r}({\bf a}_Y[1:k],\mathcal X(y),c; {\bf a}_Y)
\left[
1-f_{k,r}
(x_{k,r}\mid{\bf a}_Y[1:k],\mathcal X(x_{k,r}), c)
\right]
\Bigg],
\end{align*}
and
\[
m_t(\mathcal X(y),c; \mu, f)
=
\mu_{K,R_K}(\mathcal X(y),c)+\Delta_t(\mathcal X(y),c;f).
\]
Then
\begin{align*}
h_t(c; \mu, f)
&=
\int
m_t(\mathcal X(y),c;\mu,f)
\,dP_f^\rho(\mathcal X(y)\mid c),\\
\psi_t(P)
&=
P\{h_t(C; \mu,f)\}.
\end{align*}

The corresponding estimated quantities are obtained by replacing $\mu$ and $f$ by $\widehat\mu$ and $\widehat f$, while leaving the fixed sensitivity functions unchanged:
\begin{align*}
h_t(c;\widehat \mu, \widehat f)
=
\int
\left[
\widehat\mu_{K,R_K}(\mathcal X(y),c)
+
\Delta_{t}(\mathcal X(y),c;\widehat f)
\right]
\,dP_{\widehat f}^\rho(\mathcal X(y)\mid c).
\end{align*}
Also define
\begin{align*}
\widehat W_Y
={}&
\frac{
I\{\mathcal A(Y)={\bf a}_Y\}
}{\widehat{\bar\pi}_K({\bf a}_Y)}
\prod_{X_{k,r}\in\mathbb L}
\widehat R_{k,r}^{(1)}
\big(X_{k,r},\mathcal X(X_{k,r}), C; {\bf a}_{X_{k,r}},{\bf a}_Y[1:k]\big).
\end{align*}
The estimator can therefore be written as
\begin{align}
\widehat \psi_t
=
P_n\left[
\widehat W_Y
\{Y-\widehat\mu_{K,R_K}(\mathcal X(Y), C)\}
+
h_t(C; \widehat {\mu},\widehat f)
\right].
\label{app:eq:sens_est_compact}
\end{align}

Suppose the nuisance estimators converge in $L^2(P)$ to fixed limits, the relevant treatment probabilities and density ratios satisfy the positivity and moment conditions needed below, and the empirical-process term associated with \eqref{app:eq:sens_est_compact} is $o_p(1)$. Because the sensitivity adjustment depends on the intermediate-variable laws, we assume throughout both robustness regimes that every $\widehat f_{k,r}$ is consistent.

We first record the weighting identity used in the proof. For any integrable function $q(\mathcal X(y),c)$, correct specification of $\pi$ and $f$ gives
\begin{align}
P\left[
W_Y(\pi,f)q\{\mathcal X(Y),C\}
\right]
=
\int
q(\mathcal X(y),c)
\,dP_f^\rho(\mathcal X(y)\mid c)\,dP(c).
\label{app:eq:sens_weighting_identity}
\end{align}
Indeed, inverse weighting by the treatment mechanisms restricts the observed treatment history to ${\bf a}_Y$, while each density ratio for $X_{k,r}\in\mathbb L$ replaces its conditional law under ${\bf a}_Y[1:k]$ by its law under ${\bf a}^{(1)}[1:k]$. Variables in $\mathbb M$ require no density ratio because their node-specific assignments agree with ${\bf a}_Y$.

We now consider the two robustness regimes.

\paragraph{\textcolor{blue}{Correct outcome regression and intermediate-variable laws}}
Suppose
\[
\widehat\mu_{K,R_K}\longrightarrow\mu_{K,R_K},
\qquad
\widehat f_{k,r}\longrightarrow f_{k,r}
\quad\text{for every }(k,r),
\]
while the treatment-mechanism estimators may converge to possibly misspecified limits $\pi^*$. Conditional on $\mathcal{A}(Y)={\bf a}_Y$ and $\mathcal X(Y), C$,
\[
E\{Y-\mu_{K,R_K}(\mathcal X(Y), C)\mid \mathcal{A}(Y)={\bf a}_Y,\mathcal X(Y), C\}=0.
\]
Consequently, for any limiting weight $W_Y(\pi^*,f)$ satisfying the stated moment conditions,
\begin{align*}
P\left[
W_Y(\pi^*,f)
\{Y-\mu_{K,R_K}(\mathcal X(Y), C)\}
\right]
=0.
\end{align*}
Moreover,
\[
h_t(C; \widehat {\mu},\widehat f)
\longrightarrow
h_t(C;\mu,f).
\]
It follows that
\begin{align*}
\widehat \psi_t
&\longrightarrow
P\left[
W_Y(\pi^*,f)
\{Y-\mu_{K,R_K}(\mathcal X(Y), C)\}
+
h_t(C;\mu,f)
\right]\\
&=
P\{h_t(C;\mu,f)\}\\
&=
\psi_t(P).
\end{align*}
Thus, consistency of the outcome regression and all intermediate-variable laws is sufficient even if the treatment mechanisms are misspecified.

\paragraph{\textcolor{blue}{Correct treatment mechanisms and intermediate-variable laws}}
Suppose
\[
\widehat\pi_k\longrightarrow\pi_k
\quad\text{for every }k,
\qquad
\widehat f_{k,r}\longrightarrow f_{k,r}
\quad\text{for every }(k,r),
\]
while $\widehat\mu_{K,R_K}$ may converge to a possibly misspecified limit $\mu^*_{K,R_K}$. Taking
\[
q(\mathcal X(y),c)
=
\mu_{K,R_K}(\mathcal X(y),c)
-
\mu^*_{K,R_K}(\mathcal X(y),c)
\]
in \eqref{app:eq:sens_weighting_identity} gives
\begin{align}
&E\left[
W_Y(\pi,f)
\{Y-\mu^*_{K,R_K}(\mathcal X(Y), C)\}
\mid C=c
\right]
 =
\int
\left[
\mu_{K,R_K}(\mathcal X(y),c)
-
\mu^*_{K,R_K}(\mathcal X(y),c)
\right]
\,dP_f^\rho(\mathcal X(y)\mid c).
\label{app:eq:sens_residual_transport}
\end{align}
Because the intermediate-variable laws are consistently estimated,
\begin{align*}
h_t(c;\widehat \mu, \widehat f)
&\longrightarrow
h_t(c;\mu^*,f)\\
&=
\int
\left[
\mu^*_{K,R_K}(\mathcal X(y),c)
+
\Delta_t(\mathcal X(y),c;f)
\right]
\,dP_f^\rho(\mathcal X(y)\mid c).
\end{align*}
Adding this limit to \eqref{app:eq:sens_residual_transport} yields
\begin{align*}
&\int
\left[
\mu_{K,R_K}(\mathcal X(y),c)
-
\mu^*_{K,R_K}(\mathcal X(y),c)
\right]
\,dP_f^\rho(\mathcal X(y)\mid c)
+
h_t(c;\mu^*,f)
\\
&\qquad =
\int
\left[
\mu_{K,R_K}(\mathcal X(y),c)
+
\Delta_t(\mathcal X(y),c;f)
\right]
\,dP_f^\rho(\mathcal X(y)\mid c)
\\
&\qquad =
h_t(c;\mu,f).
\end{align*}
Therefore,
\begin{align*}
\widehat \psi_t
&\longrightarrow
P\{h_t(C;\mu,f)\}\\
&=
\psi_t(P).
\end{align*}
Thus, consistency of all treatment mechanisms and all intermediate-variable laws is sufficient even if the outcome regression is misspecified.

Combining the two cases, for each fixed collection of sensitivity functions, $\widehat \psi_t$ is consistent if all intermediate-variable laws are consistently estimated and either the outcome regression is consistent or all treatment mechanisms are consistent. The result is therefore doubly robust with respect to $\mu_{K,R_K}$ and $\pi$, conditional on correct estimation of $f$.

\newpage
\section{Additional illustrative examples}
\label{app:examples}

\subsection{Recursive definition of counterfactuals}
\label{app:ex_nested_cf}


\begin{example}
Consider the DAG in Figure~\ref{fig:graphs}(a), with structural
equations
\[
M=f_M(C,A,\epsilon_M),\qquad
L=f_L(C,A,M,\epsilon_L),\qquad
Y=f_Y(C,A,M,L,\epsilon_Y).
\]
We suppress $C$ in the expressions below. Write
$M(a)$, $L(a,m)$, and $Y(a,m,l)$ for the corresponding
counterfactuals obtained by fixing the displayed parent inputs.

Let
\[
\rho=
\{A\rightarrow M\rightarrow Y,\;
  A\rightarrow M\rightarrow L\rightarrow Y\}.
\]
The active value $a$ is assigned to both paths beginning
with $A\rightarrow M$, while the paths
$A\rightarrow L\rightarrow Y$ and $A\rightarrow Y$
receive the reference value $a'$.

The two downstream continuations for $M$ are $(Y)$ and
$(L,Y)$. Because both corresponding treatment-to-outcome
paths belong to $\rho$, the recursion gives
\[
M^{(Y)}=M(a),
\qquad
M^{(L,Y)}=M(a).
\]
The only continuation for $L$ is $(Y)$. Since
$A\rightarrow L\rightarrow Y\notin\rho$, its direct
treatment input is $a'$, giving
\[
L^{(Y)}
=
L\bigl(a',M^{(L,Y)}\bigr)
=
L\bigl(a',M(a)\bigr).
\]
Finally, $A\rightarrow Y\notin\rho$, so
\begin{align*}
Y^{\rho}
&=
Y\bigl(a',M^{(Y)},L^{(Y)}\bigr)\\
&=
Y\bigl(a',M(a),L(a',M(a))\bigr).
\end{align*}
In this example, the two versions of $M$ coincide because
their treatment assignments agree.
\end{example}

\begin{example} 
Consider the setup in the previous example. Now let
\[
\rho=\{A\rightarrow M\rightarrow Y\}.
\]
The path $A\rightarrow M\rightarrow Y$ receives $a$,
whereas $A\rightarrow M\rightarrow L\rightarrow Y$
receives $a'$. Consequently,
\[
M^{(Y)}=M(a),
\qquad
M^{(L,Y)}=M(a').
\]
The direct treatment input to $L$ is again $a'$, giving
\[
L^{(Y)}
=
L\bigl(a',M^{(L,Y)}\bigr)
=
L\bigl(a',M(a')\bigr).
\]
The resulting outcome is
\begin{align*}
Y^{\rho}
&=
Y\bigl(a',M^{(Y)},L^{(Y)}\bigr)\\
&=
Y\bigl(a',M(a),L(a',M(a'))\bigr).
\end{align*}
Thus, the occurrence of $M$ supplied directly to $Y$
uses $M(a)$, whereas the occurrence supplied through $L$
uses $M(a')$. A single unindexed counterfactual $M^\rho$
would not distinguish these two inputs. Both versions are
evaluated using the same structural error $\epsilon_M$.
\end{example}

\begin{example}
Consider the DAG in Figure~\ref{fig:graphs}(b) and let $\rho=\{A_1\rightarrow L_1\leadsto Y,\;A_1\rightarrow L_2\rightarrow Y,\;A_2\rightarrow L_2\rightarrow Y\},$ where $L_1\leadsto Y$ denotes all proper pathways from $L_1$ to $Y$. Here, edge consistency does hold, thus $Z^\rho=Z^\xi$ for all admissible continuation $\xi$ ended in $Y$.  For ${\bf a}_2=(a_1,a_2)$ and ${\bf a}'_2=(a'_1,a'_2)$, the nested counterfactual variables can be written as
$M_1^\rho =f_{M_1}(a'_1,\epsilon_{M_1})$, 
$L_1^\rho =f_{L_1}(a_1,M^{\rho}_1,\epsilon_{L_1})$,  
$M_2^\rho =f_{M_2}(a'_1,M^{\rho}_1,L_1^\rho,a'_2,\epsilon_{M_2})$, 
$L_2^\rho =f_{L_2}(a_1,M^{\rho}_1,L_1^\rho,a_2,M_2^\rho,\epsilon_{L_2})$, and 
$Y^\rho =f_Y(a'_1,M^{\rho}_1,L_1^\rho,a'_2,M_2^\rho,L_2^\rho,\epsilon_Y)$. 
\end{example} 
\begin{figure}[!htbp] 
	\begin{center}
    \scalebox{1}{
    \begin{tikzpicture}[>=stealth, node distance=1.3cm]
        \tikzstyle{format} = [thick, circle, minimum size=1.0mm, inner sep=2pt]
        \tikzstyle{square} = [draw, thick, minimum size=4.5mm, inner sep=2pt]
    \begin{scope}[xshift=0.cm, yshift=0cm]
		\path[->, thick]
		
		node[] (t) {$A$}
		node[right of=t, xshift=0.15cm] (m) {$M$}
        node[right of=m, xshift=0.15cm] (l) {$L$}
		node[above right of=m, yshift=0.2cm] (x) {$C$}
		node[right of=l, xshift=0.15cm] (y) {$Y$}
		
		(x) edge[blue, bend left=0] (t) 
		(x) edge[blue] (m) 
        (x) edge[blue] (l) 
		(x) edge[blue, bend right=0] (y) 
		(t) edge[blue] (m) 
		(m) edge[blue] (l) 
        (l) edge[blue] (y)    
        
        (t) edge[blue, out=90, in=155] (l) 
        (m) edge[blue, out=25, in=90] (y) 
        (t) edge[blue, out=-15, in=195] (y) 

        node[below right of=m, xshift=-0.2cm, yshift=-0.5cm] (a) {(a)} ;
		
	\end{scope}
    \begin{scope}[xshift=7cm, yshift=0cm]
		\path[->, thick]
		
		node[] (t1) {$A_1$}
		node[above right of=t1, xshift=0.25cm] (m1) {$M_1$}
        node[right of=m1, xshift=0.5cm] (l1) {$L_1$}
        node[below right of=l1, xshift=0cm] (t2) {$A_2$}
        node[above right of=t2, xshift=0.25cm] (m2) {$M_2$}
        node[right of=m2, xshift=0.5cm] (l2) {$L_2$}
		node[below right of=l2, xshift=0.25cm] (y) {$Y$}
		
        (t1) edge[blue] (m1) 
        (t1) edge[blue] (l1) 
        (t1) edge[blue] (t2) 
        (t1) edge[blue, out=-15, in=-90] (m2)
        (t1) edge[blue, out=-15, in=-125] (l2) 
        (t1) edge[blue, out=-15, in=200] (y) 
		
        (m1) edge[blue] (l1) 
        (m1) edge[blue] (t2)
        (m1) edge[blue, out=20, in=160] (m2)
        (m1) edge[blue, out=20, in=160] (l2) 
        (m1) edge[blue, out=-45, in=185] (y) 
         
        (l1) edge[blue] (t2)
        (l1) edge[blue] (m2)
        (l1) edge[blue, out=20, in=160] (l2)
        (l1) edge[blue, out=-25, in=170] (y) 

        (t2) edge[blue] (m2)
        (t2) edge[blue] (l2)
        (t2) edge[blue] (y)

        (m2) edge[blue] (l2)
        (m2) edge[blue, bend right=5] (y)
        
        (l2) edge[blue] (y) 
        
        node[below right of=t1, xshift=3.0cm, yshift=-0.5cm] (b) {(b)} ;
		
	\end{scope}
	\end{tikzpicture}
	}
    \vspace{-0.25cm}
	\caption{Examples of causal structures: 
    (a) one treatment and two ordered mediators; 
    (b) repeated treatments and mediators.  
    } 
	\label{fig:graphs}
	\end{center}
\end{figure}

\subsection{Minimal treatment-assignment partitions}
\label{app:ex_minimal_S}

\begin{example}
Consider a longitudinal setting with $K=3$ and observed-data ordering
$$
C,A_1,X_{1,1},X_{1,2},
A_2,X_{2,1},X_{2,2},
A_3,X_{3,1},Y.
$$
Suppose the node-specific treatment assignment map for an edge-consistent target satisfies
$$
{\bf a}_{X_{1,1}}=(1), \hspace{0.15cm}
{\bf a}_{X_{1,2}}=(0), \hspace{0.15cm}
{\bf a}_{X_{2,1}}=(1,1), \hspace{0.15cm}
{\bf a}_{X_{2,2}}=(0,0), \hspace{0.15cm}
{\bf a}_{X_{3,1}}=(1,1,1), \hspace{0.15cm}
{\bf a}_Y=(0,0,0).
$$
Completing each node-specific assignment separately does not necessarily give a minimal number of assignment vectors. For example, appending zeros to the assignments of $X_{1,1}$ and $X_{2,1}$ to full-length vectors gives
$$
(1)\longmapsto(1,0,0),\qquad
(1,1)\longmapsto(1,1,0).
$$
Together with ${\bf a}_Y=(0,0,0)$ and ${\bf a}_{X_{3,1}}=(1,1,1)$, these four vectors cover all node-specific assignments. However, the assignments  of $X_{1,1}$ and $X_{2,1}$ are already prefixes of ${\bf a}_{X_{3,1}}$, so the two additional vectors are unnecessary. The minimum is therefore $S=1$, attained by
$$
{\bf a}^{(0)}={\bf a}_Y=(0,0,0),
\qquad
{\bf a}^{(1)}=(1,1,1).
$$

{\bf Reverse-Order Prefix Grouping Procedure}. For a general assignment map, the following procedure provides one easier way to construct a minimal collection of full-length assignment vectors:  (1) Initialize the collection with ${\bf a}^{(0)}={\bf a}_Y$. Process the intermediate variables in reverse observed-data order, from $X_{K,R_K}$ to $X_{1,1}$; (2) For the current node $X_{k,r}$, check whether ${\bf a}_{X_{k,r}}$ is a prefix of an existing full-length vector. If so, assign the node to a compatible class. If several classes are compatible, choose any one and fix its membership. If no existing vector is compatible, create a new class and obtain its representative by appending $K-k$ zeros to ${\bf a}_{X_{k,r}}$. (4) Repeat steps (2) for each node in the reverse ordering. 

In the example above, $X_{3,1}$ first introduces the different vector ${\bf a}^{(1)} = (1,1,1)$. The assignments of $X_{2,1}$ and $X_{1,1}$ are subsequently absorbed into this class, while $X_{2,2}$ and $X_{1,2}$ join the class represented by ${\bf a}^{(0)}$. The same procedure is applied in the following examples.

\end{example}

\subsection{Illustrations of multiple robustness}
\label{app:ex_multiple_robustness}

\begin{example}\label{app:ex_1:robustness}
Continuing with the previous example in Section~\ref{app:ex_minimal_S}, the unique assignment partitions of mediators are as follows: 
$$
\mathbb M=\{X_{1,2},X_{2,2}\}, \qquad
\mathbb L_1=\{X_{1,1},X_{2,1},X_{3,1}\}.
$$
For the density-based estimator, the three nuisance blocks are
$$
\mathcal B_\pi =\{\pi_1,\pi_2,\pi_3\},\qquad
\mathcal B_M =\{\mu_{3,1},f_{1,2},f_{2,2}\},\qquad
\mathcal B_{L_1} =\{f_{1,1},f_{2,1},f_{3,1}\}.
$$
Corollary~\ref{cor:robustness_1} gives two-out-of-three robustness: the estimator remains consistent if any two of these three blocks are consistently estimated, while the remaining block may be misspecified.

For the regression-based estimator, the general sequential robustness conditions reduce to requiring at least one consistent estimator in each of the following pairs:
$$
(\widehat\pi_1,\widehat\mu_{1,0}), \,
(\widehat g_{1,1},\widehat\mu_{1,1}), \,
(\widehat g_{1,2},\widehat\mu_{1,2}), \,
(\widehat\pi_2,\widehat\mu_{1,2}), \,
(\widehat g_{2,1},\widehat\mu_{2,1}), \,
(\widehat g_{2,2},\widehat\mu_{2,2}), \,
(\widehat\pi_3,\widehat\mu_{2,2}), \,
(\widehat g_{3,1},\widehat\mu_{3,1}).
$$
Selected sufficient regimes from Corollaries~\ref{cor:robustness_1} and~\ref{cor:robustness_2} are summarized in Table~\ref{tab:robustness_example_minimal}. Nuisance estimators not listed as required may converge to misspecified limits, subject to the regularity conditions of the corollaries.

\begin{table}[!htbp]
\centering
\caption{Sufficient robustness regimes in Example~\ref{app:ex_1:robustness}.}
\label{tab:robustness_example_minimal}
\begin{tabular}{@{}p{0.40\linewidth}p{0.54\linewidth}@{}}
\toprule
Regime & Required consistent estimation\\
\midrule
\multicolumn{2}{@{}l}{\textit{Density-based estimator}}\\[2pt]
$\mathcal B_\pi$ misspecified
&
$\mathcal B_M$ and $\mathcal B_{L_1}$
\\
$\mathcal B_M$ misspecified
&
$\mathcal B_\pi$ and $\mathcal B_{L_1}$
\\
$\mathcal B_{L_1}$ misspecified
&
$\mathcal B_\pi$ and $\mathcal B_M$
\\[10pt]
\multicolumn{2}{@{}l}{\textit{Regression-based estimator}}\\[2pt]
All sequential regressions
&
$\widehat\mu_{1,0}$ and all $\widehat\mu_{k,r}$
\\
All treatment regressions
&
$\widehat\pi_1$, $\widehat\pi_2$, $\widehat\pi_3$,
and all $\widehat g_{k,r}$
\\
Cut at $X_{1,1}$
&
$\widehat\pi_1$,
$\widehat\mu_{1,1}$, $\widehat\mu_{1,2}$,
$\widehat\mu_{2,1}$, $\widehat\mu_{2,2}$,
and $\widehat\mu_{3,1}$
\\
Cut at $X_{1,2}$
&
$\widehat\pi_1$, $\widehat g_{1,1}$,
$\widehat\mu_{1,2}$,
$\widehat\mu_{2,1}$, $\widehat\mu_{2,2}$,
and $\widehat\mu_{3,1}$
\\
Cut at $X_{2,1}$
&
$\widehat\pi_1$, $\widehat\pi_2$,
$\widehat g_{1,1}$, $\widehat g_{1,2}$,
$\widehat\mu_{2,1}$, $\widehat\mu_{2,2}$,
and $\widehat\mu_{3,1}$
\\
Cut at $X_{2,2}$
&
$\widehat\pi_1$, $\widehat\pi_2$,
$\widehat g_{1,1}$, $\widehat g_{1,2}$,
$\widehat g_{2,1}$,
$\widehat\mu_{2,2}$, and $\widehat\mu_{3,1}$
\\
Cut at $X_{3,1}$
&
$\widehat\pi_1$, $\widehat\pi_2$, $\widehat\pi_3$,
$\widehat g_{1,1}$, $\widehat g_{1,2}$,
$\widehat g_{2,1}$, $\widehat g_{2,2}$,
and $\widehat\mu_{3,1}$
\\
\bottomrule
\end{tabular}
\end{table}

This example provides a standard illustration of our robustness results in the corollaries. Some settings may involve greater flexibility and weaker specification requirements. To illustrate this, we present two additional examples below: one with a unique assignment partition, and the other with nonunique assignment partitions.
\end{example}

\begin{example}[\textbf{Unique assignment partitions}] \label{app:ex:unique_partitions}
Revise the node-specific treatment assignment map for the observed-data structure in Section~\ref{app:ex_minimal_S} to
$$
{\bf a}_{X_{1,1}}=(0), \hspace{0.15cm}
{\bf a}_{X_{1,2}}=(0), \hspace{0.15cm}
{\bf a}_{X_{2,1}}=(1,1), \hspace{0.15cm}
{\bf a}_{X_{2,2}}=(1,0), \hspace{0.15cm}
{\bf a}_{X_{3,1}}=(1,1,1), \hspace{0.15cm}
{\bf a}_Y=(0,0,0).
$$
According to our reverse-order prefix grouping procedure, the minimum number of full-length assignment vectors covering $\mathbb L$ is $S = 2$, with ${\bf a}^{(1)}=(1,1,1)$ and ${\bf a}^{(2)}=(1,0,0)$. The unique partition of $\mathbb X$ is
$$
\mathbb M=\{X_{1,1},X_{1,2}\},\qquad
\mathbb L_1=\{X_{2,1},X_{3,1}\},\qquad
\mathbb L_2=\{X_{2,2}\}.
$$
The $S+2=4$ nuisance blocks are
$$
\mathcal B_\pi =\{\pi_1,\pi_2,\pi_3\}, \quad
\mathcal B_M=\{\mu_{3,1},f_{1,1},f_{1,2}\},\quad
\mathcal B_{L_1} =\{f_{2,1},f_{3,1}\}, \quad
\mathcal B_{L_2}=\{f_{2,2}\}.
$$
For the density-based estimator, Corollary~\ref{cor:robustness_1} gives three-out-of-four robustness. These conditions are sufficient but need not be necessary. This condition can be further relaxed by considering the temporal location of the selected misspecified block. In this example, allowing $\mathcal B_{L_2}=\{f_{2,2}\}$ to be misspecified places the last such node at $X_{2,2}$, before $A_3$. Thus, consistency of $\mathcal B_M$, $\mathcal B_{L_1}$, $\widehat\pi_1$, and $\widehat\pi_2$ is sufficient even if both $\widehat f_{2,2}$ and $\widehat\pi_3$ are misspecified. However, $\mathcal B_M$ contains the terminal regression $\mu_{3,1}$, and $\mathcal B_{L_1}$ contains $f_{3,1}$. Allowing either $\mathcal B_M$ or $\mathcal B_{L_1}$ to be misspecified therefore gives no additional relaxation of $\mathcal B_\pi$ through this argument.

These statements can be checked directly from the EIF. Write $f_{k,r}^{\bf b}$ for the conditional density of $X_{k,r}$ under treatment history ${\bf b}$, suppressing its preceding intermediate-variable history and $C$. Similarly, $\bar\pi_k({\bf b})$ denotes the product of treatment probabilities under history ${\bf b}$. The EIF is
\begin{equation}
\begin{aligned}
\Phi_1(P)(O)
={}&
\frac{I\{\mathbf A_3=(0,0,0)\}}{\bar\pi_3(0,0,0)}
\frac{f_{2,1}^{11}}{f_{2,1}^{00}}
\frac{f_{2,2}^{10}}{f_{2,2}^{00}}
\frac{f_{3,1}^{111}}{f_{3,1}^{000}}
\{Y-\mu_{3,1}\}
\\
&+
\frac{I\{\mathbf A_3=(1,1,1)\}}{\bar\pi_3(1,1,1)}
\frac{f_{1,1}^{0}}{f_{1,1}^{1}}
\frac{f_{1,2}^{0}}{f_{1,2}^{1}}
\frac{f_{2,2}^{10}}{f_{2,2}^{11}}
\{\mu_{3,1}-\mu_{2,2}\}
\\
&+
\frac{I\{\mathbf A_2=(1,0)\}}{\bar\pi_2(1,0)}
\frac{f_{1,1}^{0}}{f_{1,1}^{1}}
\frac{f_{1,2}^{0}}{f_{1,2}^{1}}
\frac{f_{2,1}^{11}}{f_{2,1}^{10}}
\{\mu_{2,2}-\mu_{2,1}\}
\\
&+
\frac{I\{\mathbf A_2=(1,1)\}}{\bar\pi_2(1,1)}
\frac{f_{1,1}^{0}}{f_{1,1}^{1}}
\frac{f_{1,2}^{0}}{f_{1,2}^{1}}
\{\mu_{2,1}-\mu_{1,2}\}
\\
&+
\frac{I\{A_1=0\}}{\pi_1(0)}
\{\mu_{1,2}-\mu_{1,1}\}
\\
&+
\frac{I\{A_1=0\}}{\pi_1(0)}
\{\mu_{1,1}-\mu_{1,0}\}
\\
&+\mu_{1,0}-\psi.
\end{aligned}
\end{equation}

Note that $\mu_{2,2} = \int \mu_{3,1}f_{3,1}^{111}\,d\nu(x_{3,1})$, so $\widehat \mu_{2,2}$ remains consistent because $\widehat\mu_{3,1}$ and $\widehat f_{3,1}$ are consistent. Only the first two rows in EIF involve $\pi_3$, and  their residuals $Y-\mu_{3,1}$ and $\mu_{3,1}-\mu_{2,2}$ have conditional mean zero under its treatment histories, respectively. So both terms retain mean zero even if $\pi_3$ is misspecified. The remaining terms do not involve $\pi_3$, and thus its misspecification will not change the consistency follows from Corollary~\ref{cor:robustness_1}.

For the regression-based estimator, Corollary~\ref{cor:robustness_2} gives the same sequential robustness regimes as in the preceding example because the observed-data ordering is unchanged. These sufficient conditions can be relaxed as well. 

Write $g_{k,r}^{\bf b}$ for the treatment-history regression evaluated at ${\bf b}$, suppressing the intermediate-variable history through $X_{k,r}$ and $C$. Similarly, in $\pi_k(a_k\mid a_1,\ldots,a_{k-1})$, the preceding intermediate-variable history and $C$ are suppressed. The regression-based representation of the EIF is

\begin{equation}
\begin{aligned}
\Phi_2(P)(O)
={}&
\frac{I\{\mathbf A_3=(0,0,0)\}}
{\pi_1(0)\pi_2(1\mid1)\pi_3(1\mid1,1)}
\frac{g_{1,2}^{0}}{g_{1,2}^{1}}
\frac{g_{2,1}^{11}}{g_{2,1}^{10}}
\frac{g_{2,2}^{10}}{g_{2,2}^{11}}
\frac{g_{3,1}^{111}}{g_{3,1}^{000}}
\{Y-\mu_{3,1}\}
\\
&+
\frac{I\{\mathbf A_3=(1,1,1)\}}
{\pi_1(0)\pi_2(1\mid1)\pi_3(1\mid1,1)}
\frac{g_{1,2}^{0}}{g_{1,2}^{1}}
\frac{g_{2,1}^{11}}{g_{2,1}^{10}}
\frac{g_{2,2}^{10}}{g_{2,2}^{11}}
\{\mu_{3,1}-\mu_{2,2}\}
\\
&+
\frac{I\{\mathbf A_2=(1,0)\}}
{\pi_1(0)\pi_2(1\mid1)}
\frac{g_{1,2}^{0}}{g_{1,2}^{1}}
\frac{g_{2,1}^{11}}{g_{2,1}^{10}}
\{\mu_{2,2}-\mu_{2,1}\}
\\
&+
\frac{I\{\mathbf A_2=(1,1)\}}
{\pi_1(0)\pi_2(1\mid1)}
\frac{g_{1,2}^{0}}{g_{1,2}^{1}}
\{\mu_{2,1}-\mu_{1,2}\}
\\
&+
\frac{I\{A_1=0\}}{\pi_1(0)}
\{\mu_{1,2}-\mu_{1,1}\}
\\
&+
\frac{I\{A_1=0\}}{\pi_1(0)}
\{\mu_{1,1}-\mu_{1,0}\}
\\
&+\mu_{1,0}-\psi.
\end{aligned}
\end{equation}

Specifically, applying Bayes' rule causes the cancellation of $g_{1,1}$ in the weights. Moreover, the two terms involving $\mu_{1,1}$ have the same weight and combine into $\frac{I\{A_1=0\}}{\pi_1(0)} \{\mu_{1,2}-\mu_{1,0}\}$. Thus, neither $g_{1,1}$ nor $\mu_{1,1}$ remains in the simplified EIF function. These cancellations also hold for the fitted functions, so neither needs to be consistently estimated, provided the other requirements of the selected regime hold. 

\begin{table}[!hbp]
\centering
\caption{Sufficient robustness regimes and additional relaxations for misspecifications in Example~\ref{app:ex:unique_partitions}, with a unique assignment partition and $S=2$.}
\label{tab:unique_robustness_example_relaxations}
\small
\resizebox{\linewidth}{!}{%
\begin{tabular}{@{}llc@{}}
\toprule
Regime & Required consistent estimation & Relaxations \\
\midrule
\multicolumn{3}{@{}l}{\textit{Density-based estimator}} \\
[2pt]
$\mathcal B_\pi$ misspecified
& $\mathcal B_M$, $\mathcal B_{L_1}$, and $\mathcal B_{L_2}$
& --- \\
$\mathcal B_M$ misspecified
& $\mathcal B_\pi$, $\mathcal B_{L_1}$, and $\mathcal B_{L_2}$
& --- \\
$\mathcal B_{L_1}$ misspecified
& $\mathcal B_\pi$, $\mathcal B_M$, and $\mathcal B_{L_2}$
& --- \\
$\mathcal B_{L_2}$ misspecified
& $\mathcal B_\pi$, $\mathcal B_M$, and $\mathcal B_{L_1}$
& $\widehat\pi_3$ \\
[10pt]
\multicolumn{3}{@{}l}{\textit{Regression-based estimator}} \\
[2pt]
All sequential regressions
& $\widehat\mu_{1,0}$ and all $\widehat\mu_{k,r}$
& $\widehat\mu_{1,1}$ \\
All treatment regressions
& $\widehat\pi_1$, $\widehat\pi_2$, $\widehat\pi_3$, and all $\widehat g_{k,r}$
& $\widehat g_{1,1}$ \\
Cut at $X_{1,1}$
& $\widehat\pi_1$, $\widehat\mu_{1,1}$, $\widehat\mu_{1,2}$, $\widehat\mu_{2,1}$, $\widehat\mu_{2,2}$, and $\widehat\mu_{3,1}$
& $\widehat\mu_{1,1}$ \\
Cut at $X_{1,2}$
& $\widehat\pi_1$, $\widehat g_{1,1}$, $\widehat\mu_{1,2}$, $\widehat\mu_{2,1}$, $\widehat\mu_{2,2}$, and $\widehat\mu_{3,1}$
& $\widehat g_{1,1}$ \\
Cut at $X_{2,1}$
& $\widehat\pi_1$, $\widehat\pi_2$, $\widehat g_{1,1}$, $\widehat g_{1,2}$, $\widehat\mu_{2,1}$, $\widehat\mu_{2,2}$, and $\widehat\mu_{3,1}$
& $\widehat g_{1,1}$ \\
Cut at $X_{2,2}$
& $\widehat\pi_1$, $\widehat\pi_2$, $\widehat g_{1,1}$, $\widehat g_{1,2}$, $\widehat g_{2,1}$, $\widehat\mu_{2,2}$, and $\widehat\mu_{3,1}$
& $\widehat g_{1,1}$ \\
Cut at $X_{3,1}$
& $\widehat\pi_1$, $\widehat\pi_2$, $\widehat\pi_3$, $\widehat g_{1,1}$, $\widehat g_{1,2}$, $\widehat g_{2,1}$, $\widehat g_{2,2}$, and $\widehat\mu_{3,1}$
& $\widehat g_{1,1}$ \\
\bottomrule
\end{tabular}%
}
\end{table}

Table~\ref{tab:unique_robustness_example_relaxations} summarizes the standard sufficient regimes and their corresponding relaxations in this example. 

\end{example}

\vspace{-2cm}
\begin{example}[\textbf{Non-unique assignment partitions}]\label{app:ex:non-unique__partitions}
We retain the observed-data ordering in Section~\ref{app:ex_minimal_S}, but revise the node-specific treatment assignments to
$$
{\bf a}_{X_{1,1}} =(0), \hspace{0.15cm}
{\bf a}_{X_{1,2}} =(1), \hspace{0.15cm}
{\bf a}_{X_{2,1}} =(0,0), \hspace{0.15cm}
{\bf a}_{X_{2,2}} =(0,1), \hspace{0.15cm}
{\bf a}_{X_{3,1}} =(1,0,0), \hspace{0.15cm}
{\bf a}_Y =(1,1,1).
$$
The reverse-order prefix grouping procedure gives $S=3$ with 
$$
{\bf a}^{(1)}=(0,0,0),\qquad
{\bf a}^{(2)}=(0,1,0),\qquad
{\bf a}^{(3)}=(1,0,0).
$$
The nodes $X_{2,1}$, $X_{2,2}$, and $X_{3,1}$ belong to $\mathbb L_1$, $\mathbb L_2$, and $\mathbb L_3$, respectively. However, $X_{1,1}$ is compatible with either $\mathbb L_1$ or $\mathbb L_2$, while $X_{1,2}$ is compatible with either $\mathbb M$ or $\mathbb L_3$. These two choices are independent, giving the four compatible partitions shown below. 

\vspace{0.3cm}
\begin{center}
\scalebox{0.8}{
\begin{tabular}{ccccc}
\toprule
Partition
& $\mathbb M$
& $\mathbb L_1$
& $\mathbb L_2$
& $\mathbb L_3$
\\
\midrule
A
& $\{X_{1,2}\}$
& $\{X_{1,1},X_{2,1}\}$
& $\{X_{2,2}\}$
& $\{X_{3,1}\}$
\\
B
& $\{X_{1,2}\}$
& $\{X_{2,1}\}$
& $\{X_{1,1},X_{2,2}\}$
& $\{X_{3,1}\}$
\\
C
& $\varnothing$
& $\{X_{1,1},X_{2,1}\}$
& $\{X_{2,2}\}$
& $\{X_{1,2},X_{3,1}\}$
\\
D
& $\varnothing$
& $\{X_{2,1}\}$
& $\{X_{1,1},X_{2,2}\}$
& $\{X_{1,2},X_{3,1}\}$
\\
\bottomrule
\end{tabular}
}
\end{center}

\vspace{0.4cm}
For each partition, the five density-based nuisance blocks with $s \in \{1,2,3\}$ are
$$
\mathcal B_\pi =\{\pi_1,\pi_2,\pi_3\}, \quad 
\mathcal B_M =\{\mu_{3,1}\} \cup\{f_{k,r}:X_{k,r}\in\mathbb M\}, \quad 
\mathcal B_{L_s} =\{f_{k,r}:X_{k,r}\in\mathbb L_s\}.
$$

All partition patterns yield identical weighting identities. Consequently, the plug-in, inverse probability-weighted, and one-step estimators are invariant to the choice of partition. Likewise, the EIF is partition-invariant, with its density-based representation given by

\vspace{-0.3cm}
\begin{equation}
\begin{aligned}
\Phi_1(P)(O)
={}&
\frac{I\{{\bf A}_3=(1,1,1)\}}{\bar\pi_3(1,1,1)}
\frac{f_{1,1}^{0}}{f_{1,1}^{1}}
\frac{f_{2,1}^{00}}{f_{2,1}^{11}}
\frac{f_{2,2}^{01}}{f_{2,2}^{11}}
\frac{f_{3,1}^{100}}{f_{3,1}^{111}}
\{Y-\mu_{3,1}\}
\\
&+
\frac{I\{{\bf A}_3=(1,0,0)\}}{\bar\pi_3(1,0,0)}
\frac{f_{1,1}^{0}}{f_{1,1}^{1}}
\frac{f_{2,1}^{00}}{f_{2,1}^{10}}
\frac{f_{2,2}^{01}}{f_{2,2}^{10}}
\{\mu_{3,1}-\mu_{2,2}\}
\\
&+
\frac{I\{{\bf A}_2=(0,1)\}}{\bar\pi_2(0,1)}
\frac{f_{1,2}^{1}}{f_{1,2}^{0}}
\frac{f_{2,1}^{00}}{f_{2,1}^{01}}
\{\mu_{2,2}-\mu_{2,1}\}
\\
&+
\frac{I\{{\bf A}_2=(0,0)\}}{\bar\pi_2(0,0)}
\frac{f_{1,2}^{1}}{f_{1,2}^{0}}
\{\mu_{2,1}-\mu_{1,2}\}
\\
&+
\frac{I\{A_1=1\}}{\pi_1(1)}
\frac{f_{1,1}^{0}}{f_{1,1}^{1}}
\{\mu_{1,2}-\mu_{1,1}\}
\\
&+
\frac{I\{A_1=0\}}{\pi_1(0)}
\{\mu_{1,1}-\mu_{1,0}\}
\\
&+\mu_{1,0}-\psi.
\end{aligned}
\label{eq:ex_nonunique_density_eif}
\end{equation}
For any fixed partition described above, Corollary~\ref{cor:robustness_1} establishes four-out-of-five robustness. Since compatible partitions yield identical weighting identities, the estimator and its second-order remainder are invariant to the choice of partition. Thus, we may choose a partition that maximizes the selected misspecified block, allowing all of its components to be misspecified while requiring consistent estimation of the remaining blocks. For example, when $\mathcal B_{L_1}$ is allowed to be misspecified, Partitions A and C yield its largest block, containing $f_{1,1}$ and $f_{2,1}$, while the remaining nuisance blocks are required to be consistently estimated. Since the last node in the largest $\mathcal B_{L_1}$ is $X_{2,1}$, which precedes $A_3$, consistency still holds when $\pi_3$ is misspecified.

Table~\ref{tab:nonunique_density_relaxations} summarizes the resulting relaxed four-out-of-five robustness regimes for the density-based estimator. Each regime permits misspecification of the largest compatible version of the selected nuisance block, along with the additional relaxation of the treatment mechanisms.

\begin{table}[!htbp]
\centering
\caption{Relaxed four-out-of-five robustness regimes for the density-based estimator under nonunique assignment partitions in Example~\ref{app:ex:non-unique__partitions}.}
\label{tab:nonunique_density_relaxations}
\setlength{\tabcolsep}{10pt}
\begin{tabular}{lll}
\toprule
Selected block
&
Allowed misspecified nuisances
&
Required consistent estimation
\\
\midrule
$\mathcal B_\pi$
&
$\pi_1,\pi_2,\pi_3$
&
$\mu_{3,1}$,
$f_{1,1}$, $f_{1,2}$,
$f_{2,1}$, $f_{2,2}$, and $f_{3,1}$
\\[3pt]
$\mathcal B_M$
&
$\mu_{3,1},f_{1,2}$
&
$\pi_1,\pi_2,\pi_3$,
$f_{1,1}$, $f_{2,1}$,
$f_{2,2}$, and $f_{3,1}$
\\[3pt]
$\mathcal B_{L_1}$
&
$f_{1,1},f_{2,1},\pi_3$
&
$\pi_1,\pi_2$,
$\mu_{3,1}$,
$f_{1,2}$, $f_{2,2}$, and $f_{3,1}$
\\[3pt]
$\mathcal B_{L_2}$
&
$f_{1,1},f_{2,2},\pi_3$
&
$\pi_1,\pi_2$,
$\mu_{3,1}$,
$f_{1,2}$, $f_{2,1}$, and $f_{3,1}$
\\[3pt]
$\mathcal B_{L_3}$
&
$f_{1,2},f_{3,1}$
&
$\pi_1,\pi_2,\pi_3$,
$\mu_{3,1}$,
$f_{1,1}$, $f_{2,1}$, and $f_{2,2}$
\\
\bottomrule
\end{tabular}
\end{table}

For the regression-based estimator, the same logic as in Corollary~\ref{cor:robustness_2} and the preceding example applies. Each cut is indexed by a node in the observed-data ordering, so neither the cut-based regimes nor the more general pairwise conditions depend on the compatible assignment partition. In this example, Bayes' rule retains all five treatment-history regressions $g_{1,1}$, $g_{1,2}$, $g_{2,1}$, $g_{2,2}$, and $g_{3,1}$ in the weights. Thus, there is no algebraic cancellation that removes an entire $g_{k,r}$ or $\mu_{k,r}$ as in the preceding unique-partition example. The general sequential robustness conditions therefore apply without that additional simplification. 
\end{example}

We emphasize that Corollaries~\ref{cor:robustness_1} and~\ref{cor:robustness_2} provide systematic sufficient conditions for consistency under nuisance functions misspecification. Specifically, Corollary~\ref{cor:robustness_1} establishes $(S+1)$-out-of-$(S+2)$ robustness for the density-based estimator, whereas Corollary~\ref{cor:robustness_2} characterizes robustness of the regression-based estimator through cut points. However, these conditions are not necessary for consistency and may be further relaxed by exploiting the flexibility of compatible assignment partitions and the structure of the EIF. For example, in the nonunique-partition setting, consistent estimation of $\pi_1$, $f_{1,2}$, $f_{2,1}$, $f_{2,2}$, $f_{3,1}$, and $\mu_{3,1}$ can also yield consistency. Our aim is therefore not to exhaustively characterize all possible misspecification regimes, but to provide a framework for examining and deriving sufficient robustness conditions.

\subsection{Sensitivity analysis examples}
\label{app:ex_sa}

We illustrate the sensitivity analysis in two settings. Case~1 considers one treatment and two ordered intermediate variables, whereas Case~2 considers two repeated treatments with two intermediate variables following each treatment occasion. Supplementary Figure~\ref{fig:sens_graphs}(a) extends Figure~\ref{fig:graphs}(a), and Supplementary Figure~\ref{fig:sens_graphs}(b) extends Figure~\ref{fig:graphs}(b). Blue bidirected arrows represent latent dependence within a common assignment class, which does not by itself create a recanting district. Red bidirected arrows connect nodes assigned different treatment histories and represent the departures indexed by the sensitivity functions.

\begin{figure}[!htbp]
\begin{center}
\scalebox{0.64}{
\begin{tikzpicture}[>=stealth, node distance=1.8cm]
\tikzstyle{format} = [thick, circle, minimum size=1.0mm, inner sep=2pt]
\tikzstyle{square} = [draw, thick, minimum size=4.5mm, inner sep=2pt]

\begin{scope}[xshift=0cm, yshift=0cm]
\path[->, thick]

node[] (t) {$A$}
node[right of=t, xshift=0.15cm] (m) {$M$}
node[right of=m, xshift=0.15cm] (l) {$L$}
node[above right of=m, yshift=0.2cm] (x) {$C$}
node[right of=l, xshift=0.15cm] (y) {$Y$}

node[below right of=m, xshift=-0.2cm, yshift=-1.0cm] (figa) {(a)}

(x) edge[gray] (t)
(x) edge[gray] (m)
(x) edge[gray] (l)
(x) edge[gray] (y)
(t) edge[gray] (m)
(m) edge[gray] (l)
(l) edge[gray] (y)

(t) edge[gray,out=90,in=155] (l)
(m) edge[gray,out=25,in=90] (y)
(t) edge[gray,out=-25,in=205] (y)

(l) edge[<->,line width=2pt,red,out=20,in=160] (y)
;
\end{scope}

\begin{scope}[xshift=8cm]
\path[->, thick]

node[] (c) {$C$}
node[right of=c] (a1) {$A_1$}
node[right of=a1] (m1) {$M_1$}
node[right of=m1] (l1) {$L_1$}
node[right of=l1] (a2) {$A_2$}
node[right of=a2] (m2) {$M_2$}
node[right of=m2] (l2) {$L_2$}
node[right of=l2] (y) {$Y$}

(c) edge[gray] (a1)
(c) edge[gray,bend right=20] (m1)
(c) edge[gray,bend left=20] (l1)
(c) edge[gray,bend right=25] (a2)
(c) edge[gray,bend left=25] (m2)
(c) edge[gray,bend right=25] (l2)
(c) edge[gray,bend left=25] (y)

(a1) edge[gray] (m1)
(a1) edge[gray,bend right=20] (l1)
(a1) edge[gray,bend left=20] (a2)
(a1) edge[gray,bend right=25] (m2)
(a1) edge[gray,bend left=27] (l2)
(a1) edge[gray,bend right=25] (y)

(m1) edge[gray] (l1)
(m1) edge[gray,bend right=18] (a2)
(m1) edge[gray,bend left=18] (m2)
(m1) edge[gray,bend right=23] (l2)
(m1) edge[gray,bend left=23] (y)

(l1) edge[gray] (a2)
(l1) edge[gray,bend right=18] (m2)
(l1) edge[gray,bend left=20] (l2)
(l1) edge[gray,bend right=23] (y)

(a2) edge[gray] (m2)
(a2) edge[gray,bend right=16] (l2)
(a2) edge[gray,bend left=20] (y)

(m2) edge[gray] (l2)
(m2) edge[gray,bend right=20] (y)

(l2) edge[gray] (y)

(m1) edge[<->,line width=2pt,blue,out=-35,in=-145] (m2)
(l1) edge[<->,line width=2pt,blue,out=-35,in=-145] (l2)
(l1) edge[<->,line width=2pt,red,out=60,in=120] (y)
(l2) edge[<->,line width=2pt,red,out=40,in=140] (y)
;

node[right of=figa, xshift=9.5cm] (figb) {(b)}
;
\end{scope}

\end{tikzpicture}
}
\vspace{-0.25cm}
\caption{Illustrative causal diagrams for sensitivity analysis.
(a) Single-treatment setting with two ordered intermediate variables.
(b) Longitudinal setting with two repeated treatments and two intermediate variables following each treatment occasion. Blue bidirected arrows represent latent dependence within a common assignment class, whereas red bidirected arrows represent departures indexed by the sensitivity functions.}
\label{fig:sens_graphs}
\end{center}
\end{figure}
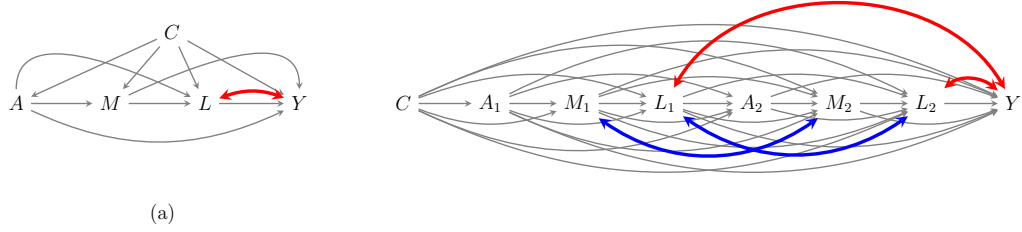

\paragraph{\textbf{Case 1: One treatment and two intermediate variables}} 

In Figure~\ref{fig:sens_graphs}(a), let $A$ indicate exposure to a school-based physical-activity promotion program, $M$ represent students' knowledge about the health benefits of regular exercise, $L$ represent their enjoyment or valuation levels of physical activity, and $Y$ denote subsequent physical-activity frequency. Consider the pathways involving the direct effect and the effects transmitted through $M$,
\[
\rho_{s1}=\{A\rightarrow M\leadsto Y,\ A\rightarrow Y\}.
\]
For $a=1$ and $a'=0$, the corresponding nested outcome is
\[
Y(a,a';\rho_{s1})
=
Y\{a,M(a),L(a',M(a))\}.
\]
Thus, the treatment assignment entering $L$ directly is held at the reference level, although $L$ may still be affected indirectly through $M(a)$.

Identification under the original model uses, among the other assumptions, for any $b \in \{0,1\}$, the cross-world restriction
\[
Y(a,m,l)\indep L(b,m)\mid M,A,C.
\]
To assess departures from this restriction, define
\begin{align}
t(b,m,l,c; a)
={}&
E\{Y(a,m,l)\mid
A=b,M=m,L=l,C=c\}
\notag\\
&-
E\{Y(a,m,l)\mid
A=b,M=m,L\neq l,C=c\}.
\label{app:eq:sens_case1_t}
\end{align}
A nonzero value of $t(b,m,l,c; a)$ implies residual cross-world dependence between $Y(a,m,l)$ and $L(b,m)$ conditional on the relevant history. Conversely, the cross-world independence restriction implies $t=0$.

Let $\mu_{1,2}(m,l,c) = E(Y\mid A=a,M=m,L=l,C=c)$. For fixed $t$, the sensitivity-indexed target is
\begin{align}
\psi_t(P)
={}&
\int
\Big[
\mu_{1,2}(m,l,c)
+
t(a',m,l,c; a)
\{1-f_L(l\mid m,a',c)\}
\notag\\
&\hspace{3.0cm}
-
t(a,m,l,c; a)
\{1-f_L(l\mid m,a,c)\}
\Big]
\notag\\
&\qquad\times
f_L(l\mid m,a',c)
f_M(m\mid a,c)
\,d\nu_L(l)\,d\nu_M(m)\,dP(c).
\label{app:eq:sens_case1_id}
\end{align}

The corresponding augmented estimator is
\begin{align}
\widehat\psi_{t}
=
P_n\bigg[
&
\frac{I(A=a)}{\widehat\pi(a\mid C)}
\frac{\widehat f_L(L\mid M,a',C)}
     {\widehat f_L(L\mid M,a,C)}
\{Y-\widehat\mu_{1,2}(M,L,C)\}
+
 h_t(C;\widehat \mu, \widehat f)
\bigg],
\label{app:eq:sens_case1_est}
\end{align}
where
\begin{align*}
 h_t(c;\widehat \mu, \widehat f)
={}&
\int
\Big[
\widehat\mu_{1,2}(m,l,c)
+
t(a',m,l,c; a)
\{1-\widehat f_L(l\mid m,a',c)\}
\\
&\hspace{2.9cm}
-
t(a,m,l,c; a)
\{1-\widehat f_L(l\mid m,a,c)\}
\Big]
\\
&\qquad\times
\widehat f_L(l\mid m,a',c)
\widehat f_M(m\mid a,c)
\,d\nu_L(l)\,d\nu_M(m).
\end{align*}

\vspace{0.5cm}
\paragraph{\textbf{Case 2: Two repeated treatments and two intermediate-variable blocks}}

Figure~\ref{fig:sens_graphs}(b) represents a longitudinal study. Let $A_1$ and $A_2$ denote smoking status at baseline and year~1, respectively. Let $M_1$ and $M_2$ denote health biomarkers and let $L_1$ and $L_2$ denote body mass index categories, measured after the corresponding treatment occasions. Let $Y$ denote blood pressure at a subsequent follow-up. We consider the pathways
\[
\rho_{s2}
=
\{
A_1\rightarrow L_1\leadsto Y,\;
A_1\rightarrow L_2\rightarrow Y,\;
A_2\rightarrow L_2\rightarrow Y
\}.
\]
Let ${\bf a}_2=(a_1,a_2)=(1,1)$ and
${\bf a}'_2=(a'_1,a'_2)=(0,0)$. Define
\begin{align*}
M_1'&=M_1(a'_1),\\
L_1^*&=L_1(a_1,M_1'),\\
M_2'&=M_2(a'_1,M_1',L_1^*,a'_2),\\
L_2^*&=L_2(a_1,M_1',L_1^*,a_2,M_2').
\end{align*}
The nested outcome can then be written compactly as
\[
Y({\bf a}_2,{\bf a}'_2;\rho_{s2})
=
Y(a'_1,M_1',L_1^*,a'_2,M_2',L_2^*).
\]
Thus, $M_1$, $M_2$, and $Y$ receive the reference assignment, whereas
$L_1$ and $L_2$ receive the active assignment, yielding $\mathbb L = \{L_1, L_2\}$.

The blue bidirected edges in Figure~\ref{fig:sens_graphs}(b) connect nodes within common assignment classes. Accordingly, latent dependence between $M_1$ and $M_2$, or between $L_1$ and $L_2$, does not by itself produce a recanting district. In contrast, latent dependence connecting $L_1$ or $L_2$ to $Y$ links nodes receiving different assignments and is indexed by two sensitivity functions.

Let ${\bf x}=(m_1,l_1,m_2,l_2)$ and $Y_{\bf x}=Y(a'_1,m_1,l_1,a'_2,m_2,l_2)$. For 
${\bf b}_2=(b_1, b_2) \in \{0, 1\}^2$, define
\begin{align}
t_1(b_1,{\bf x},c; a'_1, a'_2)
={}&
E\{Y_{\bf x}\mid
A_1=b_1,M_1=m_1,L_1=l_1,C=c\}
\notag\\
&-
E\{Y_{\bf x}\mid
A_1=b_1,M_1=m_1,L_1\neq l_1,C=c\},
\label{sens:t1}
\end{align}
and 
\begin{align}
t_2({\bf b}_2,{\bf x},c; a'_1, a'_2)
& =
E\{Y_{\bf x}\mid
A_1=b_1,M_1=m_1,L_1=l_1,
A_2=b_2,M_2=m_2,L_2=l_2,C=c\}
\notag\\
&\qquad -
E\{Y_{\bf x}\mid
A_1=b_1,M_1=m_1,L_1=l_1,
A_2=b_2,M_2=m_2,L_2\neq l_2,C=c\}.
\label{sens:t2}
\end{align}
A nonzero $t_1(b_1,{\bf x},c; a'_1, a'_2)$ implies
\[
Y_{\bf x}
\not\!\!\indep
L_1(b_1,m_1)
\mid M_1,A_1,C,
\]
and a nonzero $t_2({\bf b}_2,{\bf x},c; a'_1, a'_2)$ implies
\[
Y_{\bf x}
\not\!\!\indep
L_2(b_1,m_1,l_1,b_2,m_2)
\mid M_2,A_2,L_1,M_1,A_1,C.
\]

Define
\begin{align*}
\delta_{t_1}({\bf x},c;f)
=&
t_1(a_1,{\bf x},c; a'_1, a'_2)
\left[
1-f_{L_1}(l_1\mid a_1,m_1,c)
\right] \\
& -
t_1(a'_1,{\bf x},c; a'_1, a'_2)
\left[
1-f_{L_1}(l_1\mid a'_1,m_1,c)
\right],
\\
\delta_{t_2}({\bf x},c;f)
=&
t_2({\bf a}_2,{\bf x},c; a'_1, a'_2)\!\!
\left[
1-f_{L_2}(l_2\mid {\bf a}_2, m_1,l_1,m_2,c)
\right]\\
&-
t_2({\bf a}'_2,{\bf x},c; a'_1, a'_2) \!\!
\left[
1-f_{L_2}(l_2\mid {\bf a}'_2, m_1,l_1,m_2,c)
\right].
\end{align*}
Also let
\[
\mu_{2,2}({\bf x},c)
=
E(Y\mid
A_1=a'_1,M_1=m_1,L_1=l_1,
A_2=a'_2,M_2=m_2,L_2=l_2,C=c).
\]
The sensitivity-indexed identifying functional is
\begin{align}
\psi_{t}(P)
={}&
\int
\left[
\mu_{2,2}({\bf x},c)
+
\delta_{t_1}({\bf x},c;f)
+
\delta_{t_2}({\bf x},c;f)
\right]
\notag\\
&\quad\times
f_{L_2}(l_2\mid {\bf a}_2, m_1,l_1,m_2,c)
f_{M_2}(m_2\mid {\bf a}'_2, m_1,l_1,c)
\notag\\
&\quad\times
f_{L_1}(l_1\mid a_1,m_1,c)
f_{M_1}(m_1\mid a'_1,c)
\,d\nu({\bf x})\,dP(c).
\label{app:eq:sens_case2_id}
\end{align}

Define
\begin{align*}
\widehat W_Y
=
\frac{I(A_1=a'_1,A_2=a'_2)}
{\widehat\pi_1(a'_1\mid C)
 \widehat\pi_2(a'_2\mid
 A_1=a'_1,M_1,L_1,C)}
\times
\frac{
\widehat f_{L_1}(L_1\mid a_1, M_1,C)
}{
\widehat f_{L_1}(L_1\mid a'_1, M_1,C)
}
\frac{
\widehat f_{L_2}(L_2\mid {\bf a}_2, M_1,L_1,M_2,C)
}{
\widehat f_{L_2}(L_2\mid {\bf a}'_2, M_1,L_1,M_2,C)
}.
\end{align*}
Then
\begin{align*}
 h_{t}(c;\widehat \mu, \widehat f)
={}&
\int
\left[
\widehat\mu_{2,2}({\bf x},c)
+
\delta_{t_1}({\bf x},c;\widehat f)
+
\delta_{t_2}({\bf x},c;\widehat f)
\right]
\\
&\quad\times
\widehat f_{L_2}(l_2\mid {\bf a}_2, m_1,l_1,m_2,c)
\widehat f_{M_2}(m_2\mid {\bf a}'_2, m_1,l_1,c)
\\
&\quad\times
\widehat f_{L_1}(l_1\mid a_1, m_1,c)
\widehat f_{M_1}(m_1\mid a'_1, c)
\,d\nu({\bf x}),
\end{align*}
and the augmented estimator is
\begin{align}
\widehat\psi_{t}
=
P_n\left[
\widehat W_Y
\{Y-\widehat\mu_{2,2}(M_1,L_1,M_2,L_2,C)\}
+ h_{t}(C; \widehat \mu, \widehat f)
\right].
\label{app:eq:sens_case2_est}
\end{align}

For each fixed choice of the sensitivity functions, the estimators in \eqref{app:eq:sens_case1_est} and \eqref{app:eq:sens_case2_est} inherit the conditional double-robustness property established in Appendix~\ref{app:proofs_sa_DR}: consistency requires correct estimation of the intermediate-variable laws and either the outcome regression or the treatment mechanism.

\newpage
\section{Additional details on simulation studies}
\label{app:sims}

\subsection{Simulation 1: Confirming theoretical properties}

To confirm the asymptotic properties of the one-step estimators discussed in Theorems~\ref{thm:asymp_lin_1} and~\ref{thm:asymp_lin_2}, we generated the first simulation dataset with variables $(C_1,C_2,C_3,A_1,M_1,L_1,A_2,M_2,L_2,Y)$ according to the following data-generating process (DGP):
\begin{equation}
\label{eq:simu1_DGP}
\begin{aligned}
C_1 &\sim \mathrm{Bernoulli}(0.5), \hspace{0.15cm}
C_2 \sim \mathrm{Uniform}(-1,1), \hspace{0.15cm}
C_3 \sim \mathrm{Normal}(0,1), \hspace{0.15cm}
C_1 \indep C_2 \indep C_3 \\
A_1 &\sim \mathrm{Bernoulli}\!\left(
\expit \left(\beta_{A_1}
\begin{bmatrix}
C_1 & C_2 & C_3 & C_2^2 & C_2C_3
\end{bmatrix}^{\!T} \right)
\right), \\
M_1 &\sim \mathrm{Normal}\!\left(
\beta_{M_1}
\begin{bmatrix}
C_1 & C_2 & C_3 & A_1 & C_2^2 & C_1C_3 & C_3A_1
\end{bmatrix}^{\!T},
\,1\right), \\
L_1 &\sim \mathrm{Normal}\!\left(
\beta_{L_1}
\begin{bmatrix}
C_1 & C_2 & C_3 & A_1 & M_1 & C_2^2
\end{bmatrix}^{\!T},
\,1\right), \\
A_2 &\sim \mathrm{Bernoulli}\!\left(
\expit \left(\beta_{A_2}
\begin{bmatrix}
C_1 & C_2 & C_3 & A_1 & M_1 & L_1 & C_3^2 & C_1C_3
\end{bmatrix}^{\!T}\right)
\right), \\
M_2 &\sim \mathrm{Normal}\!\left(
\beta_{M_2}
\begin{bmatrix}
C_1 & C_2 & C_3 & A_1 & M_1 & L_1 & A_2 & C_2^2
\end{bmatrix}^{\!T},
\,1\right), \\
L_2 &\sim \mathrm{Normal}\!\left(
\beta_{L_2}
\begin{bmatrix}
C_1 & C_2 & C_3 & A_1 & M_1 & L_1 & A_2 & M_2 & C_3^2 & C_2C_3
\end{bmatrix}^{\!T},
\,1\right), \\
Y &\sim \mathrm{Normal}\!\left(
\beta_{Y}
\begin{bmatrix}
C_1 & C_2 & C_3 & A_1 & M_1 & L_1 & A_2 & M_2 & L_2 & C_3^2
\end{bmatrix}^{\!T},
\,1\right),
\end{aligned}
\end{equation}
where
\begin{equation}
\begin{aligned}
    \beta_{A_1} &= (0.31, 0.35, 0.12, -0.13, -0.08), \\
    \beta_{M_1} &= (0.47, -0.28, -0.11, 0.46, -0.19, 0.09, 0.12),\\
    \beta_{L_1} &= (0.32, 0.26, 0.19, -0.18, 0.4, 0.15),\\
    \beta_{A_2} & = (-0.18, 0.34, -0.13, 0.32, 0.22, 0.21, 0.05, -0.10), \\
    \beta_{M_2} &= (0.32, 0.35, 0.16, 0.29, 0.02, -0.13, -0.11, 0.28), \\
    \beta_{L_2} &= (0.08, -0.19, 0.37, 0.14, 0.38, 0.31, 0.29, 0.21, -0.19, 0.17), \\
    \beta_{Y} &= (-0.49, -0.23, 0.28, 0.46, -0.34, 0.73, 0.41, 0.77, 0.65, 0.16).
\end{aligned}
\end{equation}
The target estimand is
\begin{align*}
    \psi = E [ Y ( &a_1', M_1(a_1'), L_1(a_1, M_1(a_1')), a_2', M_2(a_1', M_1(a_1'), L_1(a_1, M_1(a_1')), a_2'),  \\ 
    &L_2(a_1, M_1(a_1'), L_1(a_1, M_1(a_1')), a_2, M_2(a_1', M_1(a_1'), L_1(a_1, M_1(a_1')), a_2')) ) ], 
\end{align*}
where $a_1',a_2'$ are the reference treatment levels, both set to 0, and $a_1,a_2$ are the active treatment levels, both set to 1. The corresponding simulation results are presented in Figure~\ref{fig:theoretical_estimators} of the main text.

\subsubsection{Correct specification of nuisance models via GLMs} 
\label{subsec:correct_GLMs}
In this simulation study, all nuisance functions are correctly specified using GLMs in order to evaluate the asymptotic behavior of the proposed one-step estimators. For the density-based estimator $\psi^+(\widehat Q_1)$, the primitive nuisance collection is
\[
Q_1=\{\pi_1,\pi_2,f_{1,1},f_{1,2},f_{2,1},f_{2,2},\mu_{2,2}\}.
\]
These functions can be specified directly from the DGP, and the remaining sequential regressions are recursively induced by $\mu_{2,2}$ and the fitted intermediate-variable laws. They are not fitted independently. In contrast, the regression-based estimator $\psi^+(\widehat Q_2)$ uses
\[
Q_2=\{\pi_1,\pi_2,g_{1,1},g_{1,2},g_{2,1},g_{2,2},
\mu_{2,2},\mu_{2,1},\mu_{1,2},\mu_{1,1},\mu_{1,0}\}.
\]
The treatment-history regressions $g_{1,1},g_{1,2},g_{2,1},g_{2,2}$ are not available directly from the DGP. We therefore derive their correct functional forms by expressing the density-ratio terms through Bayes' rule together with suitable factorizations of joint treatment probabilities.

\paragraph{Step 1: Bayes representations of the density ratios.}
The density ratios can be written as
\begin{align*}
& \frac{f(M_1 \mid a_1', C)}{f(M_1 \mid a_1, C)} =
    \frac{P(A_1 = 0 \mid M_1, C)}{P(A_1 = 1 \mid M_1, C)}
    \frac{P(A_1 = 1 \mid C)}{P(A_1 = 0 \mid C)}, \\
& \frac{f(L_1 \mid M_1, a_1, C)}{f(L_1 \mid M_1, a_1', C)} =
    \frac{P(A_1 = 1 \mid L_1, M_1, C)}{P(A_1 = 0 \mid L_1, M_1, C)}
    \frac{P(A_1 = 0 \mid M_1, C)}{P(A_1 = 1 \mid M_1, C)}, \\
& \frac{f(M_2 \mid a_2', L_1, M_1, a_1', C)}{f(M_2 \mid a_2, L_1, M_1, a_1, C)} =
    \frac{P(A_1 = A_2 = 0 \mid M_2, L_1, M_1, C)}{P(A_1 = A_2 = 1 \mid M_2, L_1, M_1, C)}\frac{P(A_1 = A_2 = 1 \mid L_1, M_1, C)}{P(A_1 = A_2 = 0 \mid L_1, M_1, C)} \\ 
& \frac{f(L_2 \mid M_2, a_2, L_1, M_1, a_1, C)}{f(L_2 \mid M_2, a_2', L_1, M_1, a_1', C)} =
    \frac{P(A_1 = A_2 = 1 \mid L_2, M_2, L_1, M_1, C)}{P(A_1 = A_2 = 0 \mid L_2, M_2, L_1, M_1, C)}
    \frac{P(A_1 = A_2 = 0 \mid M_2, L_1, M_1, C)}{P(A_1 = A_2 = 1 \mid M_2, L_1, M_1, C)}
\end{align*}

\paragraph{Step 2: Factorization of the joint treatment probabilities.} To obtain operational GLM specifications, we further decompose the joint treatment probabilities above.

First, by the chain rule of conditional probability,
\begin{align}
\frac{P(A_1= A_2=1 \mid L_1, M_1, C)}{P(A_1= A_2=0 \mid L_1, M_1, C)}
=\frac{P(A_2=1 \mid A_1=1, L_1, M_1, C)}{P(A_2=0 \mid A_1=0, L_1, M_1, C)} \frac{P(A_1=1 \mid L_1, M_1, C)}{P(A_1=0 \mid L_1, M_1, C)}.
\label{eq:chain-rule-factorization}
\end{align}

Second, for the joint probabilities conditional on post-second treatment  variables such as $M_2$ or $L_2$, we use the odds-ratio representation of \citet{chen07semiparametric}. This gives
\begin{align}
&\frac{P(A_1= A_2=0 \mid M_2, L_1, M_1, C)}{P(A_1 = A_2=1 \mid M_2, L_1, M_1, C)} \nonumber \\
&\hspace{1.75cm}= 
\frac{P(A_1=0 \mid A_2=0, M_2, L_1, M_1, C)}{P(A_1=1 \mid A_2=0, M_2, L_1, M_1, C)}
\frac{P(A_2=0 \mid A_1=1, M_2, L_1, M_1, C)}{P(A_2=1 \mid A_1=1, M_2, L_1, M_1, C)},
\label{eq:m2-joint-factorization}
\end{align}
and similarly,
\begin{align}
&\frac{P(A_1= A_2=1 \mid L_2, M_2, L_1, M_1, C)}{P(A_1= A_2=0 \mid L_2, M_2, L_1, M_1, C)} \nonumber \\
&\hspace{1.75cm} =\frac{P(A_1=1 \mid A_2=0, L_2, M_2, L_1, M_1, C)}{P(A_1=0 \mid A_2=0, L_2, M_2, L_1, M_1, C)}
\frac{P(A_2=1 \mid A_1=1, L_2, M_2, L_1, M_1, C)}{P(A_2=0 \mid A_1=1, L_2, M_2, L_1, M_1, C)}.
\label{eq:l2-joint-factorization}
\end{align}

Specifically,  \eqref{eq:m2-joint-factorization} and \eqref{eq:l2-joint-factorization} are derived using the odds-ratio parameterization strategy. Let $A$, $B$, and $C$ be disjoint random vectors with $A$ and $B$ being binary, and let $a_0$ and $b_0$ be reference values of $A$ and $B$, respectively. Following \citet{chen07semiparametric}, the joint conditional distribution $P(A,B \mid C)$ admits the representation
\begin{equation}
P(A,B \mid C)=\frac{1}{Z(C)}\,P(A \mid B=b_0, C)\,P(B \mid A=a_0, C)\,\mathrm{OR}(A,B \mid C),
\end{equation}
where\begin{align*}
\mathrm{OR}(A=a,B=b \mid C)=\frac{P(A=a \mid B=b, C)}{P(A=a_0 \mid B=b, C)}\cdot\frac{P(A=a_0 \mid B=b_0, C)}{P(A=a \mid B=b_0, C)},
\end{align*}
and the normalizing constant is $Z(C)=\sum_{A,B}P(A \mid B=b_0, C)\,P(B \mid A=a_0, C)\,\mathrm{OR}(A,B \mid C)$. 

\paragraph{Step 3: Logistic specification of the generalized propensity score components.}
To complete the nuisance specification for $\psi^+(\widehat Q_2)$, it remains to determine the correct functional forms of
\begin{align*}
&\hspace{3cm} P(A_1\mid C), \quad 
P(A_1 \mid  M_1,C), \quad  
P(A_1\mid L_1,M_1,C),  
\\
&\hspace{1.75cm} P(A_1\mid M_2,A_2,L_1,M_1, C), \quad 
P(A_1 \mid L_2,M_2,A_2,L_1,M_1,C),
\\
&P(A_2\mid L_1,M_1,A_1,C), \quad 
P(A_2\mid M_2,L_1,M_1,A_1,C), \quad 
P(A_2\mid L_2,M_2,L_1,M_1,A_1,C).
\end{align*}
Among these, $P(A_1 \mid C)$ and $P(A_2 \mid L_1,M_1,A_1,C)$ are available directly from the DGP and can therefore be fit using standard logistic regression. The remaining conditional treatment probabilities are also of logistic form, although this is less immediate. We illustrate the argument by deriving the correct specification of $P(A_1\mid M_1,C)$; the other cases follow by the same logic.

Under the DGP in \eqref{eq:simu1_DGP}, write
\[
A_1 \mid C \sim \mathrm{Bernoulli}\{\expit(\eta(C))\},
\qquad
M_1 \mid A_1,C \sim N\!\bigl(\nu(C)+A_1\gamma(C),\,1\bigr),
\]
where
\[
\eta(C)=\beta_{A_1}C_{A_1}^{\top}, \qquad
\nu(C)=\beta_{M_1C}C_{M_1}^{\top}, \qquad
\gamma (C)=\beta_{M_1A_1}
\begin{bmatrix}
1 & C_3
\end{bmatrix}^{\!T},
\]
with
\[
C_{A_1}=
\begin{bmatrix}
C_1 & C_2 & C_3 & C_2^2 & C_2C_3
\end{bmatrix},
\qquad
C_{M_1}=
\begin{bmatrix}
C_1 & C_2 & C_3 & C_2^2 & C_1C_3
\end{bmatrix},
\]
and $\beta_{M_1C}$ collects the coefficients of $C_{M_1}$, whereas $\beta_{M_1A_1}$ collects the coefficients of $(A_1,A_1C_3)$. Under the specified DGP, $\beta_{M_1C}=(0.47,-0.28,-0.11,-0.19,0.09), \beta_{M_1A_1}=(0.46,0.12).$

We can then show directly that $P(A_1=1\mid M_1,C)$ is of logistic form.

By Bayes' rule,
\begin{align}
P(A_1=1 \mid M_1,C)
&=
\frac{P(A_1=1 \mid C)\,f(M_1 \mid A_1=1,C)}
{P(A_1=1 \mid C)\,f(M_1 \mid A_1=1,C)+P(A_1=0 \mid C)\,f(M_1 \mid A_1=0,C)} \nonumber \\
&=
\left[
1+
\frac{P(A_1=0 \mid C)}{P(A_1=1 \mid C)}
\frac{f(M_1 \mid A_1=0,C)}{f(M_1 \mid A_1=1,C)}
\right]^{-1}.
\label{eq:A1_M1X_bayes}
\end{align}
From the Bernoulli model for $A_1 \mid C$,
\[
\frac{P(A_1=0 \mid C)}{P(A_1=1 \mid C)}
=
\exp\{-\eta(C)\}.
\]
Moreover, the ratio of the two conditional normal densities of $M_1$ is
\begin{align*}
\frac{f(M_1 \mid A_1=0,C)}{f(M_1 \mid A_1=1,C)}
&=
\exp\left\{
-\frac{1}{2}\big(M_1-\nu(C)\big)^2
+\frac{1}{2}\big(M_1-\nu(C)-\gamma(C)\big)^2
\right\} \\
&=
\exp\left\{
-\,\gamma(C)\big(M_1 - \nu(C)\big)+\frac{1}{2}\gamma(C)^2
\right\}.
\end{align*}
Substituting these expressions into \eqref{eq:A1_M1X_bayes} gives
\begin{align}
P(A_1=1 \mid M_1,C)
&=
\left[
1+\exp\left\{
-\eta(C)-\gamma(C)\big(M_1 - \nu(C)\big)+\frac{1}{2}\gamma(C)^2
\right\}
\right]^{-1} \nonumber\\
&=
\expit\!\left(
\eta(C)+\gamma(C)\big(M_1 - \nu(C)\big)-\frac{1}{2}\gamma(C)^2
\right).
\label{eq:A1_M1X_logistic_form}
\end{align}
Hence,
\begin{equation}
\logit\{P(A_1=1 \mid M_1,C)\}
=
\eta(C)+\gamma(C)\big(M_1 - \nu(C)\big)-\frac{1}{2}\gamma(C)^2.
\end{equation}

The argument above reveals the general principle underlying these nuisance specifications. In \eqref{eq:A1_M1X_bayes}, the posterior probability $P(A_1=1 \mid M_1,C)$ takes a logistic form because both the odds ratio $\frac{P(A_1=0\mid C)}{P(A_1=1\mid C)}$ and the density ratio $\frac{f(M_1\mid A_1=0,C)}{f(M_1\mid A_1=1,C)}$ inside the bracketed term admit exponential representations. Their product is therefore again of the form $\exp\{\cdot\}$, which implies that the resulting conditional treatment probability follows the logit link with a linear combination of expanded predictors. By the same reasoning, the conditional treatment models $P(A_1\mid L_1,M_1,C),
P(A_2\mid M_2,L_1,M_1,A_1,C),
P(A_2\mid L_2,M_2,L_1,M_1,A_1,C)$
are also correctly specified by logistic regressions.

\paragraph{An offset correction term.} For $P(A_1\mid M_2,A_2,L_1,M_1,C)$ and $P(A_1\mid L_2,M_2,A_2,L_1,M_1,C)$, an additional technical device is required to accommodate the conditioning on $A_2$. Specifically, we introduce an offset term that preserves the logistic regression representation.

Consider $P(A_1=1\mid M_2,A_2,L_1,M_1,C)$. By Bayes' rule,
\begin{equation}
P(A_1=1\mid M_2,A_2,L_1,M_1,C)
=
\left[
1+
\frac{P(A_1=0\mid A_2,L_1,M_1,C)}{P(A_1=1\mid A_2,L_1,M_1,C)}
\frac{f(M_2\mid A_1=0,A_2,L_1,M_1,C)}{f(M_2\mid A_1=1,A_2,L_1,M_1,C)}
\right]^{-1}.
\label{eq:A1_given_M2A2}
\end{equation}
The second factor in \eqref{eq:A1_given_M2A2} is well behaved: under the DGP, $M_2\mid A_1,A_2,L_1,M_1,C$ is Gaussian. The nontrivial term is the first factor, $\frac{P(A_1=0\mid A_2,L_1,M_1,C)}{P(A_1=1\mid A_2,L_1,M_1,C)}$.
Applying Bayes' rule again,
\begin{equation}
P(A_1=1\mid A_2,L_1,M_1,C)
=
\left[
1+
\frac{P(A_1=0\mid L_1,M_1,C)}{P(A_1=1\mid L_1,M_1,C)}
\frac{P(A_2\mid A_1=0,L_1,M_1,C)}{P(A_2\mid A_1=1,L_1,M_1,C)}
\right]^{-1}.
\label{eq:A1_given_A2L1M1X}
\end{equation}
The first ratio on the right-hand side of \eqref{eq:A1_given_A2L1M1X} can be written in exponential form because  $P(A_1\mid L_1,M_1,C)$ is specified through a logistic regression; equivalently, its conditional odds admit a logit representation with a linear predictor. In contrast, the second ratio,
\[
\frac{P(A_2\mid A_1=0,L_1,M_1,C)}{P(A_2\mid A_1=1,L_1,M_1,C)},
\]
does not, by itself, reduce directly to the exponential of a linear predictor. Consequently, this term cannot be absorbed immediately into the canonical logit-linear form, which is precisely what gives rise to the additional technical complication.

We address this issue by defining a correction term for the $A_2$ ratio as an offset term. Specifically, for the realized value of $A_2\in\{0,1\}$, define
\begin{equation}
\zeta_{A_2}(L_1,M_1,C)
=
-\log
\left\{
\frac{P(A_2\mid A_1=0,L_1,M_1,C)}
     {P(A_2\mid A_1=1,L_1,M_1,C)}
\right\}.
\label{eq:zetaA2}
\end{equation}
The quantity $\zeta_{A_2}(L_1,M_1,C)$ can be obtained directly because $P(A_2\mid A_1,L_1,M_1,C)$ can be correctly specified from the DGP. Equation \eqref{eq:A1_given_A2L1M1X} then becomes
\begin{equation}
P(A_1=1\mid A_2,L_1,M_1,C)
=
\expit\!\left\{
\logit\big(P(A_1=1\mid L_1,M_1,C)\big)
+
\zeta_{A_2}(L_1,M_1,C)
\right\}.
\label{eq:A1_given_A2_logistic}
\end{equation}
Since $\logit\{P(A_1=1\mid L_1,M_1,C)\}$ has a linear predictor in the specified covariates, $P(A_1=1\mid A_2,L_1,M_1,C)$ can again be modeled using logistic regression with offset $\zeta_{A_2}(L_1,M_1,C)$, whose coefficient is fixed at 1. Consequently, 
\[P(A_1\mid M_2,A_2,L_1,M_1,C) \quad \text{and} \quad P(A_1\mid L_2,M_2,A_2,L_1,M_1,C)
\]
can likewise be modeled using logistic regression with the corresponding offset term.

\subsection{Simulation 2: Misspecified parametric models versus flexible estimation}
\label{app:simu_flexible}

This section provides additional implementation details for Simulation~2 and presents Figure~\ref{fig:flex_est}, which compares misspecified GLMs with Super Learner nuisance estimation.

The Super Learner library contains \texttt{xgboost}, \texttt{ranger}, \texttt{dbarts}, \texttt{glm}, and \texttt{glm.interaction}. We used $1000$ trees for \texttt{ranger} and the default settings for the remaining candidate learners.

The figures show that misspecified GLMs induce persistent bias, whereas Super Learner substantially reduces this bias and produces estimates that remain much closer to the target. At the same time, the $n$-scaled variances of the two approaches are relatively similar. Thus, in this setting, nuisance-model misspecification primarily affects the centering of the estimators rather than their empirical dispersion. 

\begin{figure}[!htbp]
    \centering
    \includegraphics[width=0.82\textwidth]
    {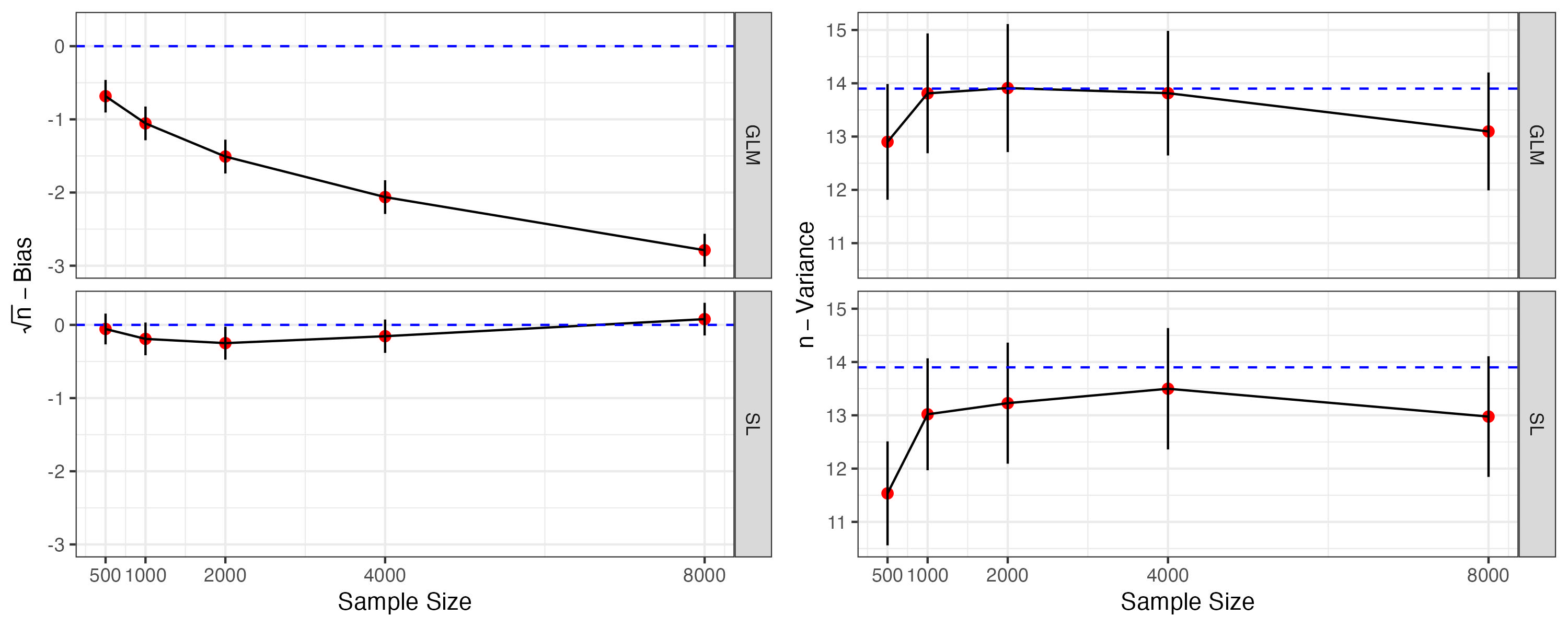}

    \vspace{0.5em}

    \includegraphics[width=0.82\textwidth]
    {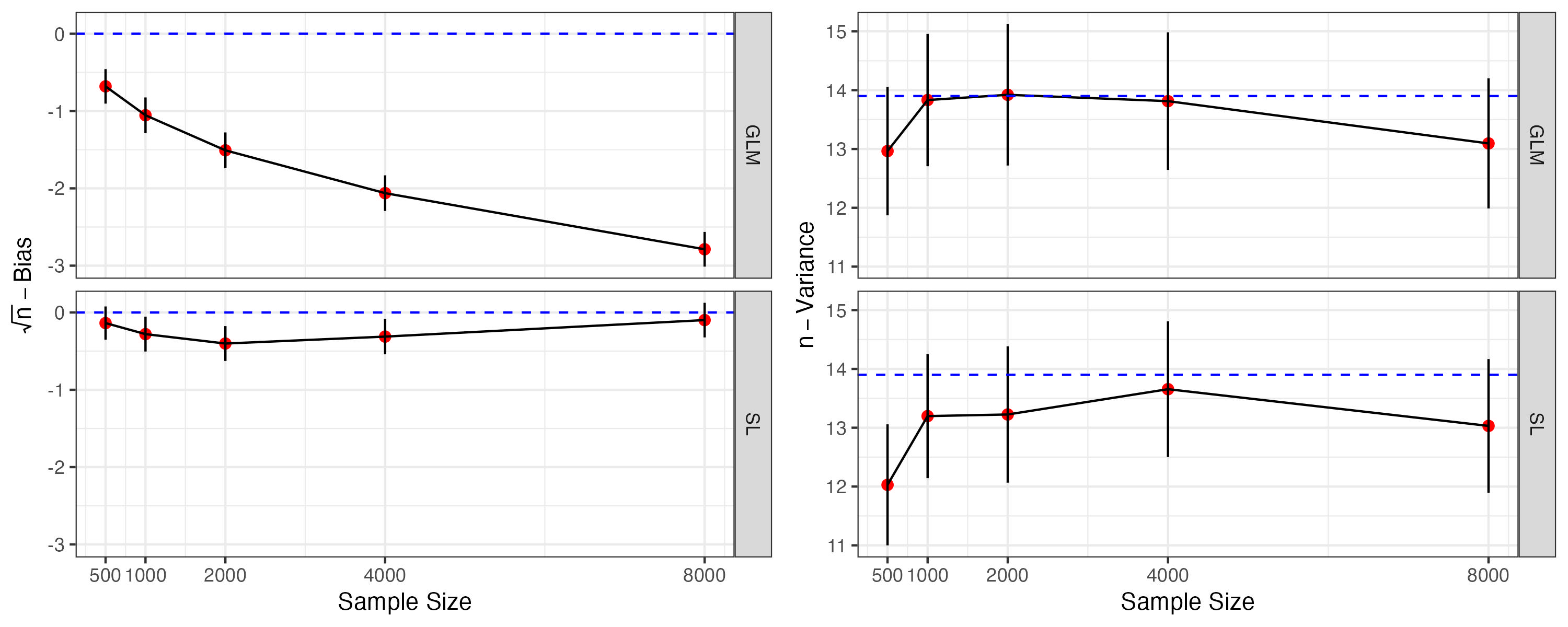}

    \caption{Root-$n$ scaled bias and $n$-scaled variance under misspecified GLMs and Super Learner nuisance estimation for the density-based estimator (upper panel) and regression-based estimator (lower panel).}
    \label{fig:flex_est}
\end{figure}

\subsection{Simulation 3: Evaluating robustness of estimators against misspecification}
\label{app:simu_misspecification}

The simulation data with variables $(C_1,C_2,C_3,C_4,A_1,M_1,L_1,A_2,M_2,L_2,Y)$ were generated to examine the robustness conditions in Corollaries~\ref{cor:robustness_1} and~\ref{cor:robustness_2} according to the following models:
{
\begin{equation}
\label{eq:simu_simplest_DGP}
\begin{aligned}
& C_1,C_2,C_3,C_4 \overset{\mathrm{iid}}{\sim} \mathrm{Uniform}(0,1),
\qquad C=\begin{bmatrix}C_1 & C_2 & C_3 & C_4\end{bmatrix}, \\
& A_1  \sim \mathrm{Bernoulli}\!\left(\expit\left( \beta_{A_1}
\begin{bmatrix}
1 & C
\end{bmatrix}^{\!T} \right) \right), \\
& M_1 \sim \mathrm{Normal}\!\left(\beta_{M_1}
\begin{bmatrix}
1 & C & A_1
\end{bmatrix}^{\!T}, \,1\right), \\
& L_1 \sim \mathrm{Normal}\!\left(
\beta_{L_1}
\begin{bmatrix}
1 & C & A_1 & M_1
\end{bmatrix}^{\!T},
\,1\right), \\
& A_2 \sim \mathrm{Bernoulli}\!\left(
\expit\left( \beta_{A_2}
\begin{bmatrix}
1 & C & A_1 & M_1 & L_1
\end{bmatrix}^{\!T}
\right) \right), \\
& M_2 \sim \mathrm{Normal}\!\left(
\beta_{M_2}
\begin{bmatrix}
1 & C & A_1 & M_1 & L_1 & A_2
\end{bmatrix}^{\!T}, \,1\right), \\
& L_2 \sim \mathrm{Normal}\!\left(
\beta_{L_2}
\begin{bmatrix}
1 & C & A_1 & M_1 & L_1 & A_2 & M_2
\end{bmatrix}^{\!T},
\,1\right), \\
& Y \sim \mathrm{Normal}\!\left(
\beta_{Y}
\begin{bmatrix}
1 & C & A_1 & M_1 & L_1 & A_2 & M_2 & L_2
\end{bmatrix}^{\!T},
\,1\right).
\end{aligned}
\end{equation}
}
The coefficient vectors are
\begin{align*}
& \beta_{A_1} = (-0.21, -0.25, 0.35, -0.26, 0.49),\\
& \beta_{M_1} =(0.50, -0.02, 0.32, 0.13, -0.14, 0.15),\\
& \beta_{L_1} =(0.33, -0.14, 0.21, 0.14, 0.46, 0.14, 0.34),\\
& \beta_{A_2} =(-0.04, 0.13, 0.26, -0.16, -0.38, 0.41, 0.12, -0.29),\\
& \beta_{M_2} =(0.24, -0.06, 0.29, 0.49, -0.15, 0.3, -0.13, 0.18, 0.48),\\
& \beta_{L_2} =(-0.11, 0.01, 0.46, 0.12, -0.04, -0.06, 0.24, 0.43, 0.45, 0.18),\\
& \beta_{Y} =(0.19, 0.55, 0.60, 0.20, 0.14, 0.65, 0.15, -0.37, 0.54, 0.66, 0.90).
\end{align*}

\subsubsection{The density-based estimator, $\psi^+(\widehat Q_1)$}
\label{app:simu_mis_density}

For $\psi^+(\widehat Q_1)$, the primitive nuisance functions can be consistently estimated via GLMs based on linear combinations of the predictors:
\begin{align*}
& \pi_1 = \expit(\alpha_1 \mt{1 & C}^{T}), 
\quad 
\pi_2 = \expit(\alpha_2 \mt{1 & C & A_1 & M_1 & L_1}^{T}), \numberthis \\
& f_{1,1} = \operatorname{dnorm}(\beta_{1,1}\mt{1 & C & A_1}^{T}, \sigma_{1,1}), 
\quad
f_{2,1} = \operatorname{dnorm}(\beta_{2,1}\mt{1 & C & A_1 & M_1 & L_1 & A_2}^{T}, \sigma_{2,1}), \\
& f_{1,2} = \operatorname{dnorm}(\beta_{1,2}\mt{1 & C & A_1 & M_1}^{T}, \sigma_{1,2}), 
\quad
f_{2,2} = \operatorname{dnorm}(\beta_{2,2}\mt{1 & C & A_1 & M_1 & L_1 & A_2 & M_2}^{T}, \sigma_{2,2}), \\
& \mu_{2,2} = \theta_{2,2}\mt{1 & C & A_1 & M_1 & L_1 & A_2 & M_2 & L_2}^{T}.
\end{align*}

The earlier sequential regressions have the population forms
\begin{align*}
\mu_{2,1} &= \theta_{2,1}\mt{1 & C & A_1 & M_1 & L_1 & A_2 & M_2}^{T}, \\
\mu_{1,2} &= \theta_{1,2}\mt{1 & C & A_1 & M_1 & L_1}^{T}, \\
\mu_{1,1} &= \theta_{1,1}\mt{1 & C & A_1 & M_1}^{T},
\qquad
\mu_{1,0} = \theta_{1,0}\mt{1 & C}^{T},
\end{align*}
but the density-based estimator does not fit these regressions independently. Instead, their estimates are obtained recursively from $\widehat\mu_{2,2}$ and the fitted intermediate-variable laws, as required by the remainder factorization underlying Corollary~\ref{cor:robustness_1}.

Let
\[
C^{\mathrm{false}}=\{C_1^2,\exp(C_2),C_1^{1/2}/C_3,\cos(10C_4)\}.
\]
Misspecified treatment mechanisms and intermediate-variable densities are obtained by replacing $C$ with $C^{\mathrm{false}}$ in the corresponding GLMs. A misspecified terminal outcome regression is similarly defined as
\[
\mu_{2,2}^{\mathrm{false}}
=
\theta_{2,2}^*
\mt{1 & C^{\mathrm{false}} & A_1 & M_1 & L_1 & A_2 & M_2 & L_2}^{T}.
\]
In every scenario, all earlier sequential regressions are induced recursively by the selected terminal outcome regression and intermediate-variable laws. We considered the following three robustness regimes.

\paragraph{Misspecified $\mathbb M$ block.}
Only $\{\pi_1,\pi_2,f_{1,2},f_{2,2}\}$ are consistently estimated. The remaining primitive nuisance functions are
\begin{align*}
f_{1,1}^{\mathrm{false}}
&=
\operatorname{dnorm}(\beta_{1,1}^*\mt{1 & C^{\mathrm{false}} & A_1}^{T},\sigma_{1,1}^*),\\
f_{2,1}^{\mathrm{false}}
&=
\operatorname{dnorm}(\beta_{2,1}^*\mt{1 & C^{\mathrm{false}} & A_1 & M_1 & L_1 & A_2}^{T},\sigma_{2,1}^*),
\end{align*}
along with $\mu_{2,2}^{\mathrm{false}}$. This corresponds to correct estimation of $(\mathcal B_\pi,\mathcal B_L)$ and misspecification of $\mathcal B_M$.

\paragraph{Misspecified treatment mechanisms.}
The terminal outcome regression and all four intermediate-variable laws are consistently estimated, while
\begin{align*}
\pi_1^{\mathrm{false}} = \expit(\alpha_1^*\mt{1 & C^{\mathrm{false}}}^{T}),
\quad & 
\pi_2^{\mathrm{false}} = \expit(\alpha_2^*\mt{1 & C^{\mathrm{false}} & A_1 & M_1 & L_1}^{T}).
\end{align*}
This corresponds to correct estimation of $(\mathcal B_M,\mathcal B_L)$ and misspecification of $\mathcal B_\pi$. Because all intermediate-variable laws and $\mu_{2,2}$ are correct, all recursively induced sequential regressions are also correct.

\paragraph{Misspecified $\mathbb L$ block.}
Only $\{\pi_1,\pi_2,f_{1,1},f_{2,1},\mu_{2,2}\}$ are consistently estimated. The misspecified $\mathbb L$-variable densities are
\begin{align*}
f_{1,2}^{\mathrm{false}}
&=
\operatorname{dnorm}(\beta_{1,2}^*\mt{1 & C^{\mathrm{false}} & A_1 & M_1}^{T},\sigma_{1,2}^*),\\
f_{2,2}^{\mathrm{false}}
&=
\operatorname{dnorm}(\beta_{2,2}^*\mt{1 & C^{\mathrm{false}} & A_1 & M_1 & L_1 & A_2 & M_2}^{T},\sigma_{2,2}^*).
\end{align*}
This corresponds to correct estimation of $(\mathcal B_\pi,\mathcal B_M)$ and misspecification of $\mathcal B_L$.

\subsubsection{The regression-based estimator, $\psi^+(\widehat Q_2)$}
\label{app:simu_mis_bayes}

For $\psi^+(\widehat Q_2)$, the nuisance functions can be consistently estimated via GLMs with linear predictors. As in Appendix~\ref{subsec:correct_GLMs}, let
$
\zeta_{A_2}
=
-\log
\left\{
\frac{P(A_2\mid A_1=0,L_1,M_1,C)}
     {P(A_2\mid A_1=1,L_1,M_1,C)}
\right\}.
$
This quantity enters the relevant $A_1$ regressions as an offset with coefficient 1. The component regressions used to construct $g_{1,1},g_{1,2},g_{2,1}$, and $g_{2,2}$, together with the treatment mechanisms and sequential outcome regressions, are
{\small 
\begin{align*}
& \pi_1  = \expit(\alpha_1 \mt{1 & C}^{T}), 
 \quad 
 \pi_2 = \expit(\alpha_2 \mt{1 & C & A_1 & M_1 & L_1}^{T}),
 \numberthis \label{eq:psi2_GLMs} \\
& g_{A_1,1,1} = \expit(\beta^{(1)}_{1,1}\mt{1 & C & M_1}^{T}), 
 \quad  
 g_{A_1,2,1} = \expit(\beta^{(1)}_{2,1}\mt{1 & C & M_1 & L_1 & A_2 & M_2 }^{T}+\zeta_{A_2}), \\
& g_{A_1,1,2} = \expit(\beta^{(1)}_{1,2}\mt{1 & C & M_1 & L_1}^{T}), 
 \quad  
 g_{A_1,2,2} = \expit(\beta^{(1)}_{2,2}\mt{1 & C & M_1 & L_1 & A_2 & M_2 & L_2 }^{T}+\zeta_{A_2}), \\
& g_{A_2,2,1} = \expit(\beta^{(2)}_{2,1}\mt{1 & C & A_1 & M_1 & L_1 & M_2}^{T}), 
 \quad 
 g_{A_2,2,2} = \expit(\beta^{(2)}_{2,2}\mt{1 & C & A_1 & M_1 & L_1 & M_2 & L_2}^{T}), \\
& \mu_{2,2} = \theta_{2,2}\mt{1 & C & A_1 & M_1 & L_1 & A_2 & M_2 & L_2}^{T},
 \quad  
 \mu_{2,1} = \theta_{2,1}\mt{1 & C & A_1 & M_1 & L_1 & A_2 & M_2}^{T}, \\
& \mu_{1,2} = \theta_{1,2}\mt{1 & C & A_1 & M_1 & L_1}^{T}, 
\quad \mu_{1,1} = \theta_{1,1}\mt{1 & C & A_1 & M_1}^{T},
\quad \mu_{1,0} = \theta_{1,0}\mt{1 & C}^{T}.
\end{align*}
}

Here $g_{A_j,k,r}$ denotes the component regression for $A_j$ entering the joint treatment-history regression $g_{k,r}$ according to equation \ref{eq:chain-rule-factorization} to \ref{eq:l2-joint-factorization}. In particular, the component regressions must be correctly specified whenever consistency of $g_{k,r}$ is required.

Let $Q_2^{\mathrm{false}}$ denote the corresponding nuisance collection obtained by replacing $C$ with
\[
C^{\mathrm{false}}=\{C_1^2,\exp(C_2),C_1^{1/2}/C_3,\cos(10C_4)\}
\]
in \eqref{eq:psi2_GLMs} and in the remaining displayed regressions. The offset $\zeta_{A_2}$ is recomputed from the treatment model used in the corresponding fit. The estimators in the different scenarios combine components from $Q_2$ and $Q_2^{\mathrm{false}}$.

We considered the following six regimes, corresponding to the two endpoints and four cut points in Corollary~\ref{cor:robustness_2}. Only the nuisance functions displayed in each regime are consistently estimated:
\begin{enumerate}
\item the regression-based endpoint,
$
\{\pi_1,\pi_2,g_{1,1},g_{1,2},g_{2,1},g_{2,2}\};
$
\item the cut at $X_{2,2}=L_2$,
$
\{\pi_1,\pi_2,g_{1,1},g_{1,2},g_{2,1},\mu_{2,2}\};
$
\item the cut at $X_{2,1}=M_2$,
$
\{\pi_1,\pi_2,g_{1,1},g_{1,2},\mu_{2,1},\mu_{2,2}\};
$
\item the cut at $X_{1,2}=L_1$,
$
\{\pi_1,g_{1,1},\mu_{1,2},\mu_{2,1},\mu_{2,2}\};
$
\item the cut at $X_{1,1}=M_1$,
$
\{\pi_1,\mu_{1,1},\mu_{1,2},\mu_{2,1},\mu_{2,2}\};
$
\item the sequential-regression endpoint,
$
\{\mu_{1,0},\mu_{1,1},\mu_{1,2},\mu_{2,1},\mu_{2,2}\}.
$
\end{enumerate}
We additionally considered all-correct and all-misspecified nuisance-model configurations. The results are presented in Figures~\ref{fig:robustness_density} and~\ref{fig:robustness_treg} of the main text. These simulations assess the consistency regimes in Corollaries~\ref{cor:robustness_1} and~\ref{cor:robustness_2}; no efficiency conclusion is implied under partial nuisance correctness.

\newpage
\section{Additional details on the real-data application}
\label{app:real_data}

The Framingham Heart Study (FHS) is a community-based longitudinal cohort study initiated in Framingham, Massachusetts, in 1948 to investigate the epidemiology of arteriosclerotic and hypertensive cardiovascular disease (CVD). The original cohort enrolled 5,209 adults, approximately 55\% of whom were women, who were examined approximately every two years. The study collected detailed information on sociodemographic characteristics, anthropometric and clinical measurements, health behaviors, and cardiovascular outcomes.

The analysis used the regression-based one-step estimator $\psi^+(\widehat Q_2)$. For each target mean, the treatment mechanisms, treatment-history regressions, and sequential outcome regressions were estimated using either generalized linear models or Super Learner. The reported 95\% Wald confidence intervals were calculated using the empirical variance of the estimated influence function for the effect, obtained as the difference between the influence functions for the two target means. No truncation was applied to the fitted treatment or treatment-history probabilities.

The analytic cohort excluded participants who died or were lost before Exam~7, those with diabetes, cancer, or CVD at baseline, and those missing analytic variables. Isolated missing BMI or smoking measurements were carried forward by one selected examination, leaving 2,850 participants in the final analytic cohort. We retained continuous intermediate variables on their original scales rather than discretizing them, since discretization can alter the target causal functional and induce non-negligible coarsening bias \citep{ou2026coarsening}.

Supplementary Table~\ref{tab:var_structure} summarizes the time-varying variable structure. The demographic and baseline health characteristics of the analytic cohort, stratified by baseline tobacco-use category, are shown in Table~\ref{app:tab:FHS_descriptive}. The cohort included 1,649 smokers, 1,023 nonsmokers, and 178 quitters. Smokers were younger and more likely to be male, whereas nonsmokers were more often female and more likely to be housewives. Quitters tended to be older and more highly educated than current smokers. Occupational and marital-status distributions also differed across groups. Baseline SBP was lowest among smokers and higher among both nonsmokers and quitters, indicating meaningful baseline heterogeneity.

\begin{table}[!htbp]
\centering
\begin{threeparttable}
\caption{Time-varying variable structure across examination cycles. The numeral following each variable name denotes the original Framingham Heart Study examination at which it was measured.}
\label{tab:var_structure}
{\small
\setlength{\tabcolsep}{4pt}
\begin{tabular}{lllll}
\toprule
& \multicolumn{4}{c}{Examination number} \\
\cmidrule(lr){2-5}
Variable type & Exam 2 & Exam 4 & Exam 5 & Exam 7 \\
\midrule
Covariates ($\overline M$)
& $M_0$: SBP1, TC2
& $M_1$: SBP2, TC4, HRX4
& $M_2$: SBP4, TC5, HRX5
& $M_3$: SBP5, TC7, HRX7 \\
Exposure ($\bf A$)
& $A_1$: SMOKE2
& $A_2$: SMOKE4
& $A_3$: SMOKE5
& $A_4$: SMOKE7 \\
Mediator ($\mathbb L$)
& $L_1$: BMI2
& $L_2$: BMI4
& $L_3$: BMI5
& $L_4$: BMI7 \\
Outcome
& & & & $Y$: SBP7 \\
\bottomrule
\end{tabular}
\begin{tablenotes}
\footnotesize
\item \textit{Note:} SBP = systolic blood pressure; TC = total cholesterol;
HRX = antihypertensive treatment; SMOKE = current smoking status; BMI = body mass index. For the target $\psi_{\mathrm{BMI}}$, $\mathbb L=\{L_1,L_2,L_3,L_4\}$ and $\mathbb M=\{M_1,M_2,M_3\}$. The full baseline covariates $C$ in our general framework comprise the baseline covariates $C_0$ and $M_0$.
\end{tablenotes}
}
\end{threeparttable}
\end{table}

\begin{table}[!htbp]
\caption{Baseline characteristics of the analytic cohort by tobacco-use category.}
\label{app:tab:FHS_descriptive}
\begin{tabular*}{\linewidth}{@{\extracolsep{\fill}}lcccc}
\toprule
\textbf{Characteristic}
& \shortstack{\textbf{Overall}\\$\boldsymbol{n=2,850}$}
& \shortstack{\textbf{Quitters}\\$\boldsymbol{n=178}$}
& \shortstack{\textbf{Nonsmokers}\\$\boldsymbol{n=1,023}$}
& \shortstack{\textbf{Smokers}\\$\boldsymbol{n=1,649}$} \\
\midrule
Sex & & & & \\
\quad Male
& 1,328 (46.6\%) & 129 (72.5\%) & 182 (17.8\%) & 1,017 (61.7\%) \\
\quad Female
& 1,522 (53.4\%) & 49 (27.5\%) & 841 (82.2\%) & 632 (38.3\%) \\
Age (years)
& $43.1\pm8.3$ & $45.6\pm8.5$ & $44.6\pm8.6$ & $41.8\pm7.9$ \\
Height (inches)
& $65.1\pm3.6$ & $66.8\pm3.5$ & $63.4\pm3.2$ & $65.9\pm3.4$ \\
Education level & & & & \\
\quad $<$ High school
& 739 (25.9\%) & 49 (27.5\%) & 319 (31.2\%) & 371 (22.5\%)  \\
\quad High school
& 1,298 (45.5\%) & 59 (33.1\%) & 428 (41.8\%) & 811 (49.2\%) \\
\quad College or higher
& 813 (28.5\%) & 70 (39.3\%) & 276 (27.0\%) & 467 (28.3\%)  \\
Occupation & & & & \\
\quad Clerical
& 173 (6.1\%) & 12 (6.7\%) & 70 (6.8\%) & 91 (5.5\%) \\
\quad Executive/supervisory
& 409 (14.4\%) & 39 (21.9\%) & 71 (6.9\%) & 299 (18.1\%) \\
\quad Housewife
& 1,028 (36.1\%) & 32 (18.0\%) & 561 (54.8\%) & 435 (26.4\%) \\
\quad Laborer
& 748 (26.2\%) & 49 (27.5\%) & 210 (20.5\%) & 489 (29.7\%) \\
\quad Professional
& 203 (7.1\%) & 17 (9.6\%) & 66 (6.5\%) & 120 (7.3\%) \\
\quad Sales
& 112 (3.9\%) & 8 (4.5\%) & 18 (1.8\%) & 86 (5.2\%) \\
\quad Technical
& 177 (6.2\%) & 21 (11.8\%) & 27 (2.6\%) & 129 (7.8\%) \\
Marital status & & & & \\
\quad Single
& 222 (7.8\%) & 11 (6.2\%) & 122 (11.9\%) & 89 (5.4\%) \\
\quad Married
& 2,479 (87.0\%) & 157 (88.2\%) & 828 (80.9\%) & 1,494 (90.6\%) \\
\quad Divorced or widowed
& 149 (5.2\%) & 10 (5.6\%) & 73 (7.1\%) & 66 (4.0\%) \\
Baseline SBP (mm Hg)
& $132.7\pm19.4$ & $135.4\pm19.5$ & $135.8\pm21.8$
& $130.5\pm17.5$ \\
\bottomrule
\end{tabular*}
\begin{minipage}{\linewidth}
\footnotesize
\textit{Note:} Values are $n$ (\%) for categorical variables and mean $\pm$ standard deviation for continuous variables.
\end{minipage}
\end{table}


\label{supp:lastpage}

\endgroup

\begin{thebibliography}{}

\bibitem[\protect\citeauthoryear{Avin, Shpitser, and Pearl}{Avin et~al.}{2005}]{avin2005identifiability}
Avin, C., Shpitser, I., and Pearl, J. (2005).
\newblock Identifiability of path-specific effects.
\newblock In {\em Proceedings of the IJCAI}, pages 357--363.

\bibitem[\protect\citeauthoryear{Bai, Wu, Sun, and Cui}{Bai et~al.}{2026}]{bai2026proximal}
Bai, Y., Wu, S., Sun, B., and Cui, Y. (2026).
\newblock Proximal path-specific inference.
\newblock {\em arXiv preprint arXiv:2605.09462} .

\bibitem[\protect\citeauthoryear{Benkeser and Ran}{Benkeser and Ran}{2021}]{benkeser2021nonparametric}
Benkeser, D. and Ran, J. (2021).
\newblock Nonparametric inference for interventional effects with multiple mediators.
\newblock {\em JCI} {\bf 9,} 172--189.

\bibitem[\protect\citeauthoryear{Daniel, De~Stavola, Cousens, and Vansteelandt}{Daniel et~al.}{2015}]{daniel2015causal}
Daniel, R.~M., De~Stavola, B.~L., Cousens, S.~N., and Vansteelandt, S. (2015).
\newblock Causal mediation analysis with multiple mediators.
\newblock {\em Biometrics} {\bf 71,} 1--14.

\bibitem[\protect\citeauthoryear{D{\'\i}az and Hejazi}{D{\'\i}az and Hejazi}{2020}]{diaz2020causal}
D{\'\i}az, I. and Hejazi, N.~S. (2020).
\newblock Causal mediation analysis for stochastic interventions.
\newblock {\em JRSS Series B: Statistical Methodology} {\bf 82,} 661--683.

\bibitem[\protect\citeauthoryear{D{\'\i}az, Williams, and Rudolph}{D{\'\i}az et~al.}{2023}]{diaz2023efficient}
D{\'\i}az, I., Williams, N., and Rudolph, K.~E. (2023).
\newblock Efficient and flexible mediation analysis with time-varying mediators, treatments, and confounders.
\newblock {\em JCI} {\bf 11,} 20220077.

\bibitem[\protect\citeauthoryear{Domingo-Relloso, Zhang, Wang, Suchy-Dicey, Buchwald, Navas-Acien, Schwartz, Berhane, Coull, and Valeri}{Domingo-Relloso et~al.}{2026}]{domingo2026path}
Domingo-Relloso, A., Zhang, Y., Wang, Z., Suchy-Dicey, A.~M., Buchwald, D.~S., Navas-Acien, A., Schwartz, J., Berhane, K., Coull, B.~A., and Valeri, L. (2026).
\newblock A path-specific effect approach to mediation analysis with time-varying mediators and time-to-event outcomes accounting for competing risks.
\newblock {\em Statistics in Medicine} {\bf 45,} e70425.

\bibitem[\protect\citeauthoryear{Lin, Young, Logan, Tchetgen~Tchetgen, and VanderWeele}{Lin et~al.}{2017}]{lin2017parametric}
Lin, S.-H., Young, J., Logan, R., Tchetgen~Tchetgen, E.~J., and VanderWeele, T.~J. (2017).
\newblock Parametric mediational g-formula approach to mediation analysis with time-varying exposures, mediators, and confounders.
\newblock {\em Epidemiology} {\bf 28,} 266--274.

\bibitem[\protect\citeauthoryear{Malinsky, Shpitser, and Richardson}{Malinsky et~al.}{2019}]{malinsky2019potential}
Malinsky, D., Shpitser, I., and Richardson, T. (2019).
\newblock A potential outcomes calculus for identifying conditional path-specific effects.
\newblock In {\em AISTATS}, pages 3080--3088. PMLR.

\bibitem[\protect\citeauthoryear{Miles, Kanki, Meloni, and Tchetgen~Tchetgen}{Miles et~al.}{2017}]{miles2017partial}
Miles, C., Kanki, P., Meloni, S., and Tchetgen~Tchetgen, E.~J. (2017).
\newblock On partial identification of the natural indirect effect.
\newblock {\em Journal of Causal Inference} {\bf 5,} 20160004.

\bibitem[\protect\citeauthoryear{Miles}{Miles}{2023}]{miles2023causal}
Miles, C.~H. (2023).
\newblock On the causal interpretation of randomised interventional indirect effects.
\newblock {\em JRSS Series B: Statistical Methodology} {\bf 85,} 1154--1172.

\bibitem[\protect\citeauthoryear{Miles, Shpitser, Kanki, Meloni, and Tchetgen~Tchetgen}{Miles et~al.}{2017}]{miles2017quantifying}
Miles, C.~H., Shpitser, I., Kanki, P., Meloni, S., and Tchetgen~Tchetgen, E.~J. (2017).
\newblock Quantifying an adherence path-specific effect of antiretroviral therapy in the {Nigeria PEPFAR} program.
\newblock {\em JASA} {\bf 112,} 1443--1452.

\bibitem[\protect\citeauthoryear{Miles, Shpitser, Kanki, Meloni, and Tchetgen~Tchetgen}{Miles et~al.}{2020}]{miles2020semiparametric}
Miles, C.~H., Shpitser, I., Kanki, P., Meloni, S., and Tchetgen~Tchetgen, E.~J. (2020).
\newblock On semiparametric estimation of a path-specific effect in the presence of mediator-outcome confounding.
\newblock {\em Biometrika} {\bf 107,} 159--172.

\bibitem[\protect\citeauthoryear{Mittinty and Vansteelandt}{Mittinty and Vansteelandt}{2020}]{mittinty2020longitudinal}
Mittinty, M.~N. and Vansteelandt, S. (2020).
\newblock Longitudinal mediation analysis using natural effect models.
\newblock {\em American Journal of Epidemiology} {\bf 189,} 1427--1435.

\bibitem[\protect\citeauthoryear{Nabi and Benkeser}{Nabi and Benkeser}{2026}]{nabi2026formulating}
Nabi, R. and Benkeser, D. (2026).
\newblock Formulating cross-world mediation estimands through single-world mixtures.
\newblock {\em arXiv preprint arXiv:2609.27075} .

\bibitem[\protect\citeauthoryear{Pearl}{Pearl}{2001}]{pearl2001direct}
Pearl, J. (2001).
\newblock Direct and indirect effects.
\newblock In {\em Proceedings of the UAI}, page 411–420.

\bibitem[\protect\citeauthoryear{Pearl}{Pearl}{2009}]{pearl2009causality}
Pearl, J. (2009).
\newblock {\em Causality: Models, Reasoning, and Inference}.
\newblock Cambridge University Press, 2nd edition.

\bibitem[\protect\citeauthoryear{Robins and Greenland}{Robins and Greenland}{1992}]{robins1992identifiability}
Robins, J.~M. and Greenland, S. (1992).
\newblock Identifiability and exchangeability for direct and indirect effects.
\newblock {\em Epidemiology} {\bf 3,} 143--155.

\bibitem[\protect\citeauthoryear{Robins and Richardson}{Robins and Richardson}{2010}]{robins2010alternative}
Robins, J.~M. and Richardson, T.~S. (2010).
\newblock Alternative graphical causal models and the identification of direct effects.
\newblock {\em Causality and psychopathology: Finding the determinants of disorders and their cures} {\bf 84,} 103--158.

\bibitem[\protect\citeauthoryear{Shpitser}{Shpitser}{2013}]{shpitser2013counterfactual}
Shpitser, I. (2013).
\newblock Counterfactual graphical models for longitudinal mediation analysis with unobserved confounding.
\newblock {\em Cognitive science} {\bf 37,} 1011--1035.

\bibitem[\protect\citeauthoryear{Shpitser and Tchetgen~Tchetgen}{Shpitser and Tchetgen~Tchetgen}{2016}]{shpitser2016causal}
Shpitser, I. and Tchetgen~Tchetgen, E.~J. (2016).
\newblock Causal inference with a graphical hierarchy of interventions.
\newblock {\em Annals of statistics} {\bf 44,} 2433.

\bibitem[\protect\citeauthoryear{Steen, Loeys, Moerkerke, and Vansteelandt}{Steen et~al.}{2017}]{steen2017flexible}
Steen, J., Loeys, T., Moerkerke, B., and Vansteelandt, S. (2017).
\newblock Flexible mediation analysis with multiple mediators.
\newblock {\em American journal of epidemiology} {\bf 186,} 184--193.

\bibitem[\protect\citeauthoryear{Stensrud, Young, Didelez, Robins, and Hern{\'a}n}{Stensrud et~al.}{2022}]{stensrud2022separable}
Stensrud, M., Young, J., Didelez, V., Robins, J., and Hern{\'a}n, M. (2022).
\newblock Separable effects for causal inference in the presence of competing events.
\newblock {\em JASA} {\bf 117,} 175--183.

\bibitem[\protect\citeauthoryear{Stensrud, Hern{\'a}n, Tchetgen~Tchetgen, Robins, Didelez, and Young}{Stensrud et~al.}{2021}]{stensrud2021generalized}
Stensrud, M.~J., Hern{\'a}n, M.~A., Tchetgen~Tchetgen, E.~J., Robins, J.~M., Didelez, V., and Young, J.~G. (2021).
\newblock A generalized theory of separable effects in competing event settings.
\newblock {\em Lifetime data analysis} {\bf 27,} 588--631.

\bibitem[\protect\citeauthoryear{Tai, Lin, Chu, Yu, Puhan, and VanderWeele}{Tai et~al.}{2023}]{tai2023causal}
Tai, A.-S., Lin, S.-H., Chu, Y.-C., Yu, T., Puhan, M.~A., and VanderWeele, T. (2023).
\newblock Causal mediation analysis with multiple time-varying mediators.
\newblock {\em Epidemiology} {\bf 34,} 8--19.

\bibitem[\protect\citeauthoryear{Tchetgen~Tchetgen and Phiri}{Tchetgen~Tchetgen and Phiri}{2014}]{tchetgen2014bounds}
Tchetgen~Tchetgen, E.~J. and Phiri, K. (2014).
\newblock Bounds for pure direct effect.
\newblock {\em Epidemiology} {\bf 25,} 775--776.

\bibitem[\protect\citeauthoryear{Tchetgen~Tchetgen and Shpitser}{Tchetgen~Tchetgen and Shpitser}{2012}]{tchetgen2012semiparametric}
Tchetgen~Tchetgen, E.~J. and Shpitser, I. (2012).
\newblock Semiparametric theory for causal mediation analysis: efficiency bounds, multiple robustness, and sensitivity analysis.
\newblock {\em Annals of statistics} {\bf 40,} 1816.

\bibitem[\protect\citeauthoryear{VanderWeele, Vansteelandt, and Robins}{VanderWeele et~al.}{2014}]{vanderweele2014effect}
VanderWeele, T., Vansteelandt, S., and Robins, J. (2014).
\newblock Effect decomposition in the presence of an exposure-induced mediator-outcome confounder.
\newblock {\em Epidemiology} {\bf 25,} 300--306.

\bibitem[\protect\citeauthoryear{VanderWeele and Tchetgen~Tchetgen}{VanderWeele and Tchetgen~Tchetgen}{2017}]{vanderweele2017mediation}
VanderWeele, T.~J. and Tchetgen~Tchetgen, E.~J. (2017).
\newblock Mediation analysis with time varying exposures and mediators.
\newblock {\em JRSS Series B: Statistical Methodology} {\bf 79,} 917--938.

\bibitem[\protect\citeauthoryear{Vansteelandt and Daniel}{Vansteelandt and Daniel}{2017}]{vansteelandt2017interventional}
Vansteelandt, S. and Daniel, R.~M. (2017).
\newblock Interventional effects for mediation analysis with multiple mediators.
\newblock {\em Epidemiology} {\bf 28,} 258--265.

\bibitem[\protect\citeauthoryear{Wu, Bai, and Cui}{Wu et~al.}{2026}]{wu2026proximal}
Wu, S., Bai, Y., and Cui, Y. (2026).
\newblock Proximal mediation analysis with hidden recanting witnesses.
\newblock {\em arXiv preprint arXiv:2606.17600} .

\bibitem[\protect\citeauthoryear{Zheng and van~der Laan}{Zheng and van~der Laan}{2017}]{zheng2017longitudinal}
Zheng, W. and van~der Laan, M. (2017).
\newblock Longitudinal mediation analysis with time-varying mediators and exposures, with application to survival outcomes.
\newblock {\em JCI} {\bf 5,} 20160006.

\bibitem[\protect\citeauthoryear{Zhou}{Zhou}{2022}]{zhou2022semiparametric}
Zhou, X. (2022).
\newblock Semiparametric estimation for causal mediation analysis with multiple causally ordered mediators.
\newblock {\em JRSS Series B: Statistical Methodology} {\bf 84,} 794--821.

\bibitem[\protect\citeauthoryear{Zhou and Yamamoto}{Zhou and Yamamoto}{2023}]{zhou2023tracing}
Zhou, X. and Yamamoto, T. (2023).
\newblock Tracing causal paths from experimental and observational data.
\newblock {\em The Journal of Politics} {\bf 85,} 250--265.

\end{thebibliography}

\vspace{2cm}
\begin{thebibliography}{}

\bibitem[\protect\citeauthoryear{Chen}{Chen}{2007}]{chen07semiparametric}
Chen, H.~Y. (2007).
\newblock A semiparametric odds ratio model for measuring association.
\newblock {\em biometrics} {\bf 63,} 413--421.

\bibitem[\protect\citeauthoryear{Ou and Nabi}{Ou and Nabi}{2026}]{ou2026coarsening}
Ou, X. and Nabi, R. (2026).
\newblock Coarsening bias from variable discretization in causal functionals.
\newblock In {\em Proceedings of the 42nd Conference on Uncertainty in Artificial Intelligence}, volume 337, pages 5138--5167. PMLR.

\end{thebibliography}
\end{document}